\documentclass[proceedings]{JHEP37}
\usepackage{eurosym}
\usepackage{amssymb}
\usepackage{amsfonts,cite}
\usepackage{amsmath}
\usepackage{graphicx}
\usepackage{multirow}
\usepackage{epsfig}
\usepackage[utf8]{inputenc}
\usepackage{multicol}
\usepackage{graphicx}
\usepackage{wrapfig}
\usepackage{placeins}
\usepackage{slashed}
\usepackage{lipsum}
\usepackage{array}
\usepackage[thinlines]{easytable}
\usepackage{mathrsfs}
\usepackage[export]{adjustbox}
\usepackage{blindtext}

\newbox\mybox

\newcommand\fverb{\setbox\mybox=\hbox\bgroup\verb}
\newcommand\fverbdo{\egroup\medskip\noindent\fbox{\unhbox\mybox}\ }
\newcommand\fverbit{\egroup\item[\fbox{\unhbox\mybox}]}
\conference{Higher derivative Hamiltonians with benign ghosts from  affine Toda lattices }
\abstract{We provide further evidence for Smilga's conjecture that higher charges of integrable systems are suitable candidates for higher derivative theories that possess benign ghost sectors in their parameter space. As concrete examples we study the properties of the classical phase spaces for a number of affine Toda lattices theories related to different types of Kac-Moody algebras. We identify several types of scenarios for theories with higher charge Hamiltonians: some that possess oscillatory, divergent, benign oscillatory and benign divergent behaviour when ghost sectors are present in the quantum theory. No divergent behaviour was observed for which the trajectories reach a singularity in finite time. For theories based on particular representations for the Lie algebraic roots we found an extreme sensitivity towards the initial conditions governed by the Poisson bracket relations between the centre-of-mass coordinate and the charges.}

\title{Higher derivative Hamiltonians with benign ghosts from  affine Toda lattices}
\author{Andreas Fring and Bethan Turner\\
 Department of Mathematics, City, University of London, Northampton Square,\\ London EC1V 0HB, UK \\
a.fring@city.ac.uk, bethan.turner.2@city.ac.uk}

\input{tcilatex}
\begin{document}

\section{Introduction}
Higher derivative Lagrangian theories, i.e. those that include derivative terms of the coordinates of order larger than one, arise naturally in a number of different contexts. For instance, in some approaches to theories of everything (TOE) that include gravity besides all the other known fundamental forces consist of embedding the standard (3+1)-dimensional universe into a higher dimensional space. In doing so, and demanding in addition these theories to be renormalizable, one is automatically led to higher derivative Lagrangian theories by simple scaling arguments. Unfortunately these theories are generally plagued \cite{raidal2017quantisation} by so-called ghosts states that possess negative norms, thus leading to collapse and/or a violation of unitarity. This is the main reason why they are usually discarded and in comparison only very few explicit studies of these theories have been carried out to a full extent. For instance, in the field of gravity and cosmology they have been proposed as a resolution of the cosmological singularity problem  \cite{biswas2010towards} and some of their black holes solutions have been studied \cite{mignemi1992black}. Furthermore, for some cases the
BRST symmetries have been identified \cite{rivelles2003triviality}, they were explored in a massless particle description of bosons and fermions \cite{Mpl}
and also some supersymmetric versions have been studied \cite{dine1997comments}.

However, in general such types of theories remain to be regarded as undesirable, or at least unpopular, for the above mentioned reason and it remains unclear which theories deserve further considerations. In a recent series of papers \cite{smilga2005benign,smilga2021exactly,Smilga6,smilga2021benign} Smilga and collaborators addressed this question and gathered evidence to suggest that the dismissal of higher derivative theories might be too premature. The central idea in these studies is to distinguish between benign and malevolent ghost states in the sense that the latter states are genuinely unphysical while the former are solutions that might not be bounded from below, but are oscillatory in character, hence allowing for a unitary evolution. The next question is then of course how to identify theories that have such features and possess sectors in their parameter space with benign ghost solutions that in addition might be stable against small perturbations. Very recently Smilga  \cite{smilga2005benign} proposed that higher charges of integrable systems might be suitable candidates for such types of higher derivative Lagrangian theories. Here our main goal is to gather further evidence for this conjecture by considering a particular class of integrable systems and interpret their charges as Hamiltonians for higher derivative theories. We will analyse their classical phase spaces in the hope that benign classical systems will also lead to benign quantum systems as conjectured in \cite{smilga2005benign}.  

Here we will exclusively focus on higher derivatives in space rather than time, that however, may or may not have ghost states in their spectrum. As different types of terminology can be found in the literature we state here our nomenclature. For theories with no ghost states we encounter the usual {\em oscillatory} and {\em divergent} behaviour in phase space, i.e. trajectories that are confined in phase space and run to infinity in infinite time, respectively. For theories that possess ghost states we distinguish between two different types of divergent behaviour, {\em benign divergent} and {\em malevolent}, where the former describe trajectories that run to infinity in infinite time and the latter trajectories that reach a singularity in finite time.   

In general, we will be considering here a prototype integrable theory that is affine Toda lattices with Hamiltonians of the form
\begin{equation}
	 H_{\bf{g}} =  \sum_{i=1}^\ell \frac{ p_i^2}{2} + \sum_{i=0}^r   n_i e^{\alpha_i \cdot q}  ,  \label{H1}
\end{equation}	
where $q=(q_1, \ldots, q_{\ell})$ are the coordinates, $p=(p_1, \ldots, p_{\ell})$ are the momenta, $\bf{g}$ is a semi-simple Lie algebra, $r$ the rank of this algebra, $\alpha_i$ for $i=1, \ldots , r$ are the simple roots of the root space $\Delta_{\bf{g}}$ represented in an $\ell$-dimensional space, $\alpha_0 =- \sum_{i=1}^r n_i \alpha_i $ and $n_i \in \mathbb{N}$ are positive integers with $n_0=1$. The choice of $\alpha_0$ ensures that the minimum of the potential of the theory is at $q=(q_1, \ldots, q_{\ell})=(0,\ldots,0)$, i.e. all first order terms in the $q_i$ vanish. Often $\alpha_0$ is taken to be the negative of the highest root, so that the integers $n_i$ are the Kac labels, but this need not be the case and is a mere convention. The inclusion of the $\alpha_0$-root means that the associated algebra becomes a Kac-Moody algebra rather than a semi-simple Lie algebra. Thus we are not considering here theories of the type $ H_{\bf{g}}$ with the sum in the potential starting at $i=1$, which are conformally invariant and do not possess minima in the potentials at finite values of the coordinates.

It is well known \cite{OP2,mikhailov1981two,OP4} that these type of theories are integrable in the Liouville sense, that is they possess as many conserved charges as degrees of freedom. It is these charges that we will be using as potential candidates for higher order derivative theories. The key question we will be addressing here is whether the classical trajectories in phase space associated to the Hamiltonian systems of these charges will be benign or malevolent according to the characterisation put forward by Smilga in \cite{smilga2005benign,smilga2021exactly,Smilga6} and specified in more detail above. The initial assumption is that the benign nature on the classical level is inherited in the quantum theory. Naturally, this supposition needs further investigation, which we leave for future studies.

Our manuscript is organised as follows: In section 2 we recall the constructions of the conserved classical charges for the $A_n$-affine Toda lattice theories, with a particular focus on $A_2$ and $A_6$ for different types, i.e. dimensions, of representations of the roots in (\ref{H1}). Interpreting these charges as Hamiltonians we  numerically study their classical solutions in phase space. In section 3 and 4 we carry out similar type of studies for the $B_3$ and $G_2$-affine Toda lattice theories, respectively. We construct relevant charges from a reduction/folding procedure of the corresponding root systems or by direct computation. In section 5 we investigate the stability of the benign solutions with regard to the sensitivity of the initial conditions and to strong deformations by harmonic oscillator potentials. Our conclusions are stated in section 6. 

\section{Higher derivative Hamiltonians from $A_n$-affine Toda lattice charges}
The expressions for the higher charges are central to our investigations and therefore we will provide here their explicit construction. All higher charges that will be considered are for theories associated with Hamiltonians of the general form in equation (\ref{H1}) with $\bf{g}$ taken to be $A_n$. Using the standard Lax approach for classical integrable systems \cite{Lax} we employ the Lax pair given by the two operators in form of $(n+1) \times (n+1)$-matrices
\begin{equation}
	L=\left(
	\begin{array}{ccccccc}
		p_1 & W_1 & 0 & \cdots  & \cdots  & 0 & W_0 \\
		W_1 & p_2 & W_2 &  &   & 0 & 0 \\
		0 & W_2 & p_3 & \ddots &   &   & \vdots  \\
		\vdots  &   & \ddots & \ddots & \ddots &   & \vdots  \\
		\vdots  &   &   & \ddots & \ddots & \ddots & 0 \\
		0 & 0 &  &   & \ddots & p_n & W_n \\
		W_0 & 0 & \cdots  & \cdots  & 0 & W_n & p_{n+1} \\
	\end{array}
	\right),  \qquad M=\left(
	\begin{array}{ccccccc}
		0 & W_1 & 0 & \cdots  & \cdots  & 0 & -W_0 \\
		-W_1 & 0 & W_2 &   &   & 0 & 0 \\
		0 & -W_2 & 0 & \ddots &   &   & \vdots  \\
		\vdots  &   & \ddots & \ddots & \ddots &  & \vdots  \\
		\vdots  &   &   & \ddots & \ddots & \ddots & 0 \\
		0 & 0 &   &   & \ddots & 0 & W_n \\
		W_0 & 0 & \cdots  & \cdots  & 0 & -W_n & 0 \\
	\end{array}
	\right), \label{LM}
\end{equation}
where we abbreviated $W_i:=\exp(\alpha_i \cdot q)/2$, with $\alpha_i \in \mathbb{R}^{n+1}$, $i=1,\ldots, n$ denoting the simple roots of $A_n$, $\alpha_0 = -\sum_{i=1}^n \alpha_i$ the negative of the highest $A_n$-root, $q=(q_1, \ldots, q_{n+1})$ the coordinates and $p=(p_1, \ldots, p_{n+1})$ the momenta. The dimension of the phase space is therefore $(n+1) \times (n+1)$ at this point. 

By definition of the Lax operators, the equations of motion are then equivalent to the Lax pair equation
\begin{equation}
 \dot{L}+[M,L] =0, \quad \Leftrightarrow \quad 
	\dot{p}_i+W_i^2-W_{i-1}^2  =0, \quad  \alpha_i \cdot \dot{q} = p_i -p_{i+1},    \qquad i=1, \ldots , n+1 , \label{laxp}
\end{equation}
where we formally identified $W_{n+1}=W_0$. As usual we denote here derivatives with respect to time by overdots. Taking $\ell = n+1$ in (\ref{H1}) these equations also correspond to Hamilton's equations $\dot{q}_i = \partial H / \partial p_i$, $\dot{p}_i = -\partial H / \partial q_i$ as we will show below. By construction, it then follows immediately that all quantities $Q_k:= Tr(L^k)/k$ are conserved in time, i.e. $\dot{Q}=0$. Given the expressions in (\ref{LM}) we easily construct all of these charges. Taking $n=6$ as an example and interpreting the summation indices modulo 7, e.g. $W_7=W_0$, $p_8=p_1$, etc, we obtain
\begin{eqnarray}
	Q_1 &=& \sum_{i=1}^7 p_i, \label{chargeq1}\\
	Q_2 &=&H= \sum_{i=1}^7 \left( \frac{p_i^2}{2} + W_i^2 \right), \label{chargeq2} \\
    Q_3 &=& \sum_{i=1}^7 \left[ \frac{p_i^3}{3} + W_i^2 (p_i + p_{i+1}) \right], \label{chargeq3}\\
    Q_4 &=& \sum_{i=1}^7 \left[ \frac{p_i^4}{4}  + \frac{1}{2} W_i^4 + W_i^2 (p_i^2 + p_i p_{i+1} + p_{i+1}^2) + W_i^2 W_{i+1}^2 \right] \label{chargeq4} \\
    Q_5 &=& \sum_{i=1}^7 \left[ \frac{p_i^5}{5} + W_i^4 (p_i + p_{i+1}) 
          + W_i^2 (p_i^3 + p_i p_{i+1}^2 + p_i^2 p_{i+1} + p_{i+1}^3 ) \right. \label{chargeq5} \\
          && \left. \qquad  +  W_i^2 W_{i+1}^2(p_{i-1} + 2 p_i + 2 p_{i+1}) \right] , \notag\\ 
    Q_6 &=& \sum_{i=1}^7 \left[ \frac{p_i^6}{6}  + \frac{1}{2} W_i^6 + W_i^4 \left( \frac{3}{2} p_i^2+ \frac{3}{2} p_{i+1}^2 + 2 p_i p_{i+1} \right)
    + W_i^2 \left( p_i^4 + p_i^3 p_{i+1}  +p_i^2 p_{i+1}^2   \right)  \right. , \qquad \label{chargeq6}\\
     &&  \qquad  +  W_{i-1}^2 \left( p_i^4 + p_i^3 p_{i-1}    \right) 
     +  W_i^2 W_{i+1}^2 \left(  p_i^2 + 2 p_i p_{i-1} + 3 p_{i+1}^2 + p_i p_{i+2} + 2 p_{i+1} p_{i+2} +  p_{i+2}^2 \right) \notag \\
      && \left. \qquad  +  W_{i}^4 \left( W_{i-1}^2 + W_{i+1}^2    \right) + W_{i}^2 W_{i+1}^2  W_{i+2}^2 \right] \notag \\
 Q_7 &=&\sum_{i=1}^7 \left[ \frac{p_i^7}{7} + W_i^6 (p_i + p_{i+1})   
       + W_{i-1}^2 W_{i}^2 W_{i+1}^2(p_{i-1}+2 p_i + 2 p_{i+1} + p_{i+2})  \right. \label{chargeq7} \\ 
 &&  \qquad  + W_i^4 W_{i-1}^2   (p_{i-1} +3 p_i + 2 p_{i+1} )
 + W_{i-1}^4 W_{i}^2 (2p_{i-1}+ 3 p_i +  p_{i+1}) \notag \\
  &&  \qquad  +  W_{i-1}^2 W_{i}^2 (p_{i-1}^3 + 4 p_i^3 + 3 p_i^2 p_{i+1}+ 2 p_i p_{i+1}^2 + p_{i+1}^3)  \notag\\
   &&  \qquad  +  W_{i-1}^2 W_{i}^2 (p_{i-1}^3 (p_{i-1}^2 (2 p_i + p_{i+1}) + p_{i-1} (3 p_i^2 + 2 p_i p_{i+1} + p_{i+1}^2) ) \notag\\
   &&  \qquad  \left. +  W_{i}^2 ( p_i^5 +p_i^4 p_{i+1} + p_i^3 p_{i+1}^3 + p_i^2 p_{i+1}^3 + p_i p_{i+1}^4 +p_{i+1}^5) \right] + 2\prod_{i=1}^7 W_i  . \notag
\end{eqnarray}
These charges, and versions thereof, will be our potential candidates for higher derivative theories with regard to time when interpreted as Hamiltonians. 

\subsection{Higher derivative Hamiltonians from the 3 particle $A_2$-affine Toda lattice}
Next we evaluate the expressions of the charges for the $A_2$-theory more explicitly. First we notice that the second equation in (\ref{laxp}) is simply solved by taking $q=(q_1,q_2,q_3)$, so that we obtain $\dot{q}_i = p_i$ when the roots are represented as $\alpha_1 = (1,-1,0)$, $\alpha_2 = (0,1,-1)$ and $\alpha_0 =-\alpha_1-\alpha_2 =(-1,0,1)$. This is the standard three dimensional representation for the $A_2$-roots, see for instance \cite{bourbaki1968groupes}.
The charges (\ref{chargeq1})-(\ref{chargeq4}) then acquire the form
\begin{eqnarray}
	Q_1&=& p_1 +p_2 + p_3, \label{Q1}\\
	Q_2&=&H= \frac{1}{2} \left(p_1^2 +p_2^2 + p_3^2 \right)+ V_{12}+ V_{23}+ V_{31}= \sum_{i=1}^3 \left(\frac{ p_i^2}{2} +    e^{\alpha_i \cdot q}  \right)   , \label{QH} \\
	Q_3&=& \frac{1}{3} \left(p_1^3 +p_2^3 + p_3^3 \right) +
	p_1 \left( V_{12}+ V_{31} \right) + p_2 \left( V_{12}+ V_{23} \right) + p_3 \left( V_{23}+ V_{ 31} \right) +2, \label{Q3}\\
	&=&  \sum_{i=1}^3 \left[\frac{ p_i^3}{3} + p_i \left(    e^{\alpha_i \cdot q} + e^{\alpha_{i-1} \cdot q} \right)  \right] +2, \\
	Q_4&=&\frac{Q_1^4}{24}-\frac{Q_1^2 Q_2}{2}+Q_1 Q_3+\frac{Q_2^2}{2},\\
	Q_5&=& \frac{1}{5}tr(L^5)= \frac{Q_1^5}{80}-\frac{Q_1^3 Q_2}{12}-\frac{Q_1 Q_2^2}{4}+\frac{Q_1 Q_4}{2}+Q_2 Q_3,
\end{eqnarray}
where we introduced the new abbreviation $V_{ij}:=\exp(q_i-q_j)$. We notice that only the first three conserved quantities are independent, as $Q_4$ and $Q_5$ can be constructed from combinations of them. In fact, this property will persist for higher charges and all $Q_i$ for $i > 3 $ can be build from combinations of $Q_1$, $Q_2$ and $Q_3$.    

 Moreover, one may easily verify that all charges, in particular the independent ones $Q_1$, $Q_2$ and $Q_3$, are in involution, i.e. their mutual Poisson brackets vanish
 \begin{equation}
 	\left\{ Q_i, Q_j  \right\}:= \sum_{k=1}^3\frac{\partial Q_i}{\partial q_k}  \frac{\partial Q_j}{\partial p_k} -\frac{\partial Q_i}{\partial p_k}  \frac{\partial Q_j}{\partial q_k} =0, \qquad \text{for} \,i,j = 1,2,3. 
 \end{equation}
Interpreting $Q_{n+1}$ as Hamiltonian and defining the centre-of-mass coordinate $\chi:=q_1+q_2+q_3$, we observe that
\begin{eqnarray}
	\frac{d \chi}{dt}= \left\{ \chi , Q_{n+1}   \right\} = n Q_n , \quad n=1,2,3,4, \dots  \label{Poisson}
\end{eqnarray}
This means that $\chi$ grows linearly in time with a slope $ n Q_n$. Convergence or oscillatory behaviour can be achieved when choosing initial values for which $Q_n$ vanishes.  

Recalling the $\mathbb{Z}_2$-symmetry $\tau : \alpha_i \leftrightarrow \alpha_{n+1-i}$ of the $A_n$ Dynkin diagram, this amounts to $\tau : q_i \leftrightarrow -q_{n+2-i} $ for the coordinates with the roots taken in the standard representation. Appealing now briefly to the quantum theory and adapting the argument in \cite{smilga2021benign}, this means when representing this symmetry by an operator $\tau$ it anti-commutes with the odd charges and commutes with the even charges, i.e. $\{ \tau, Q_{2n+1} \}=0$ and $[ \tau, Q_{2n} ]=0$, respectively. This implies that the spectrum for the odd charges is not bounded from below, i.e. it will have ghost states, as for each eigenvalue $\lambda$ of $Q_{2n+1}$ with eigenstate $\psi$ also $-\lambda$ will be an eigenvalue with eigenstate $\tau \psi$, since $  Q_{2n+1} \tau \psi = -\tau Q_{2n+1} \psi = - \lambda \tau \psi$.

We now solve the equations of motion for different Hamiltonians. As indicated in (\ref{QH}), at first we identify as usual the charge $Q_2$ with the standard Hamiltonian so that the classical equations of motion result from Hamilton's equations to 
\begin{equation}
	\dot{q}_1 = p_1, \quad 	\dot{q}_2 = p_2, \quad \dot{q}_3 = p_3 \quad 
 \dot{p}_1 = V_{31} - V_{12}, \quad \dot{p}_2 = V_{12} - V_{23}, \quad \dot{p}_3 = V_{23} - V_{31} , \label{eomst}
\end{equation}
which are identical to the equations compatible with the Lax pair equation (\ref{laxp}).

In the first instance we solve these equations numerically. 

\begin{figure}[h]
	\centering         
	\begin{minipage}[b]{0.49\textwidth}           \includegraphics[width=\textwidth]{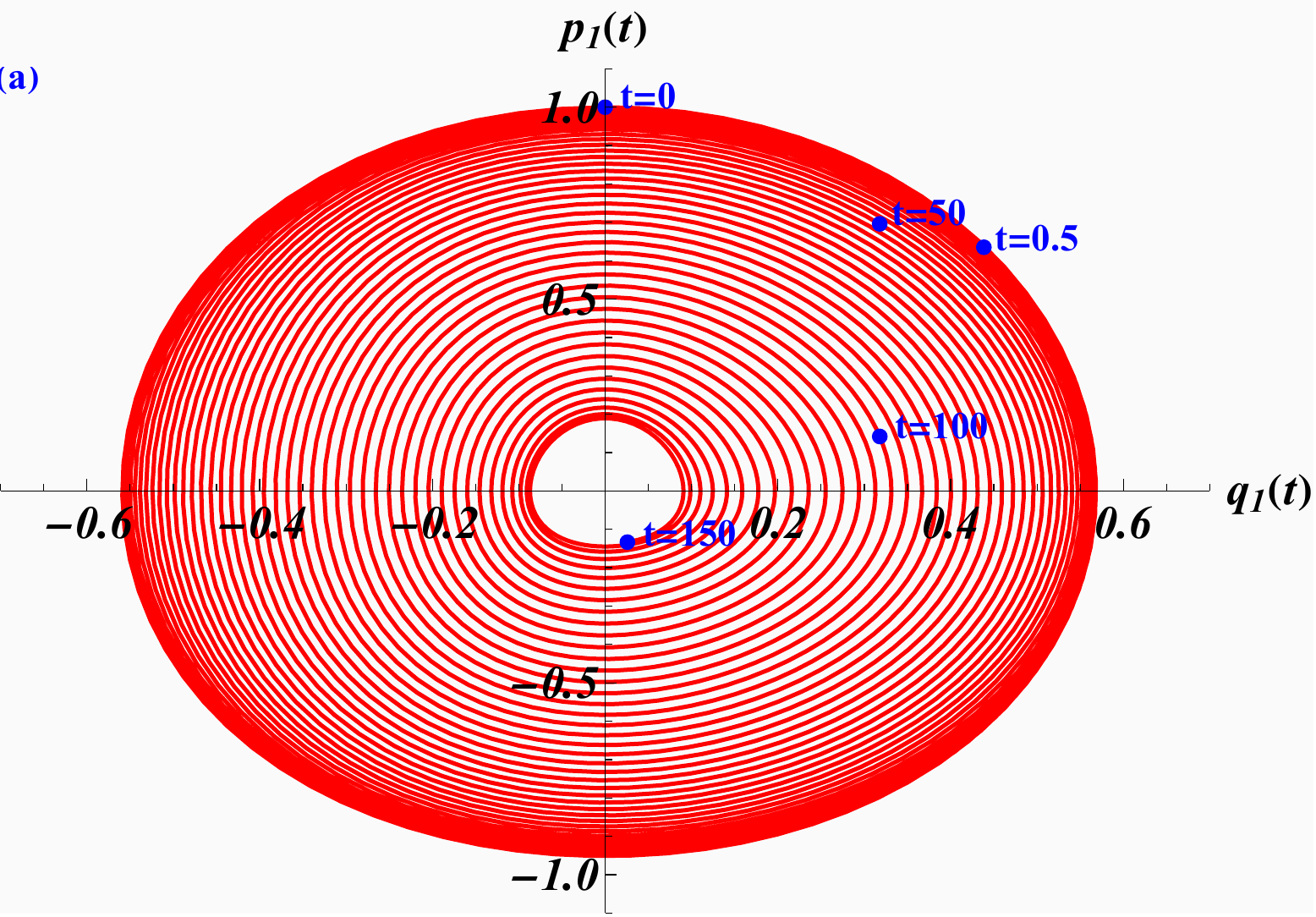}
	\end{minipage}   
	\begin{minipage}[b]{0.49\textwidth}           
		\includegraphics[width=\textwidth]{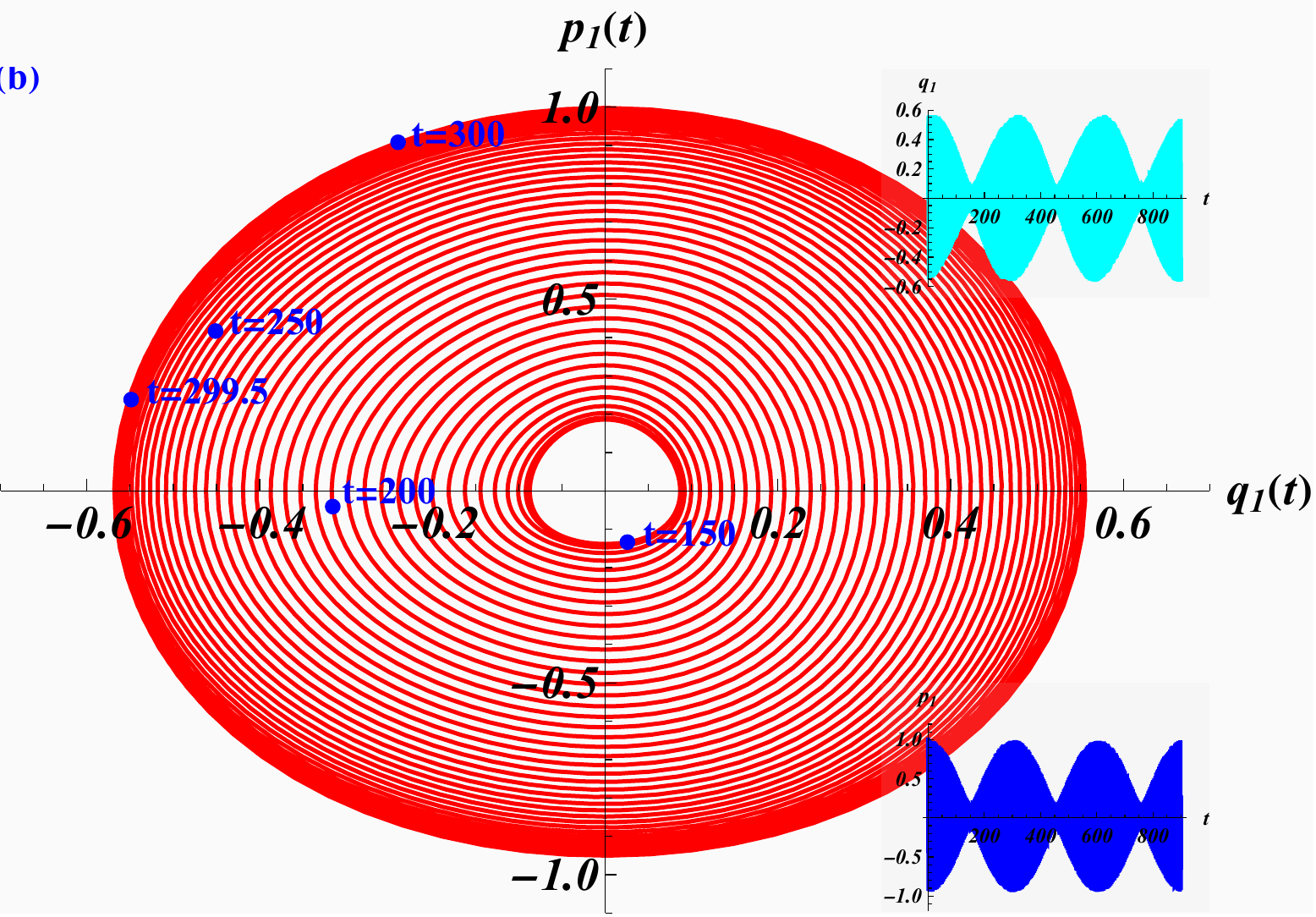}
	\end{minipage}  
	\caption{Phase space $(q_1,p_1)$ for the $A_2$-affine Toda lattice Hamiltonian with three particles. Panel (a): inward spiralling trajectories from time $t=0$ to $t=150$. Panel (b): outward spiralling trajectories from time $t=150$ to $t=300$.  The initial conditions are taken as $q_1(0)=q_2(0)=q_3(0)=0$, $p_1(0)=1$ and $p_2(0)=p_3(0)=-1/2$, i.e. $Q_1=0$. The insets in panel (b) show $x_1$ and  $p_1$ as functions of time $t$.} 
	\label{Phase1}
\end{figure}

In figure \ref{Phase1} we depict the solutions to the three particle equations of motion (\ref{eomst}) in phase space, observing confined orbits that periodically spiral inward and outward, a behaviour that continues beyond the time shown in the figure. The insets in figure \ref{Phase1} panel (b) demonstrate how the small period $\tau_s \approx 1.778$ is modulated by a larger period $\tau_l \approx 300.2$, with $\tau_s$ governing the quasiperiodic elliptic motion and $\tau_l$ the period of the inward/outward pulsation. We stress here that after each period we observe a small offset and therefore these solutions are not exactly periodic and only quasiperiodic, i.e. $f(x+\tau) = g(x,f(x))$ with $g$ being a simpler function than $f$ or almost periodic in the sense of \cite{bohr1925theorie}. Almost periodic is here to be understood in the sense that we have a small offset after one period, i.e. $\vert f(t) - f(t+ \tau) \vert \leq \varepsilon$. We may adapt these observations more rigorously to the strict sense of the definition of almost periodic functions by H. Bohr \cite{bohr1925theorie} and adjust the values of $\tau_H$ for a pre-selected $\epsilon$. For the other directions in phase space $(q_2,p_2)$ and $(q_3,p_3)$ we obtain similar types of periodic behaviour. 

\begin{figure}[h]
	\centering         
	\begin{minipage}[b]{0.68\textwidth}           \includegraphics[width=\textwidth]{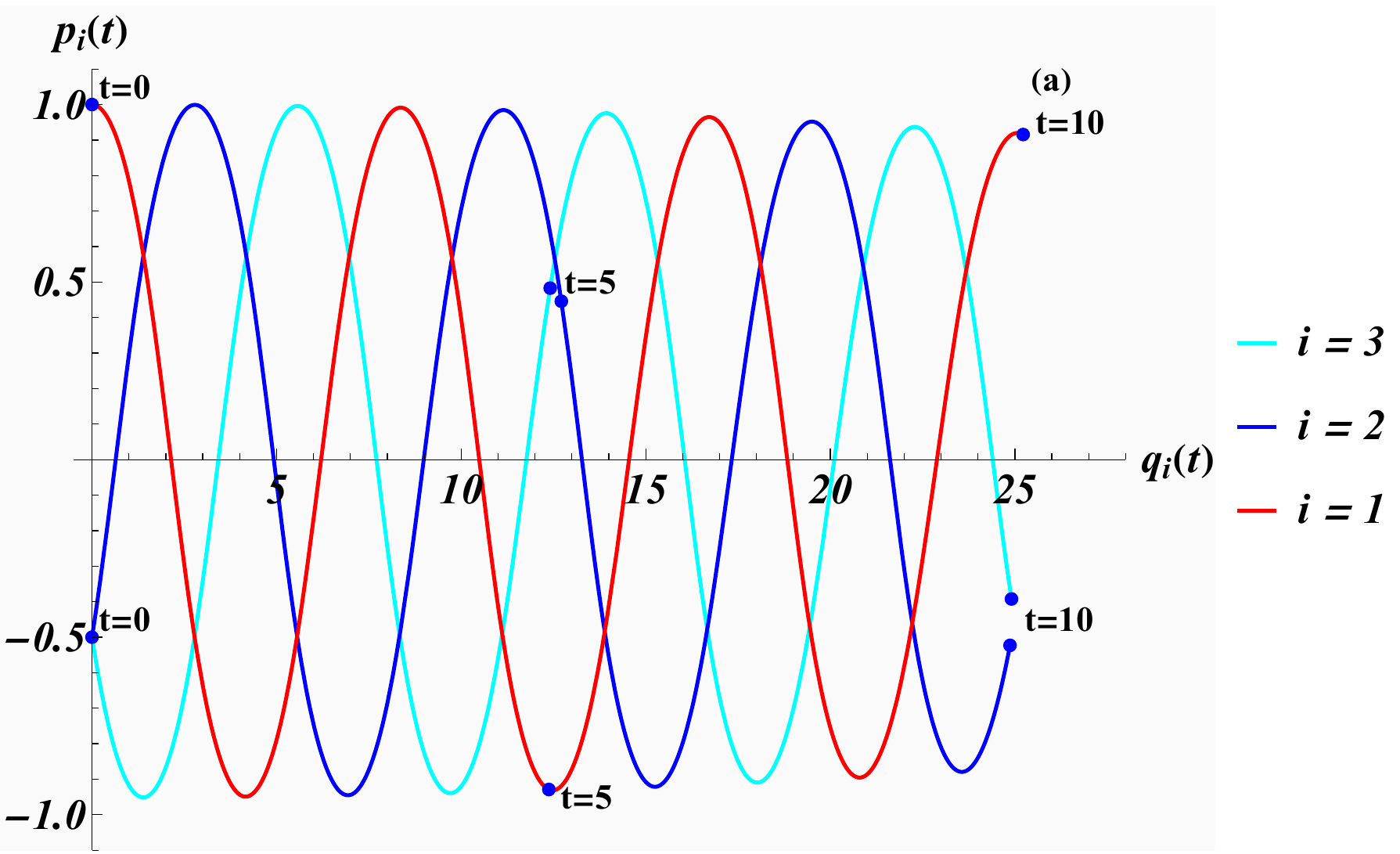}
	\end{minipage}   
	\begin{minipage}[b]{0.28\textwidth}        
		\includegraphics[width=\textwidth,height=3.1cm]{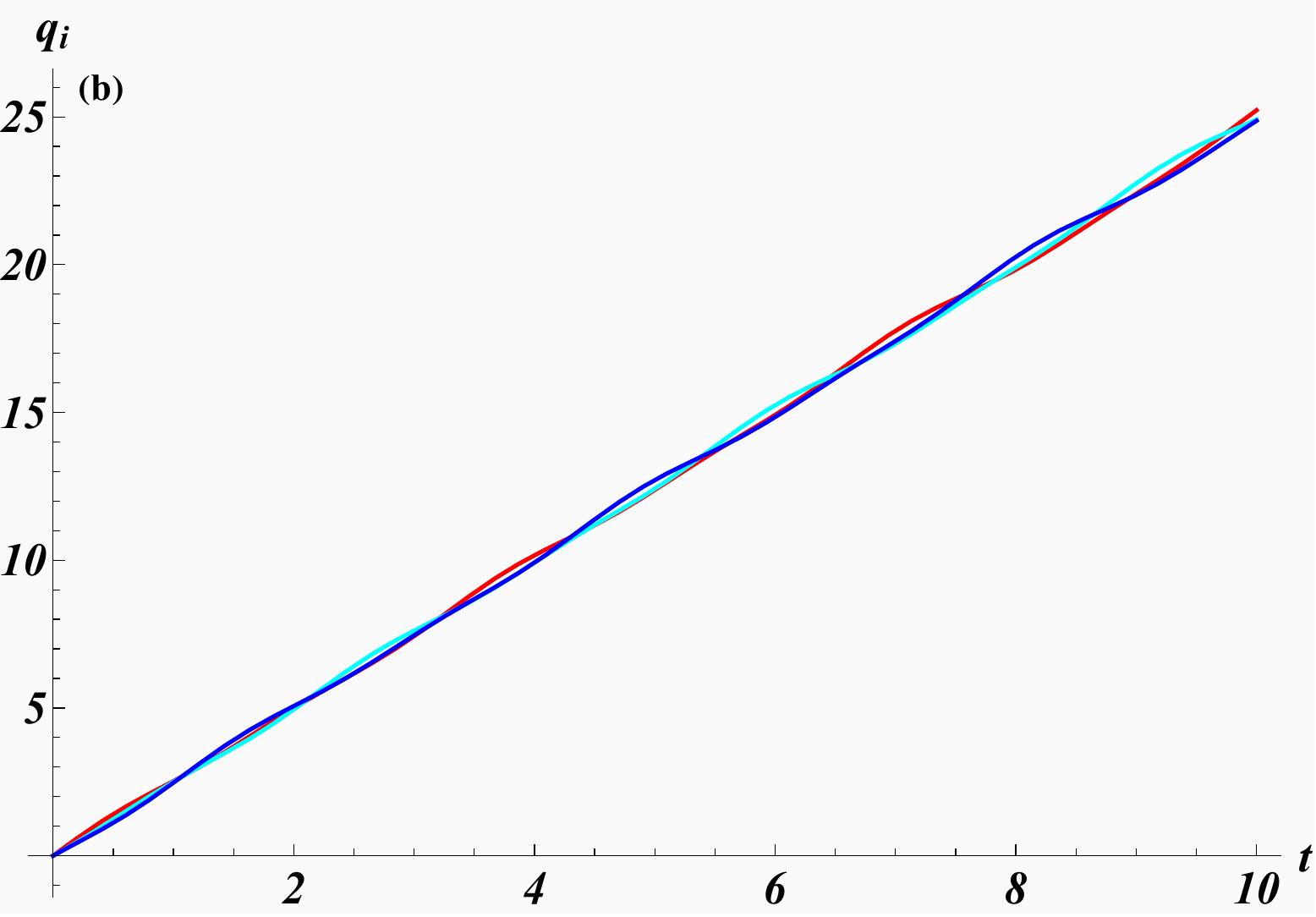}
		\includegraphics[width=\textwidth,height=3.1cm]{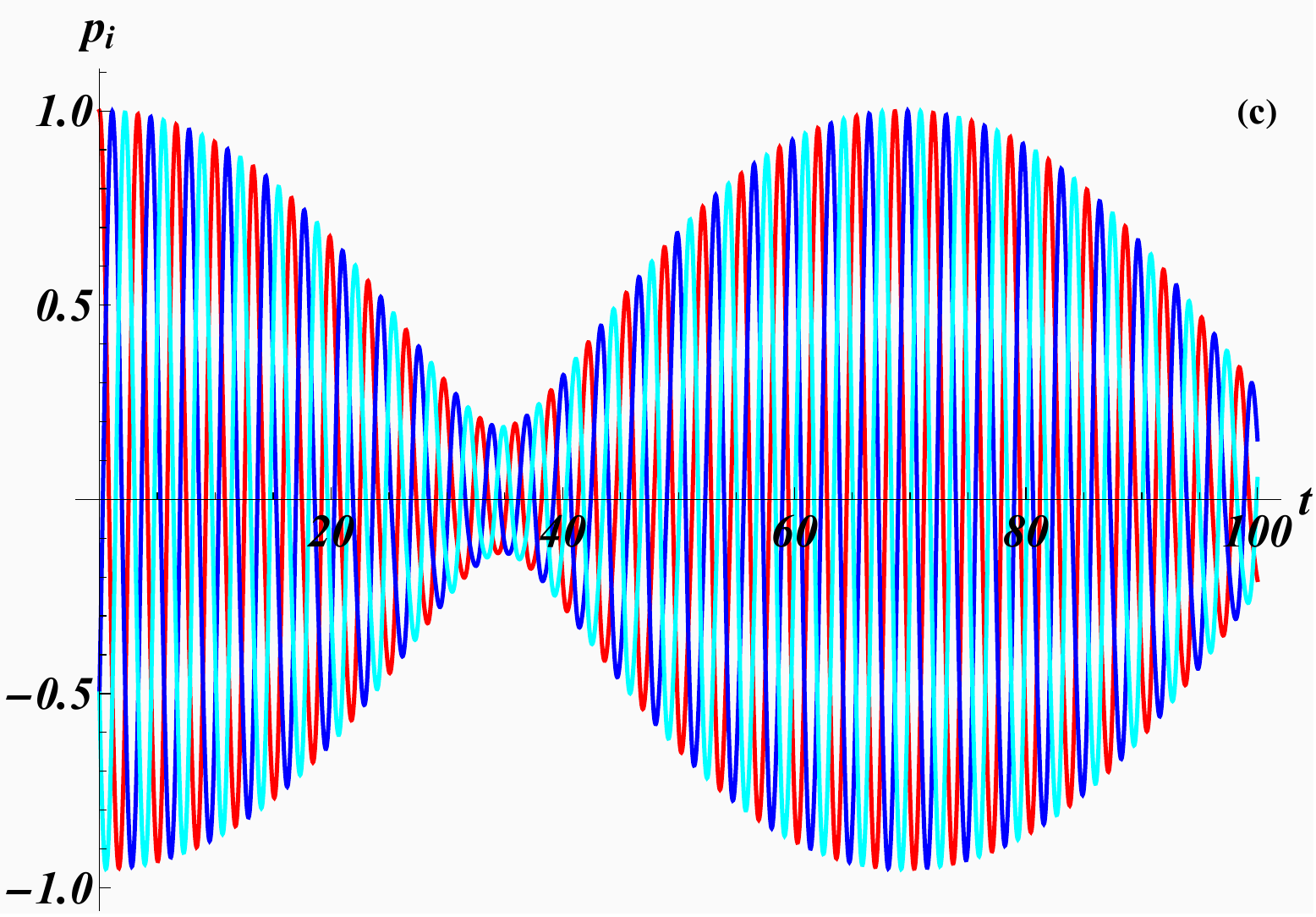}
	\end{minipage}   
	\caption{Panel (a): Phase space $(q_i,p_i)$, $i=1,2,3$ for the $A_2$-affine Toda lattice with unrestricted $Q_3$-Hamiltonian, with initial conditions $q_1(0)=q_2(0)=q_3(0)=0$, $p_1(0)=1$ and $p_2(0)=p_3(0)=-1/2$. Panel (b) and (c): $x_i$ and $p_i$ as functions of time $t$, respectively.} 
	\label{Phase2}
\end{figure}

Now we come to the key point in this approach and interpret the higher charges as Hamiltonians following the suggestion in \cite{smilga2021exactly,Smilga6}. Thus we take here the charge $Q_3$ as the Hamiltonian. Deriving the new set of equations of motion from  $\dot{q}_i = \partial Q_3 / \partial p_i$, $\dot{p}_i = -\partial Q_3 / \partial q_i$ we obtain 
\begin{eqnarray}
	\dot{q}_1 &=& p_1^2 + V_{12} +V_{31} , \qquad 	\dot{p}_1 = (p_1 +p_3) V_{31} - (p_1+p_2) V_{12}, \label{q31} \\
	\dot{q}_2 &=& p_2^2 + V_{12} +V_{23}, \qquad  \dot{p}_2 = (p_1 +p_2) V_{12} - (p_2 +p_3)V_{23} ,   \\
	\dot{q}_3 &=& p_3^2 + V_{23} +V_{31}, \qquad 
	\dot{p}_3 = (p_2 +p_3) V_{23} - (p_1 +p_3) V_{31} , \label{q33}
\end{eqnarray}
which are identical to the equations previously considered in \cite{smilga2021exactly}. Once more we solve these equations, (\ref{q31})-(\ref{q33}), numerically and depict the solutions in figure \ref{Phase2}. 

We observe that while the momenta are bounded as $-1 \leq p_i \leq 1$, the individual coordinate components $q_i$ oscillate as functions of time around the line of centre-of-mass-coordinate $\chi$ with equal contributions to the slope $1/3 \times 2 H(t=0)=5/2$ as predicted by the Poisson bracket relation (\ref{Poisson}). We note that since the Hamiltonian is positive definite, we can not find a set of initial values so that according to (\ref{Poisson}) we achieve zero growth.

 Thus, the trajectories do not close in phase space and the system with $Q_3$ taken as a Hamiltonian is always divergent. As argued above, this system also contains ghosts, but as the singularity is not reached in finite time they are still of a benign nature. The divergence is, however, due to the fact that we have treated the $A_2$-system as a three rather than a two particle system. In the next section we will represent the roots in a lower dimensional space and consequently re-define the coordinates and momenta of the model in the dual space. The effect will be that the trajectories become confined and quasi-oscillatory in phase space.  

\subsection{Higher derivative Hamiltonians from the 2 particle $A_2$-affine Toda lattice}
We will now constrain the $(3 \times 3)$-dimensional phase space to a $(2 \times 2)$-dimensional one. Recalling that root systems are isomorphic to each other as long as they reproduce the same Cartan matrix $K_{ij} = 2 \alpha_i \alpha_j / \alpha_j^2$ we may achieve this by defining a new set of simple roots $\beta_i$ in a two-dimensional representation through an orthogonal transformation that preserve $K$ of the $A_2$ root system obtained from the roots $\alpha_i$ in the standard representation. This means we have to solve 
  \begin{equation}
  K_{ij}= \alpha_i \cdot \alpha_j = A \beta_i \cdot A\beta_j  =  \beta_i \cdot \beta_j= \left(  \begin{array}{cc}
  2 & -1  \\
  -1 & 2 
  \end{array}   \right)_{ij}, \, \quad   \beta_i=A^{-1} \alpha_i, \, A^{-1}=A^\intercal, \, i,j=1,2, \label{redmat}
  \end{equation}
for the orthogonal matrix $A$ and the roots $\beta_i$. We find the solutions 
\begin{equation}
A= \left(
\begin{array}{ccc}
	\frac{1}{\sqrt{6}} & \frac{1}{\sqrt{2}} & \frac{1}{\sqrt{3}} \\
	-\sqrt{\frac{2}{3}} & 0 & \frac{1}{\sqrt{3}} \\
		\frac{1}{\sqrt{6}} & -\frac{1}{\sqrt{2}} & \frac{1}{\sqrt{3}} 
\end{array}
\right), \quad 
\beta_1 = \left(  \sqrt{ \frac{3}{2} },\frac{1}{\sqrt{2}},0 \right)  \quad 
\beta_2 = \left( -\sqrt{ \frac{3}{2} },\frac{1}{\sqrt{2}},0 \right). \label{ortho}
\end{equation}
The negative of the highest root is therefore $\beta_0 =-\beta_1-\beta_2 =(0,-\sqrt{2},0)$. 

Having reduced the dimension of the representation space for the roots from 3 to 2, we shift this reduction now to the dual space of the roots, i.e. the coordinates and the momenta. For this we define a new set of dynamical variables $(\zeta,\eta)$ in the dual space of the roots by
 \begin{eqnarray}
	\alpha_i \cdot q &=& A \beta_i \cdot A \zeta  =  \beta_i \cdot \zeta, \qquad \text{for} \quad \zeta=A^{-1} q,  \,\, i=1,2,  \label{222} \\
		\alpha_i \cdot p &=& A \beta_i \cdot A \eta  =  \beta_i \cdot \eta, \qquad \text{for} \quad \eta=A^{-1} p,  \,\, i=1,2. \label{223}
\end{eqnarray}
 With $A$ as identified in (\ref{ortho}) we have
 \begin{eqnarray}
 	q&=&\left( \frac{\zeta_1}{\sqrt{6}}+\frac{\zeta_2}{\sqrt{2}}, -\sqrt{\frac{2}{3}} \zeta_1 , \frac{\zeta_1}{\sqrt{6}}-\frac{\zeta_2}{\sqrt{2}}  \right)=(q_1,q_2,q_3), \label{newq} \\ 
 	p&=& \left( \frac{\eta_1}{\sqrt{6}}+\frac{\eta_2}{\sqrt{2}}, -\sqrt{\frac{2}{3}} \eta_1 , \frac{\eta_1}{\sqrt{6}}-\frac{\eta_2}{\sqrt{2}}  \right)=(p_1,p_2,p_3),
 \end{eqnarray}
 or when inverted
\begin{eqnarray}
\zeta&=&\left(\frac{q_1-2 q_2+q_3}{\sqrt{6}},\frac{q_1-q_3}{\sqrt{2}},\frac{q_1+q_2+q_3}{\sqrt{3}}\right)=(\zeta_1,\zeta_2,0), \label{newv} \\ 
\eta&=& \left(\frac{p_1-2 p_2+p_3}{\sqrt{6}},\frac{p_1-p_3}{\sqrt{2}},\frac{p_1+p_2+p_3}{\sqrt{3}}\right)=(\eta_1,\eta_2,0). \label{newv2}
\end{eqnarray}
From the last component in (\ref{newv}) and (\ref{newv2}) we observe that we can interpret the new $(2 \times 2)$-dimensional phase space $(\zeta, \eta)$ as the old $(3 \times 3)$-dimensional phase space $(q,p)$ in the centre of mass frame with additional constraints. We stress that this property is not imposed, but the conditions $q_1+q_2+q_3=0$ and $p_1+p_2+p_3=0$ are automatically satisfied with the definitions of the new variables in (\ref{newq}), which in turn results from representing the roots in a lower dimensional space. 

\begin{figure}[h]
	\centering         
	\begin{minipage}[b]{0.49\textwidth}           \includegraphics[width=\textwidth]{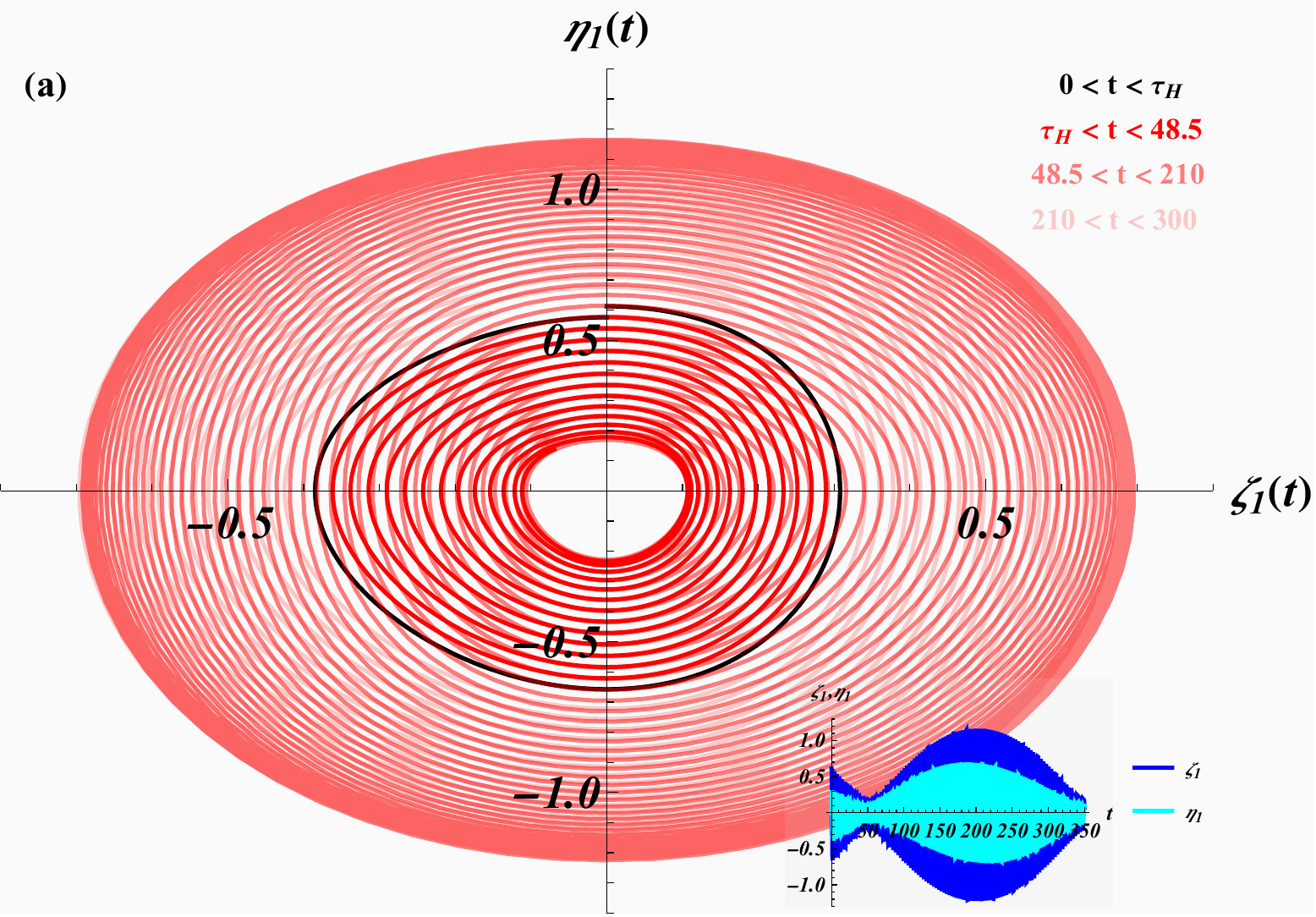}
	\end{minipage}   
	\begin{minipage}[b]{0.49\textwidth}           
		\includegraphics[width=\textwidth]{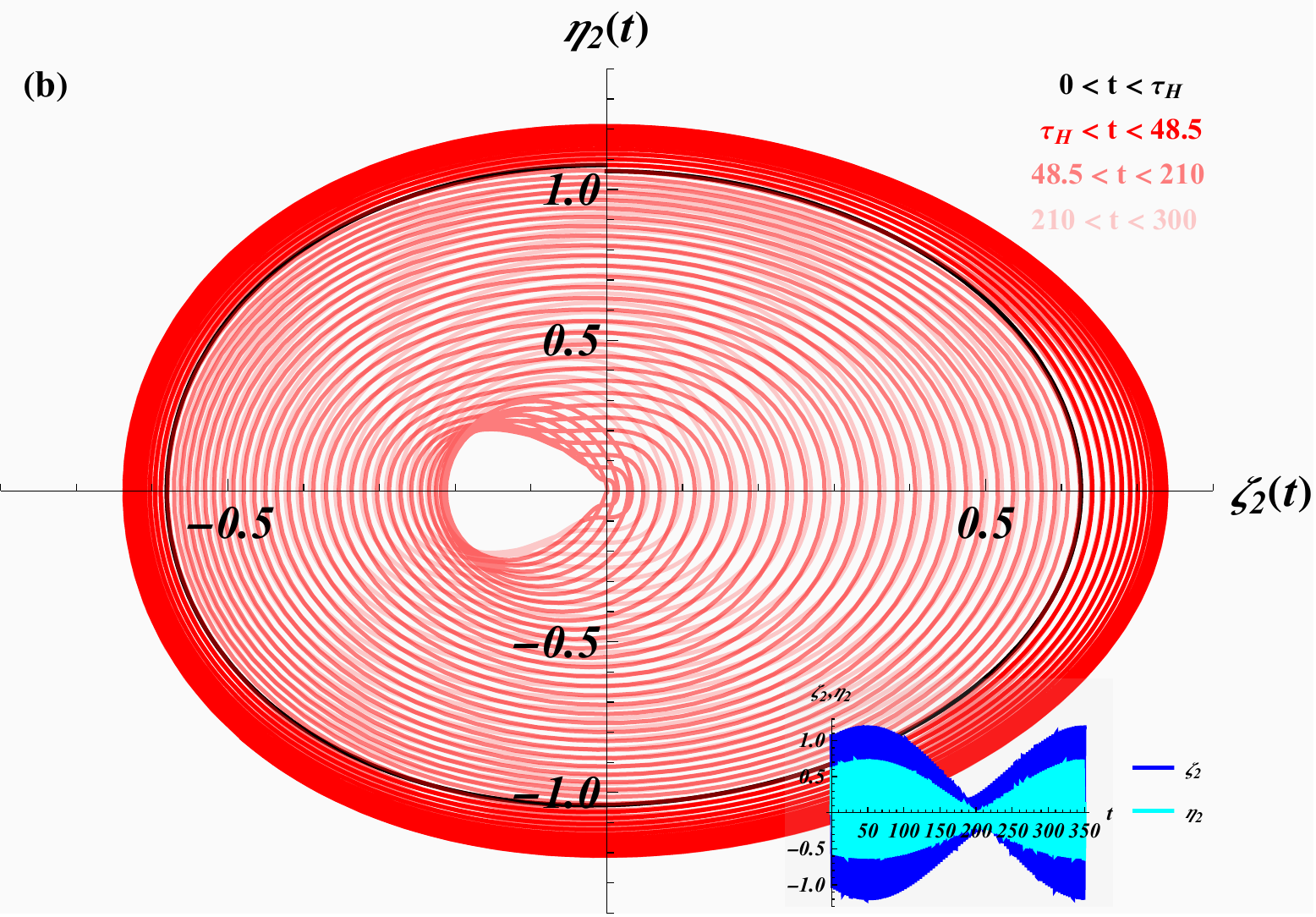}
	\end{minipage}  
	\caption{Phase spaces $(\zeta_i,\eta_i)$, $i=1,2$ for the standard Hamiltonian of the reduced two particle $A_2$-affine Toda lattice with initial conditions $\zeta_1(0)=\zeta_2(0)=0$, $\eta_1(0)=\sqrt{3}/2 \sqrt{2}$ and $\eta_2(0)=3/2 \sqrt{2}$ ($\equiv$ $p_1(0)=1$, $p_2(0)=p_3(0)=-1/2$) for times $t=0$ to $t=300$ with ``almost period" $\tau_H \approx 3.543$. The insets in panels (a) and (b) show $\zeta_1,\eta_1$ and $\zeta_2,\eta_2$ as functions of time, respectively.} 
	\label{Phase3}
\end{figure}

The conserved quantities $Q_1,Q_2,Q_3$ in (\ref{Q1})-(\ref{Q3}) can now also be transformed to the new variables as
\begin{eqnarray}
	Q_1 &=&  0, \notag \\
	Q_2 &=& H(\zeta,\eta)=  \frac{1}{2} \left( \eta_1^2  +\eta_2^2   \right)+ e^{-\sqrt{2} \zeta _2}+2 e^{\frac{\zeta _2}{\sqrt{2}}} \cosh \left(\sqrt{\frac{3}{2}} \zeta _1\right),  \label{A22cons}  \\
	Q_3 &=& \frac{\eta _1 \left(6 e^{-\sqrt{2} \zeta _2}-\eta _1^2+3 \eta _2^2\right)}{3 \sqrt{6}}- \sqrt{2} e^{\frac{\zeta _2}{\sqrt{2}}}
	\left[ \frac{\eta _1}{\sqrt{3}} \cosh \left(\sqrt{\frac{3}{2}} \zeta _1\right)- \eta _2 \sinh \left(\sqrt{\frac{3}{2}} \zeta _1\right)\right]+2. \notag
\end{eqnarray}
The equations of motion resulting from the standard Hamiltonian $ H(\zeta,\eta)$ become
\begin{eqnarray}
	\dot{\zeta}_1& =& \eta_1, \quad 	\dot{\zeta}_2 = \eta_2, \label{eq12}\\
	\dot{\eta}_1 &=& -\sqrt{6} e^{\frac{\zeta _2}{\sqrt{2}}} \sinh \left(\sqrt{\frac{3}{2}} \zeta _1\right), \quad \dot{\eta}_2 =\sqrt{2} e^{-\sqrt{2} \zeta _2} \left[1-e^{\frac{3 \zeta _2}{\sqrt{2}}} \cosh \left(\sqrt{\frac{3}{2}} \zeta _1\right)\right] , \label{eq13}
\end{eqnarray}
whereas the equations resulting from taking $Q_3(\zeta, \eta)$ interpreted as the Hamiltonian are
\begin{eqnarray}
	\dot{\zeta}_1& =& \frac{2 e^{-\sqrt{2} \zeta _2}-2 e^{\frac{\zeta _2}{\sqrt{2}}} \cosh \left(\sqrt{\frac{3}{2}} \zeta _1\right)-\eta _1^2+\eta _2^2}{\sqrt{6}},\label{e1}  \\
		\dot{\zeta}_2 &=& \sqrt{2} e^{\frac{\zeta _2}{\sqrt{2}}} \sinh \left(\sqrt{\frac{3}{2}} \zeta _1\right)+\sqrt{\frac{2}{3}} \eta _1 \eta _2 , 
\end{eqnarray}
\begin{eqnarray}	
	\dot{\eta}_1 &=&e^{\frac{\zeta _2}{\sqrt{2}}} \left[\eta _1 \sinh \left(\sqrt{\frac{3}{2}} \zeta _1\right)-\sqrt{3} \eta _2 \cosh \left(\sqrt{\frac{3}{2}}
	\zeta _1\right)\right], \\ \dot{\eta}_2&=& \frac{2 e^{-\sqrt{2} \zeta _2} \eta _1}{\sqrt{3}}+\frac{1}{3} e^{\frac{\zeta _2}{\sqrt{2}}} \left[\sqrt{3} \eta _1 \cosh
	\left(\sqrt{\frac{3}{2}} \zeta _1\right)-3 \eta _2 \sinh \left(\sqrt{\frac{3}{2}} \zeta _1\right)\right] . \label{e4}
\end{eqnarray}
The phase space trajectories obtained from the standard equations of motion for the Hamiltonian, (\ref{eq12}) and (\ref{eq13}), are still confined to a finite region in phase space as seen from the numerical solutions figure \ref{Phase3}. 
We may still identify a small period $\tau_H$ that governs one turn, up to a small displacement, and a larger period controlling the inward/outward motion.

\begin{figure}[h]
	\centering         
	\begin{minipage}[b]{0.49\textwidth}           \includegraphics[width=\textwidth]{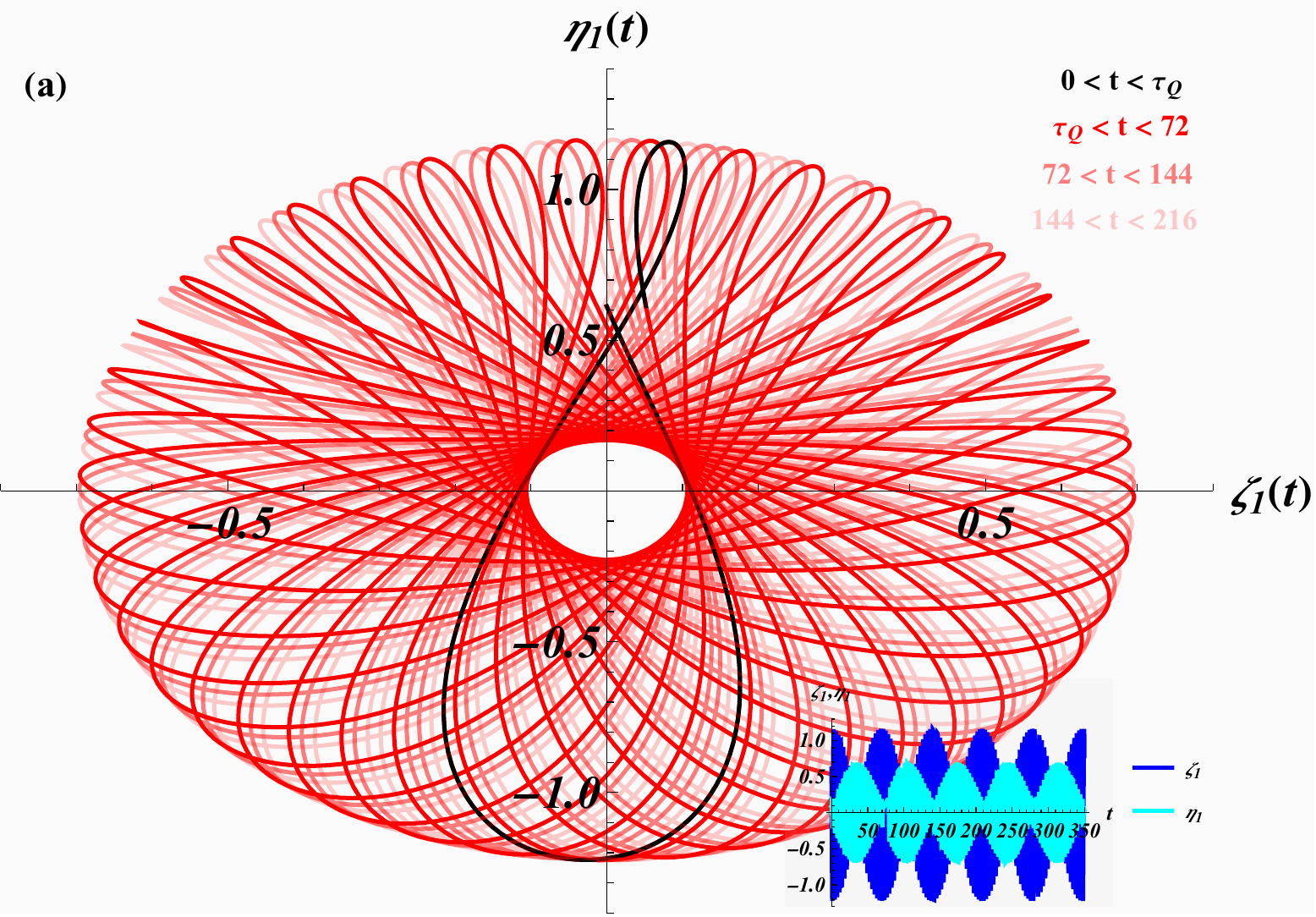}
	\end{minipage}   
	\begin{minipage}[b]{0.49\textwidth}           
		\includegraphics[width=\textwidth]{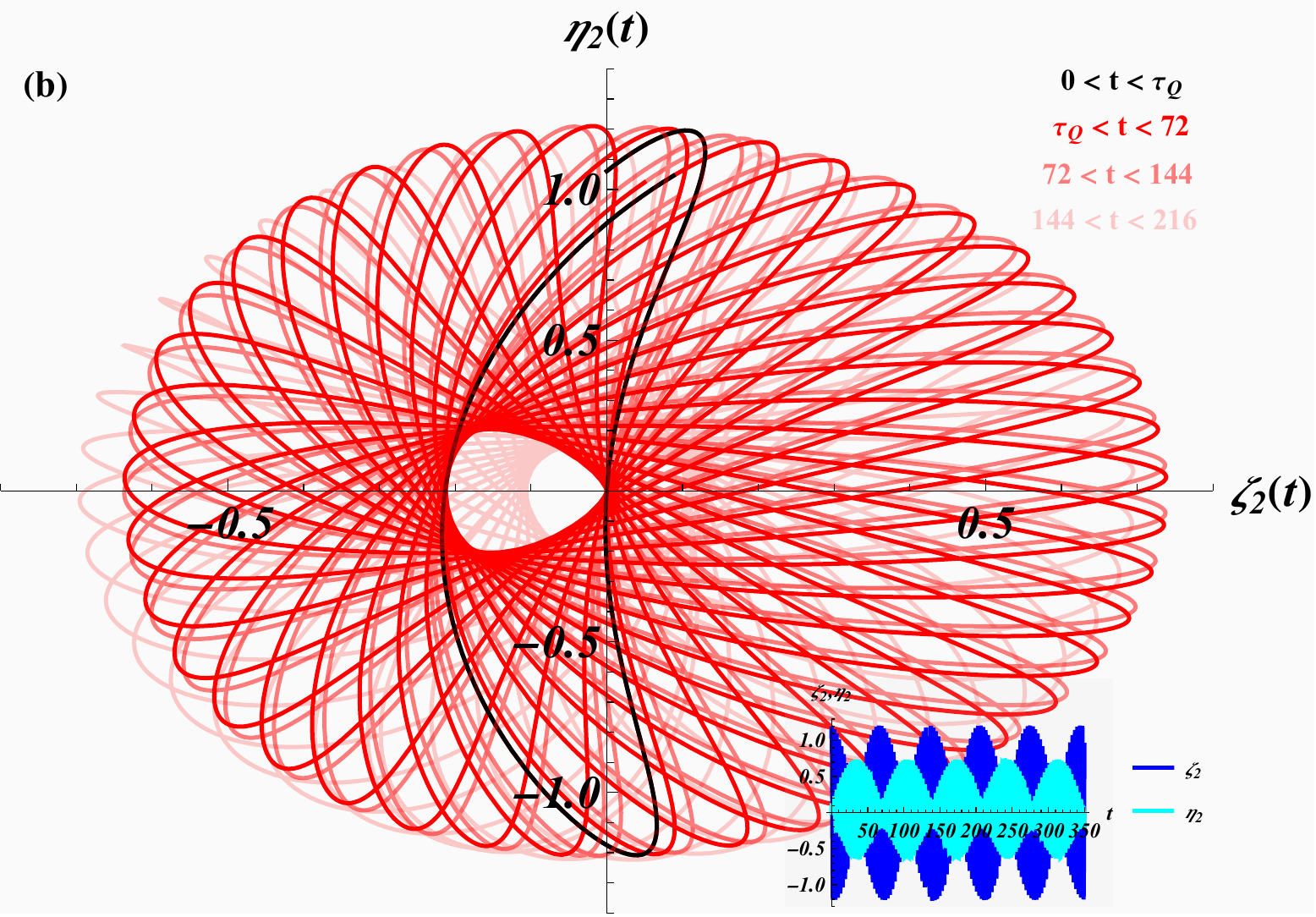}
	\end{minipage}  
	\caption{Phase space $(\zeta_i,\eta_i)$, $i=1,2$ for the $Q_3$-Hamiltonian of the reduced two particle $A_2$-Toda lattice  with initial conditions $\zeta_1(0)=\zeta_2(0)=0$, $\eta_1(0)=\sqrt{3}/2 \sqrt{2}$ and $\eta_2(0)=3/2 \sqrt{2}$ for times $t=0$ to $t=216$ with $\tau_Q \approx 3.347$. The insets in panels (a) and (b) show $\zeta_1,\eta_1$ and $\zeta_2,\eta_2$ as functions of time, respectively.} 
	\label{Phase4}
\end{figure}

In figure \ref{Phase4} we depict the numerical solutions to the equations (\ref{e1})-(\ref{e4}) obtained as equations of motions from the third order derivative $Q_3$-Hamiltonian. We determine an almost period $\tau_Q$ for the small intersecting almost closed loops. The larger period now governs the rotation of these loops that due to the repeated offset fill in the phase space regions that appear to be identical to the regions identified for the Hamiltonian $H$.  

Thus while the trajectories resulting from the three and two particle $A_2$-Hamiltonians are all confined in phase space, this behaviour is different for those derived from the higher $Q_3$-charge where only the trajectories for the reduced model are confined. The divergent behaviour was already reported in \cite{smilga2021exactly}, where it was also conjectured that in the centre of mass system convergence might be achieved. Here we have shown explicitly that this conjecture is partially correct, in the sense that the system can be interpreted as being in the centre of mass, but the more accurate statement is to view the system as the reduction from three to two particles along the change of the dimensions of the representation space of the roots. One should say that the two particle picture of the $A_2$-theory is the more natural one as for instance also in the closely related affine Toda quantum field theory the number of particles always equals the rank of the semi-simple Lie algebra \cite{BCDS,FO}. The mismatch between rank and particles simply results form the higher dimensional representation space of the simple roots. In \cite{chen2013higher} a similar reduction procedure was carried out by imposing additional constraints in order to ``exorcise" Ostrogradski’s ghosts. One may view the centre-of-mass condition as such a constraint, although here we have not employed Lagrange multipliers is to implement them.

\subsection{Higher derivative Hamiltonians from the $A_6$-affine Toda lattice}
Next we consider a system that possesses more than one independent higher charge. Specifying the general Lax operator in (\ref{LM}) to $n=6$ and computing the traces over the products of this operator we calculate the seven independent charges (\ref{chargeq1})-(\ref{chargeq7}). For the seven particle system with the roots taken in the fundamental representation we obtain the explicit expressions for the charges by replacing $W_i^2 \rightarrow V_{i,i+1}$. For instance, when taking all the roots in the standard representation the Hamiltonian acquires the form 
\begin{equation}
H  =\frac{1}{2} \sum_{i=1}^7 p_i^2+\sum_{i=1}^6 e^{q_i-q_{i+1}} + e^{q_7-q_1} , \label{Q2A6} 
 \end{equation}
with $\alpha_7$ taken as the negative of the highest root. We also convince ourselves that all mutual Poisson brackets vanish. Moreover, also in this case the Poisson bracket relation (\ref{Poisson}) involving the centre-of-mass coordinate, now $\chi := \sum_{i=1}^7 q_i$,  still holds.

\begin{figure}[h]
	\centering         
	\begin{minipage}[b]{0.49\textwidth}           \includegraphics[width=\textwidth]{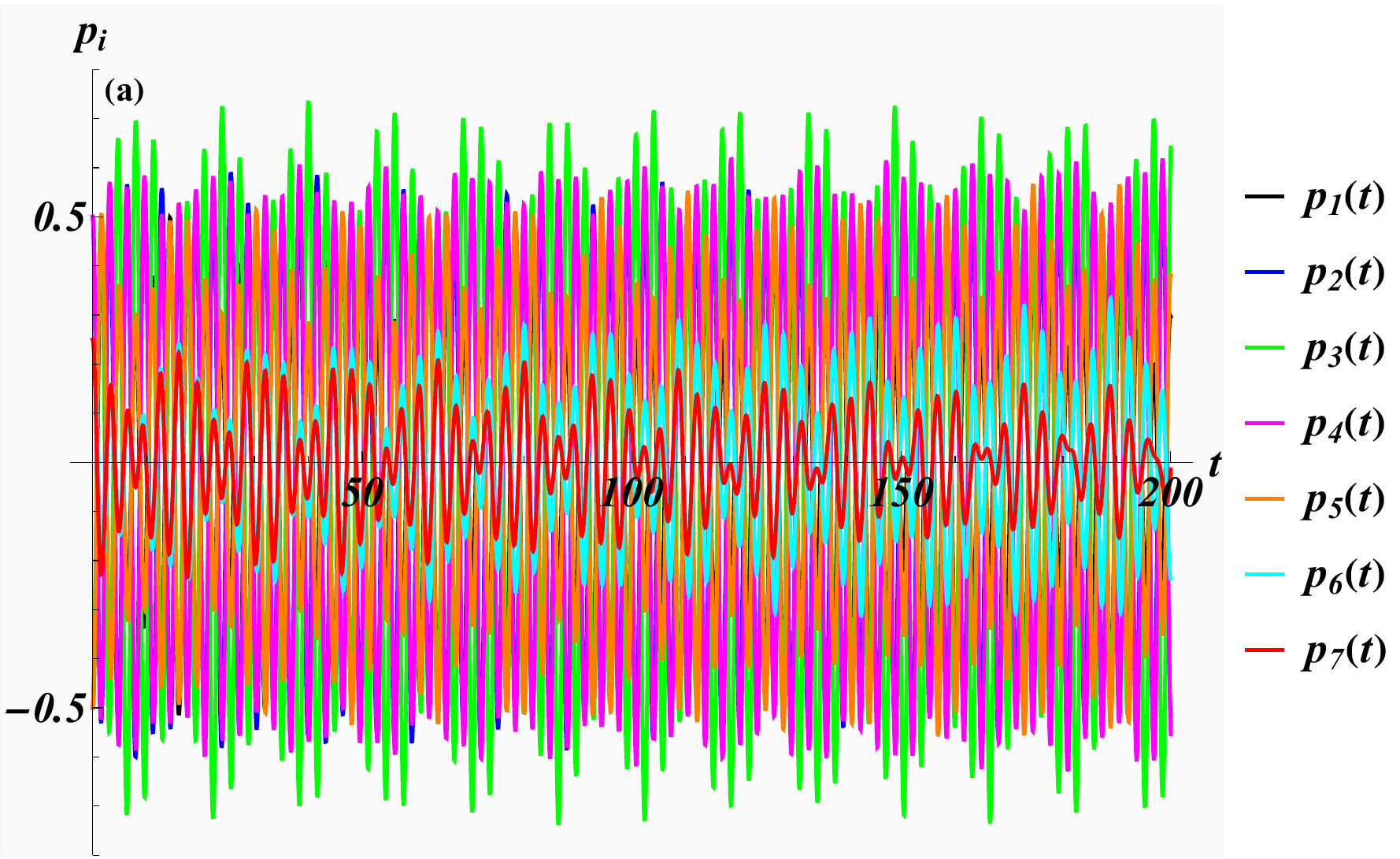}
	\end{minipage}   
	\begin{minipage}[b]{0.49\textwidth}           
		\includegraphics[width=\textwidth]{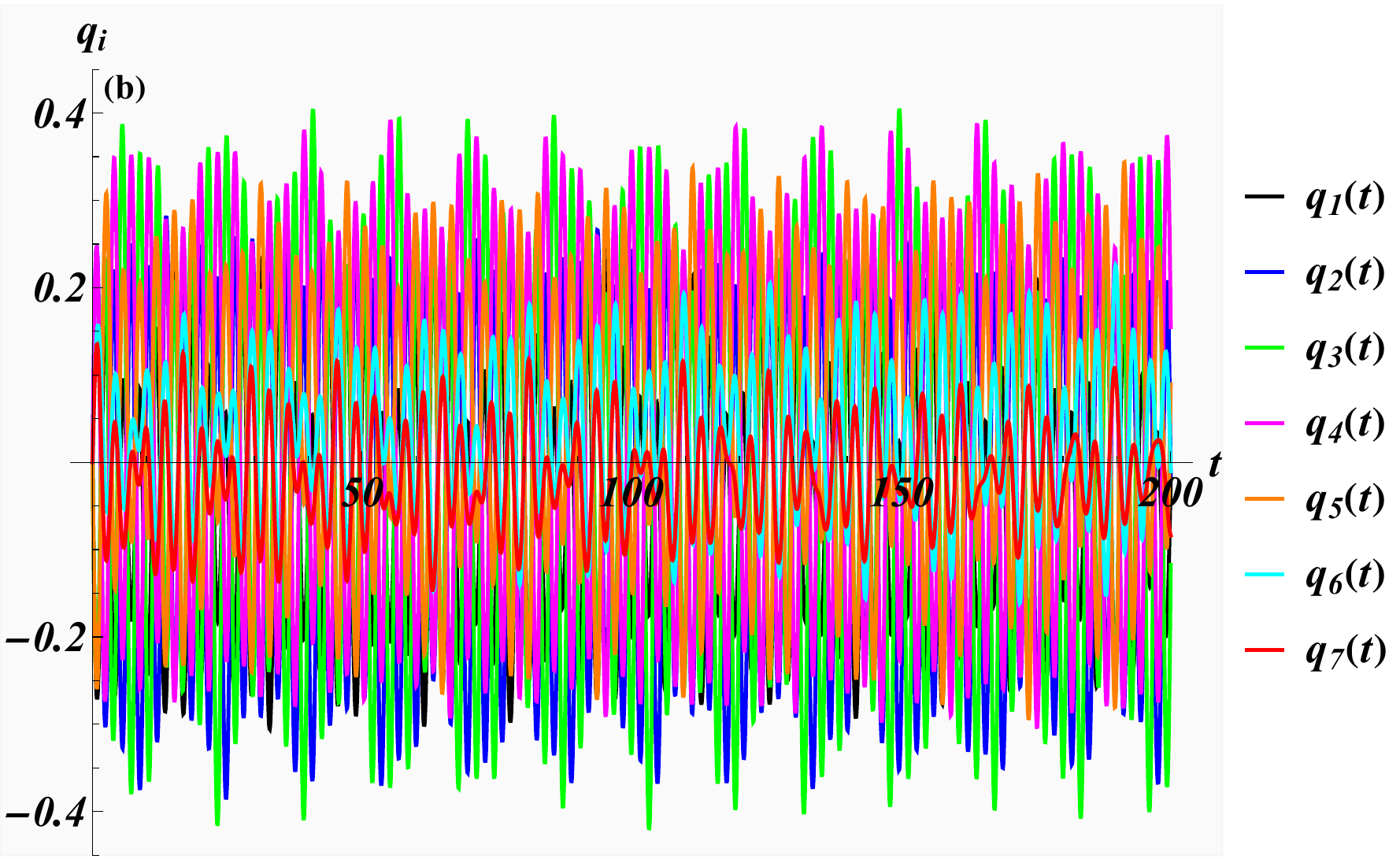}
	\end{minipage}  
	\begin{minipage}[b]{0.49\textwidth}           \includegraphics[width=\textwidth]{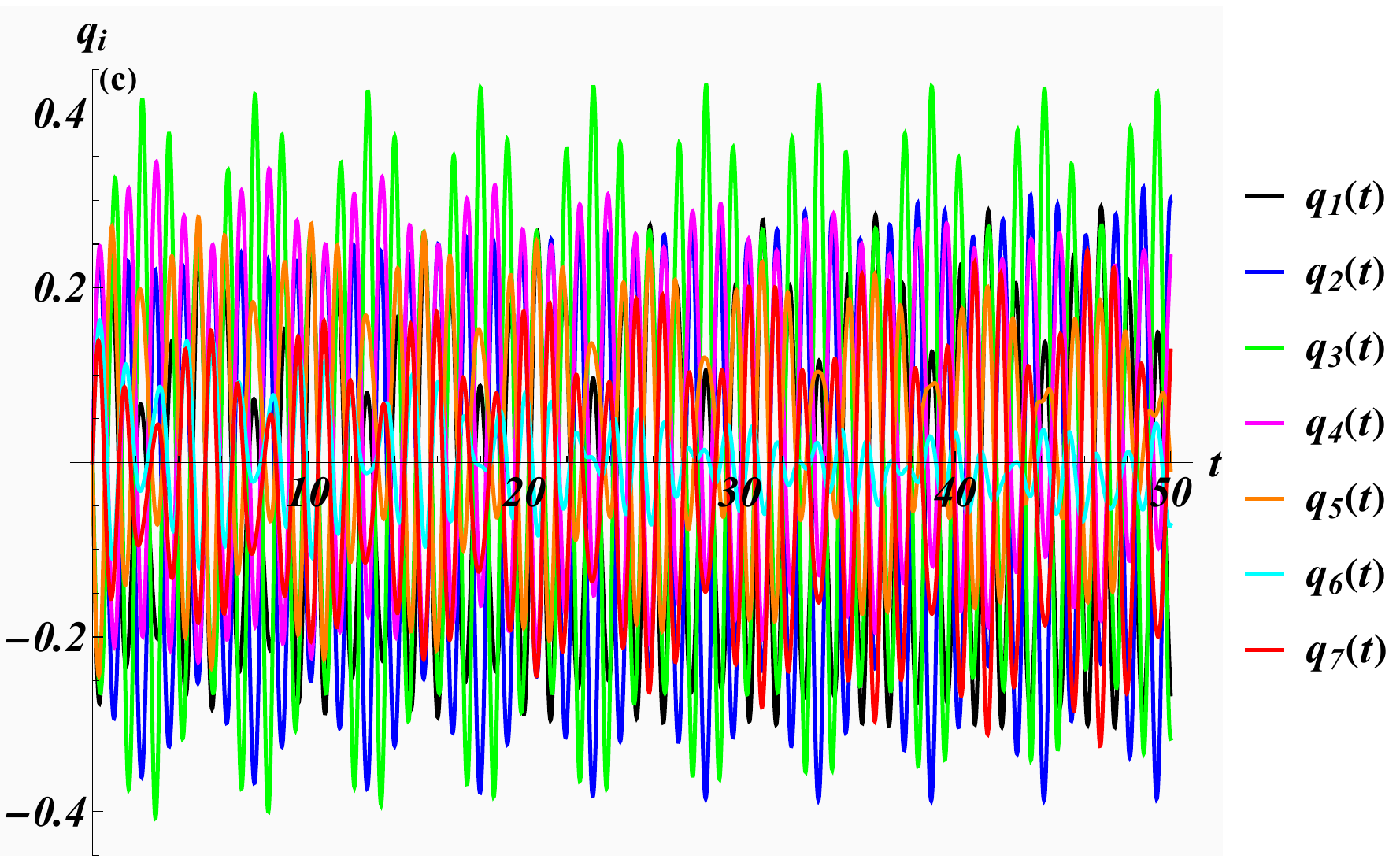}
	\end{minipage}   
	\begin{minipage}[b]{0.49\textwidth}           
		\includegraphics[width=\textwidth]{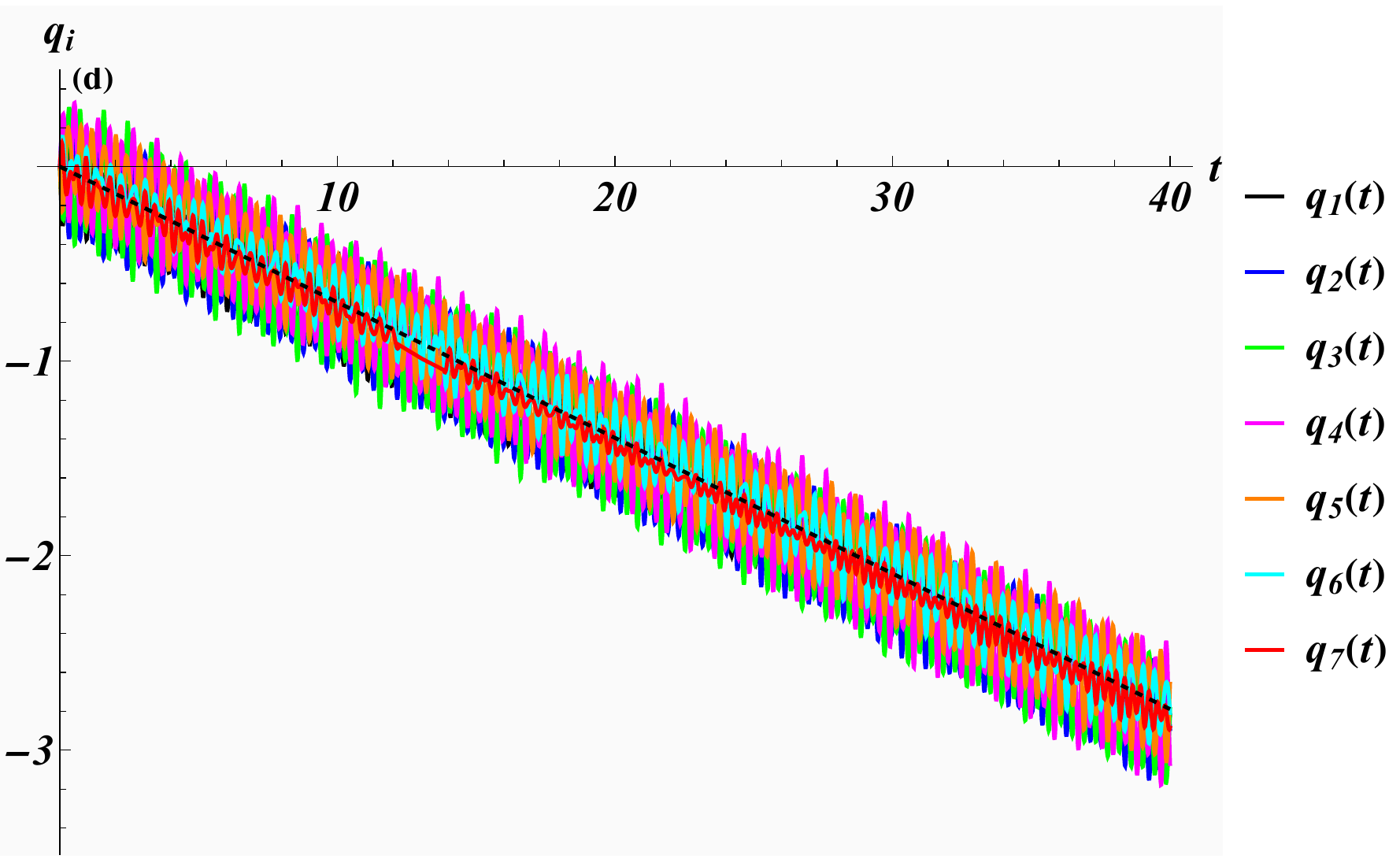}
	\end{minipage}  
	\caption{$A_6$-affine Toda lattice phase spaces as functions of time $t$ of the Hamiltonian, panels (a), (b), the $Q_4$-charge Hamiltonian panel (c) and the $Q_6$-charge Hamiltonian panel (d), with seven particles. In panels (a), (b) and (d) the initial conditions are taken as $Q_1=0$ with $q_i=0$, $i=1, \ldots, 7$ and $p_1=-p_2=p_3=-p_4=p_5=-2p_6=2p_7=-1/2$. In panels (c) the initial conditions is taken so that $Q_3=0$ with $q_i=0$, $i=1, \ldots, 7$ and $p_1=-0.486068$, $-p_2=p_3=-p_4=p_5=-2p_6=2p_7=-1/2$. The dashed line has slope $1/7 \times 5 Q_5=-0.0711496$.} 
	\label{PhaseA6seven}
\end{figure}

We proceed now in the same manner as for the $A_3$-system by interpreting all of the charges as Hamiltonians and solve their respective Hamilton's equations. For the seven particle system we find periodic solutions for the momenta and coordinates in the phase space of the standard Hamiltonian with initial conditions taken such that $Q_1=0$, as seen in figure \ref{PhaseA6seven}. However, for all higher charges only the momenta remain periodic whereas the coordinates diverge when keeping these initial conditions. In figure \ref{PhaseA6seven} we present as sample solution for the phase space of the higher charges the one for the  $Q_6$-charge Hamiltonian. In panel (d) we observe the divergence of all coordinates. As expected from the Poisson bracket relations the coordinates oscillate around the straight (dashed) line with slope $1/7 \times 5 Q_5=-0.0711496$. In panel (c) we present the solutions for the coordinates when taking the $Q_4$-charge as Hamiltonian with initial conditions $Q_3(t-0)=0$. These characteristic behaviours are shared by the solutions for all the other higher charge Hamiltonians which we do not represent here. The classical behaviour does not distinguish between odd and even charges, that is between systems that contains ghost states and those which do not.

\begin{figure}[h]
  		\centering         
  		\begin{minipage}[b]{0.49\textwidth}           \includegraphics[width=\textwidth]{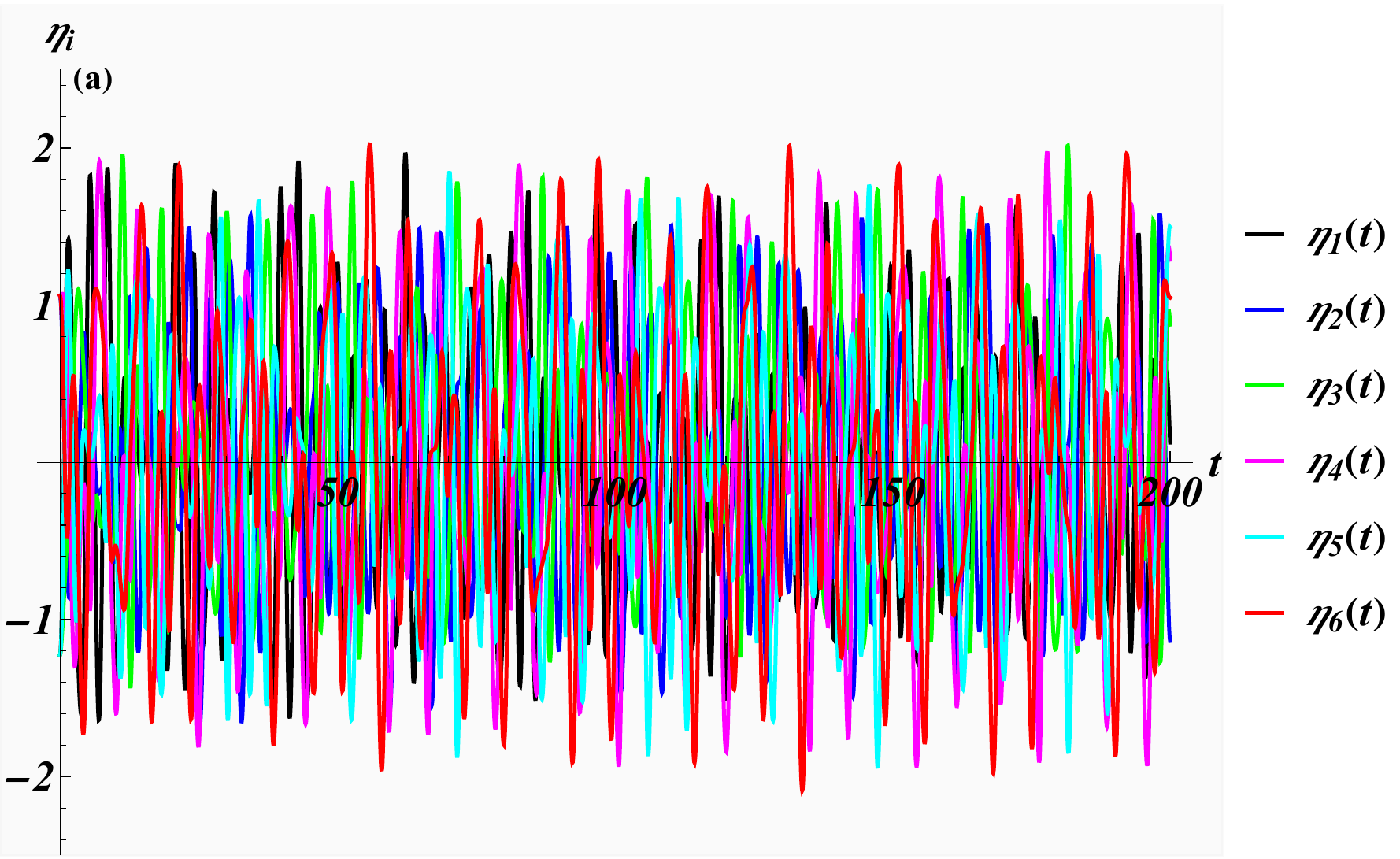}
  		\end{minipage}   
  		\begin{minipage}[b]{0.49\textwidth}           
  			\includegraphics[width=\textwidth]{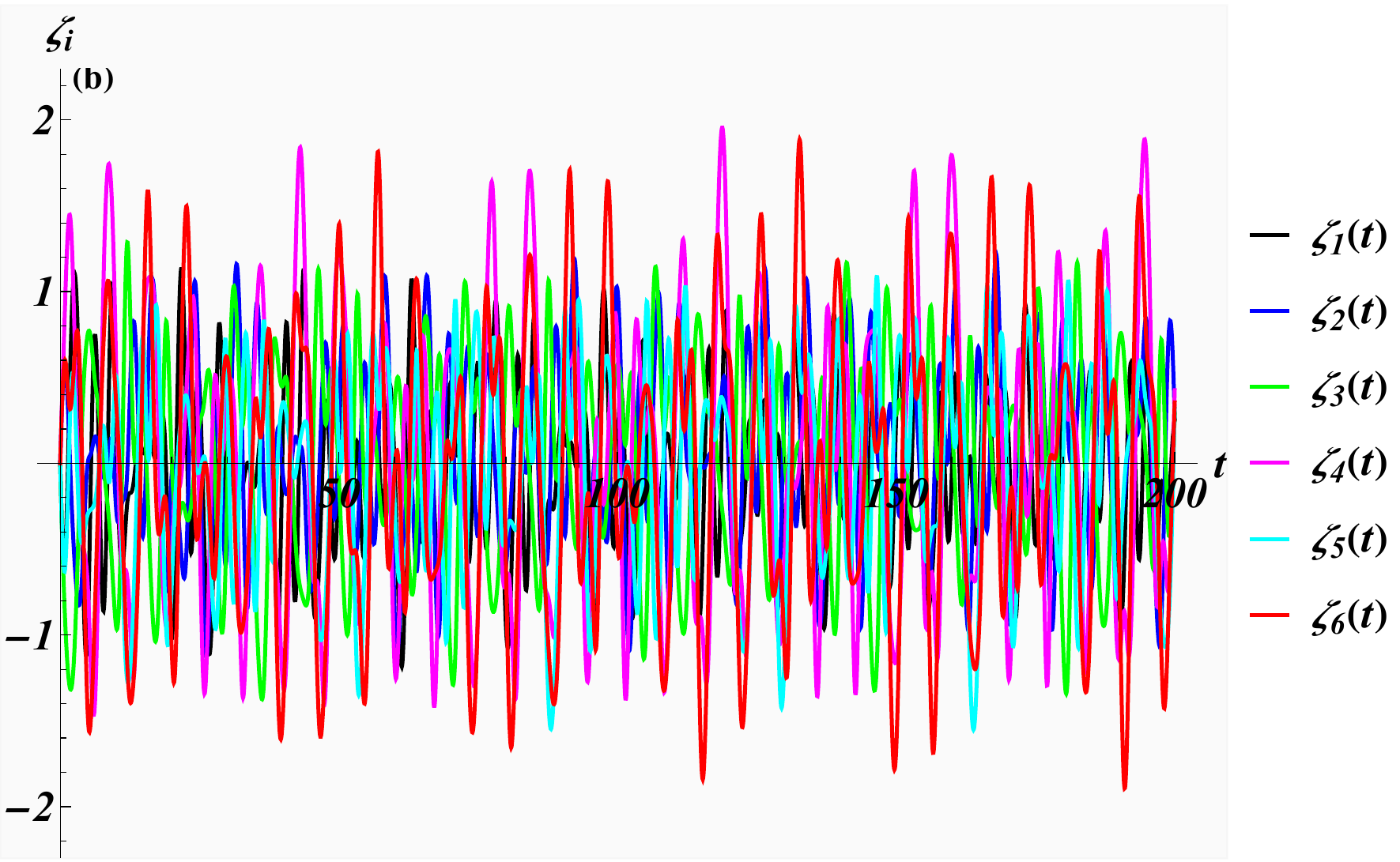}
  		\end{minipage}  
  		\begin{minipage}[b]{0.49\textwidth}           \includegraphics[width=\textwidth]{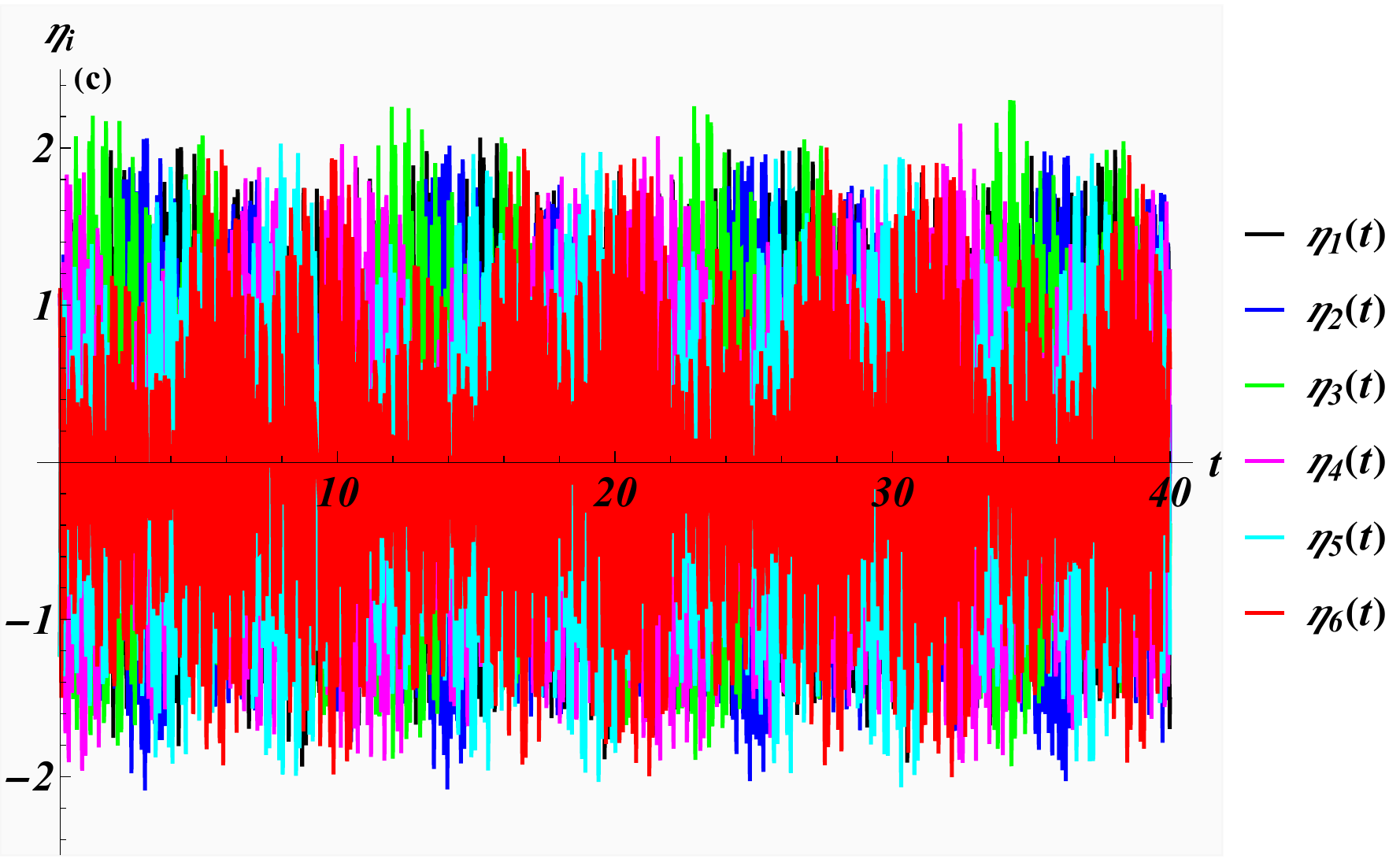}
  		\end{minipage}   
  		\begin{minipage}[b]{0.49\textwidth}           
  			\includegraphics[width=\textwidth]{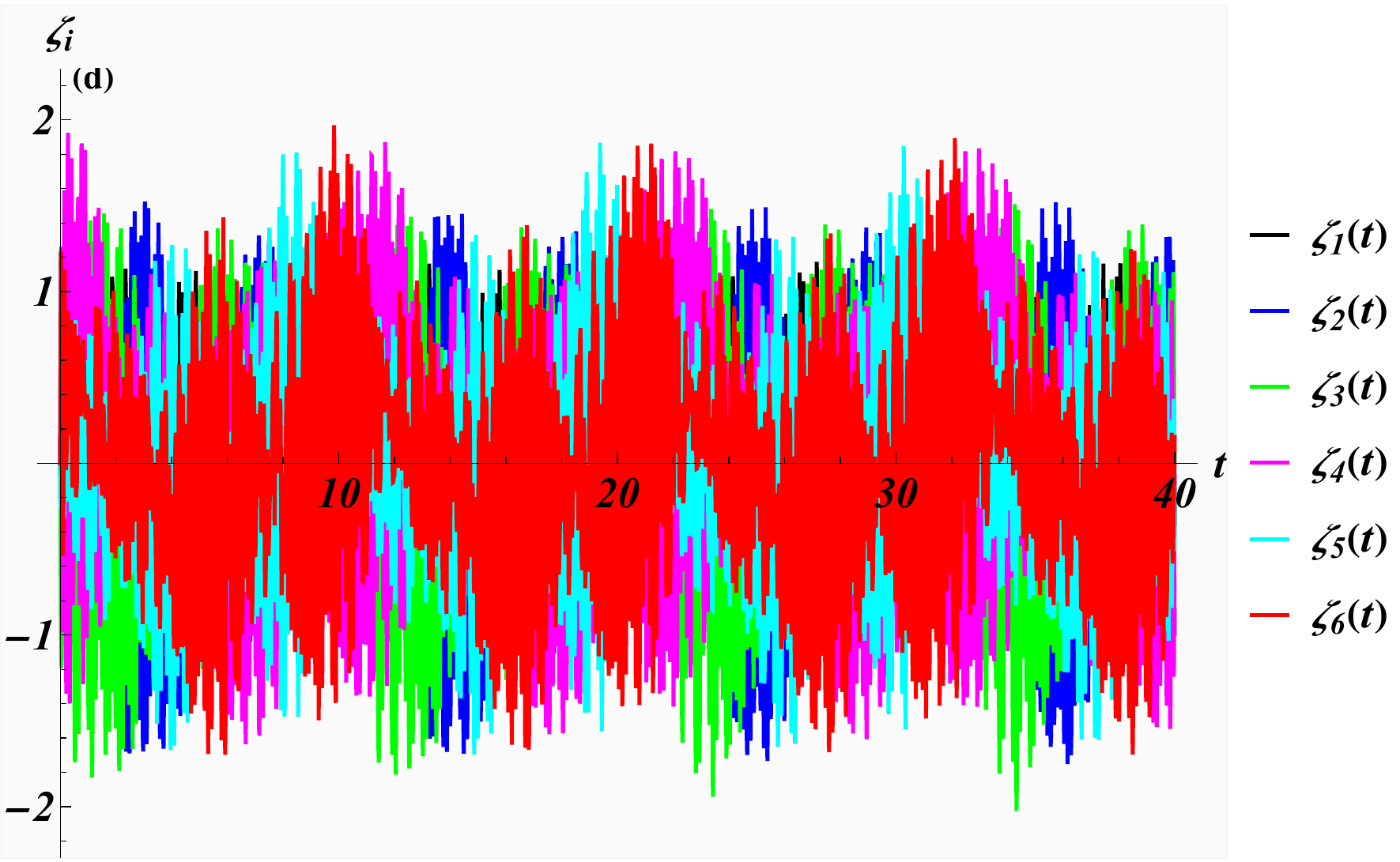}
  		\end{minipage}  
  		\caption{$A_6$-affine Toda lattice phase spaces as functions of time $t$ of the Hamiltonian, panels (a), (b), and the $Q_6$-charge Hamiltonian, panels (c), (d), with six particles. The initial conditions are taken in both cases as $\zeta_i=0$, $i=1, \ldots, 6$ and $\eta_1=\eta_2=\eta_3=-3/\sqrt{6}$, $\eta_2=\eta_4=\eta_6=3/2\sqrt{2}$.} 
  		\label{PhaseA6six}
  	\end{figure}

Similarly as for the $A_2$-case, we attempt to eliminate the divergence by reducing the number of particles to the rank, that is from seven to six. For this purpose we solve the analogue to the equation (\ref{redmat}) with the $A_6$-Cartan matrix instead. Taking the $\alpha$-roots in the standard representation we find an orthogonal matrix as 
\begin{equation}
	A=\left(
	\begin{array}{ccccccc}
		\frac{1}{\sqrt{6}} & \frac{1}{\sqrt{2}} & \frac{1}{2 \sqrt{3}} & -\frac{1}{2 \sqrt{5}} & -\frac{1}{\sqrt{42}} & -\frac{1}{\sqrt{30}} & -\frac{1}{\sqrt{7}} \\
		-\sqrt{\frac{2}{3}} & 0 & \frac{1}{2 \sqrt{3}} & -\frac{1}{2 \sqrt{5}} & -\frac{1}{\sqrt{42}} & -\frac{1}{\sqrt{30}} & -\frac{1}{\sqrt{7}} \\
		\frac{1}{\sqrt{6}} & -\frac{1}{\sqrt{2}} & \frac{1}{2 \sqrt{3}} & -\frac{1}{2 \sqrt{5}} & -\frac{1}{\sqrt{42}} & -\frac{1}{\sqrt{30}} & -\frac{1}{\sqrt{7}} \\
		0 & 0 & -\frac{\sqrt{3}}{2} & -\frac{1}{2 \sqrt{5}} & -\frac{1}{\sqrt{42}} & -\frac{1}{\sqrt{30}} & -\frac{1}{\sqrt{7}} \\
		0 & 0 & 0 & \frac{2}{\sqrt{5}} & -\frac{1}{\sqrt{42}} & -\frac{1}{\sqrt{30}} & -\frac{1}{\sqrt{7}} \\
		0 & 0 & 0 & 0 & -\frac{1}{\sqrt{42}} & \sqrt{\frac{5}{6}} & -\frac{1}{\sqrt{7}} \\
		0 & 0 & 0 & 0 & \sqrt{\frac{6}{7}} & 0 & -\frac{1}{\sqrt{7}} \\
	\end{array}
	\right),
	\end{equation}
	together with the new six dimensional roots $\beta_i= A^{-1} \alpha_i$
	\begin{equation}
		\begin{array}{ll}
		 \beta_1=\left( \sqrt{\frac{3}{2}},\frac{1}{\sqrt{2}},0,0,0,0,0\right),  &\beta_2=\left(-\sqrt{\frac{3}{2}},\frac{1}{\sqrt{2}},0,0,0,0,0\right), \\  
		\beta_3 = \left( \frac{1}{\sqrt{6}},-\frac{1}{\sqrt{2}},\frac{2}{\sqrt{3}},0,0,0,0\right),  \qquad \qquad
		&\beta_4= \left( 0,0,-\frac{\sqrt{3}}{2},-\frac{\sqrt{5}}{2},0,0,0\right), \\
		\beta_5 =\left( 0,0,0,\frac{2}{\sqrt{5}},0,-\sqrt{\frac{6}{5}},0\right),   
		&\beta_6= \left( 0,0,0,0,-\sqrt{\frac{7}{6}},\sqrt{\frac{5}{6}},0\right).
		\end{array}
	\end{equation}
	 The corresponding coordinate transformations resulting from this are
	\begin{eqnarray}
		q&=& (q_1,q_2,q_3,q_4,q_5,q_6,q_7) \\ 
		&=& \left(\frac{\zeta _1}{\sqrt{6}}+\frac{\zeta _2}{\sqrt{2}}+\frac{\zeta _3}{2 \sqrt{3}}-\frac{\zeta _4}{2 \sqrt{5}}-\frac{\zeta _5}{\sqrt{42}}-\frac{\zeta
			_6}{\sqrt{30}},-\sqrt{\frac{2}{3}} \zeta _1+\frac{\zeta _3}{2 \sqrt{3}}-\frac{\zeta _4}{2 \sqrt{5}}-\frac{\zeta _5}{\sqrt{42}}-\frac{\zeta
			_6}{\sqrt{30}}, \right. \notag \\
		&& \quad \frac{\zeta _1}{\sqrt{6}}-\frac{\zeta _2}{\sqrt{2}}+\frac{\zeta _3}{2 \sqrt{3}}-\frac{\zeta _4}{2 \sqrt{5}}-\frac{\zeta
			_5}{\sqrt{42}}-\frac{\zeta _6}{\sqrt{30}},-\frac{1}{2} \sqrt{3} \zeta _3-\frac{\zeta _4}{2 \sqrt{5}}-\frac{\zeta _5}{\sqrt{42}}-\frac{\zeta
			_6}{\sqrt{30}}, \notag \\
		&& \quad \left. \frac{2 \zeta _4}{\sqrt{5}}-\frac{\zeta _5}{\sqrt{42}}-\frac{\zeta _6}{\sqrt{30}},\sqrt{\frac{5}{6}} \zeta _6-\frac{\zeta
			_5}{\sqrt{42}},\sqrt{\frac{6}{7}} \zeta _5\right), \notag
	\end{eqnarray}	
and 
	  \begin{eqnarray}
	  	\zeta&=& (\zeta_1,\zeta_2,\zeta_3,\zeta_4,\zeta_5,\zeta_6,0)  \label{A6zeta}    \\
	    &=& 	\left(\frac{q_1-2 q_2+q_3}{\sqrt{6}},\frac{q_1-q_3}{\sqrt{2}},\frac{q_1+q_2+q_3-3 q_4}{2 \sqrt{3}},-\frac{q_1+q_2+q_3+q_4-4 q_5}{2
	  		\sqrt{5}}, \right.  \notag \\
	  	&&  \left. -\frac{q_1+q_2+q_3+q_4+q_5+q_6-6 q_7}{\sqrt{42}},-\frac{q_1+q_2+q_3+q_4+q_5-5
	  		q_6}{\sqrt{30}},-\frac{\sum_{i=1}^7 q_i}{\sqrt{7}}\right). \notag 
	  	\end{eqnarray}
  	Since the last entry for $\zeta$ in (\ref{A6zeta}) is zero, we note that once again the new coordinates transform the old ones to the centre-of-mass frame. The momenta are transformed in the same way, with $p_i \rightarrow \eta_i$. The Hamiltonian now acquires the form  
  	\begin{eqnarray}
  		H  &=& \frac{1}{2}\sum_{i=1}^6  \eta_i^2 +e^{\frac{\zeta _2-\sqrt{3} \zeta _1}{\sqrt{2}}}+e^{\frac{\sqrt{3} \zeta _1+\zeta _2}{\sqrt{2}}}+ e^{\frac{\zeta _1}{\sqrt{6}}-\frac{\zeta
  				_2}{\sqrt{2}}+\frac{2 \zeta _3}{\sqrt{3}}}+e^{-\frac{1}{2} \sqrt{3} \zeta _3-\frac{\sqrt{5} \zeta _4}{2}} \label{Q2A6xx}\\
  		&& +e^{\sqrt{\frac{5}{6}} \zeta _6-\sqrt{\frac{7}{6}}
  			\zeta _5}+e^{\frac{2 \zeta _4-\sqrt{6} \zeta _6}{\sqrt{5}}} +e^{ \frac{1}{30} \left(-5 \sqrt{6} \zeta _1-15 \sqrt{2} \zeta _2-5 \sqrt{3} \zeta _3+3 \sqrt{5} \zeta _4+5 \sqrt{42} \zeta _5+\sqrt{30} \zeta
  		_6\right)} . \notag
  	\end{eqnarray}
  	In this reduced space all trajectories become oscillatory as we observe in figure \ref{PhaseA6six}. We recognise once more that each of the solutions is made up of superposition of various quasi/almost periodic functions.

  Thus all five higher charges of the $A_6$-affine Toda lattice theory when interpreted as Hamiltonians for a six particle system possess solutions of oscillatory type in their classical phase spaces.

\section{Higher derivative Hamiltonians from the $B_3$ affine Toda lattice}
Many physical systems based on non-simply laced algebras display quite different behaviour from those based on simply laced ones. To find out whether this also holds for higher order derivative theories we will investigate some sample representative theories based on non-simply laced algebras. We first recall how to obtain the latter.  

\subsection{Reduction of the root spaces and charges}
It is well known that non-simply laced Lie algebras can be obtained from a folding procedure of the associated Dynkin diagrams for a simply laced Lie algebra along a non-trivial automorphism \cite{DIO,bordner1999calogero,fring2005affine,fring2006g}. Here we use a reduction from the $A_6$ root space $\Delta_{A_6}$ to the $B_3$ root space $\hat{\Delta}_{B_3}$ with a subsequent reduction to the $G_2$ root space $\tilde{\Delta}_{G_2}$ previously constructed in \cite{fring2006g}. Denoting the corresponding simple roots as $\alpha_i \in \Delta_{A_6}  $, $ i =1,\ldots, 6$,  $\hat{\alpha}_i \in \hat{\Delta}_{B_3} $, $ i =1,2,3$ and $\tilde{\alpha}_i \in \tilde{\Delta}_{G_2} $, $ i =1,2$ we define the following reduction maps and their inverses 
\begin{eqnarray}
	\omega &:& \Delta_{A_6} \rightarrow \hat{\Delta}_{B_3}, \qquad \alpha_i \mapsto \omega(\alpha_i) = \left\{ \begin{array}{ll} 
	\hat{\alpha}_i  \,\, \quad \quad \text{for} \,\, i=1,2,3 \\  
   \hat{\alpha}_{7-i}  \,\, \quad \text{for} \,\, i=4,5,6 
\end{array}      \right. ,\\
	\omega^{-1} &:& \hat{\Delta}_{B_3} \rightarrow \Delta_{A_6}, \qquad \hat{\alpha}_i \mapsto \omega^{-1}(\hat{\alpha}_i) = \alpha_i + \alpha_{7-i}  \,\, \quad \text{for} \,\, i=1,2,3 , \label{redb3} \\
	\hat{\omega} &:& \hat{\Delta}_{B_3} \rightarrow \tilde{\Delta}_{G_2}, \qquad \hat{\alpha}_i \mapsto \hat{\omega}(\hat{\alpha}_i) = \left\{ \begin{array}{ll} 
		\tilde{\alpha}_1  \,\, \quad  \text{for} \,\, i=1,3 \\  
		\tilde{{\alpha}}_{2}  \,\, \quad \text{for} \,\, i=2 
	\end{array}      \right.  , 
\end{eqnarray}
\begin{eqnarray}
\hat{\omega}^{-1} &:&  \tilde{\Delta}_{G_2} \rightarrow \hat{\Delta}_{B_3} , \qquad \tilde{\alpha}_i \mapsto \tilde{\omega}^{-1}(\tilde{\alpha}_i) = \left\{ \begin{array}{ll} 
	\hat{\alpha}_1+ 2 \hat{\alpha}_3  \,\, \quad  \text{for} \,\, i=1 \\  
	3 \hat{\alpha}_1  \,\, \qquad \quad \,\, \text{for} \,\, i=2 
\end{array}      \right. \label{redg21}.
\end{eqnarray}
One may verify that the roots involved reproduce the respective Cartan matrices. The associated  charges are then reduced by the appropriate actions of the coordinates and momenta according to 
\begin{equation}
Q_n^{A_6}(q,p) \rightarrow  \hat{Q}_n^{B_3}({\hat{q},\hat{p}}) = Q_n^{A_6}[\omega^{-1}{(\hat{q})},\omega^{-1}({\hat{p}})]\rightarrow
\tilde{Q}_n^{G_2}({\tilde{q},\tilde{p}}) = \hat{Q}_n^{B_3}[\tilde{\omega}^{-1}{(\tilde{q})},\tilde{\omega}^{-1}({\tilde{p}})] .  \label{redcharges}
\end{equation}	
We will employ the root systems from above, but will construct the $G_2$-charges in a different manner.
	Let us now see in detail how the consecutive steps are carried out.
	\subsection{Higher derivative Hamiltonians from $B_3$ affine Toda lattice theory}
	In order to define the reduced charges according to equation (\ref{redcharges}) we expand the coordinates of the  $B_3$-system as $\hat{q} =  \hat{q}_1 \hat{\alpha}_1 + (\hat{q}_1 +\hat{q}_2) \hat{\alpha}_2 + (\hat{q}_1 +\hat{q}_2 +\hat{q}_3) \hat{\alpha}_3 $ and compute $\omega^{-1}(\hat{q})$ using the defining relation for this map in (\ref{redb3}). We expand the momenta in a similar fashion. Representing the $A_6$-roots in the standard seven dimensional Euclidean space as specified in section 2.2, we obtain in this manner the reduction of the coordinates and momenta
	\begin{eqnarray}
	q &\rightarrow& \omega^{-1}(\hat{q})= (\hat{q}_1,\hat{q}_2,\hat{q}_3,0,-\hat{q}_3,-\hat{q}_2,-\hat{q}_1), \label{red1} \\
		p &\rightarrow& \omega^{-1}(\hat{p})= (\hat{p}_1,\hat{p}_2,\hat{p}_3,0,-\hat{p}_3,-\hat{p}_2,-\hat{p}_1), \label{red2}
	\end{eqnarray}
respectively.
 We notice that when employing the new phase space variables we obtain another solutions of the second equation in Lax pair equations (\ref{laxp}) with $\hat{p}_i = (\hat{x}_i)_t$ for $i=1,2,3$. It is easily seen from (\ref{chargeq1})-(\ref{chargeq7}) that with the replacements (\ref{red1}) and (\ref{red2}) the charges of odd order vanish 
\begin{equation}
		Q_1 \rightarrow	\hat{Q}_1 =0, \qquad Q_3 \rightarrow 	\hat{Q}_3 =0, \qquad Q_5 \rightarrow	\hat{Q}_5 =0,
\end{equation}
and the remaining $B_3$-charges acquire the forms
\begin{eqnarray}
	Q_2 &\rightarrow& 	\hat{Q}_2 = \hat{H} = \sum_{i=1}^3 \hat{p}_i^2 + 2 e^{\hat{q}_1-\hat{q}_2} +2  e^{\hat{q}_2-\hat{q}_3} + 2 e^{\hat{q}_3}+ e^{-2\hat{q}_1}\\
	&& \qquad \qquad  = \sum_{i=1}^3  \hat{p}_i^2 + \sum_{i=1}^3  2e^{\hat{\alpha}_i \cdot \hat{q}}  + 
	 e^{-(\hat{\gamma}+\hat{\alpha}_1) \cdot \hat{q}}  \label{Ham}\\ 
	Q_4 &\rightarrow& 	\hat{Q}_4 =\frac{\hat{p}_1^4}{2}+\frac{\hat{p}_2^4}{2}+\frac{\hat{p}_3^4}{2}+\hat{p}_1^2 e^{-2 \hat{q}_1}+2 \hat{p}_1^2 e^{\hat{q}_1-\hat{q}_2}+2 \hat{p}_2 \hat{p}_1 e^{\hat{q}_1-\hat{q}_2}+2 \hat{p}_2^2
	e^{\hat{q}_1-\hat{q}_2}\\
	&& +2 \hat{p}_3^2 e^{\hat{q}_2-\hat{q}_3}+2 \hat{p}_3^2 e^{\hat{q}_3}+2 \hat{p}_2
	\hat{p}_3 e^{\hat{q}_2-\hat{q}_3}+\frac{1}{2} e^{-4 \hat{q}_1}+e^{2 \hat{q}_1-2
		\hat{q}_2}+2 e^{-\hat{q}_1-\hat{q}_2}+2 e^{\hat{q}_2}  \notag\\
	&&+e^{2 \hat{q}_2-2 \hat{q}_3} +2 \hat{p}_2^2 e^{\hat{q}_2-\hat{q}_3}+2 e^{\hat{q}_1-\hat{q}_3}+2 e^{2 \hat{q}_3} \notag \\ 
	Q_6 &\rightarrow& 	\hat{Q}_6 =\frac{\hat{p}_1^6}{3}+\frac{\hat{p}_2^6}{3}+\frac{\hat{p}_3^6}{3}+\frac{1}{3} e^{-6
		\hat{q}_1}+2 e^{\hat{q}_1}+2 e^{-3 \hat{q}_1-\hat{q}_2}+\frac{2}{3} e^{3 \left(\hat{q}_1-\hat{q}_2\right)}+\frac{2}{3} e^{3
		\left(\hat{q}_2-\hat{q}_3\right)} \\
	&&+\hat{p}_1^4 e^{-2 \hat{q}_1}+\hat{p}_1^2 \left(e^{-4 \hat{q}_1}+4 e^{-\hat{q}_1-\hat{q}_2}+3 e^{2 \left(\hat{q}_1-\hat{q}_2\right)}\right)+\hat{p}_2
	\hat{p}_1 \left(2 e^{-\hat{q}_1-\hat{q}_2}+4 e^{2 \left(\hat{q}_1-\hat{q}_2\right)}\right) \notag \\
	&&+2 \hat{p}_3^4 e^{\hat{q}_3}+\hat{p}_2^2 \left(2
	e^{-\hat{q}_1-\hat{q}_2}+3 e^{2 \left(\hat{q}_1-\hat{q}_2\right)}+2 e^{\hat{q}_2}+3 e^{2 \left(\hat{q}_2-\hat{q}_3\right)}\right)  +\hat{p}_2 \hat{p}_3 \left(4 e^{\hat{q}_2}+4 e^{2
		\left(\hat{q}_2-\hat{q}_3\right)}\right) \notag \\
	&&+2 \left(\hat{p}_1^4+\hat{p}_2 \hat{p}_1^3+\hat{p}_2^2
	\hat{p}_1^2+\hat{p}_2^3 \hat{p}_1+\hat{p}_2^4\right) e^{\hat{q}_1-\hat{q}_2}+2 \left(\hat{p}_2^4+\hat{p}_3 \hat{p}_2^3+\hat{p}_3^2 \hat{p}_2^2+\hat{p}_3^3
	\hat{p}_2+\hat{p}_3^4\right) e^{\hat{q}_2-\hat{q}_3} \notag \\
	&&
	+2 \left(\hat{p}_1^2+\left(2 \hat{p}_2+\hat{p}_3\right) \hat{p}_1+3 \hat{p}_2^2+\hat{p}_3^2+2
	\hat{p}_2 \hat{p}_3\right) e^{\hat{q}_1-\hat{q}_3} +\hat{p}_3^2
	\left(6 e^{\hat{q}_2}+3 e^{2 \left(\hat{q}_2-\hat{q}_3\right)}+4 e^{2 \hat{q}_3}\right)  \notag
\end{eqnarray}
\begin{eqnarray}
	&& +3 e^{-2 \hat{q}_2}+2
	e^{\hat{q}_1+\hat{q}_2-2 \hat{q}_3}+2 e^{-\hat{q}_1-\hat{q}_3}+2 e^{2 \hat{q}_1-\hat{q}_2-\hat{q}_3}+2 e^{2 \hat{q}_2-\hat{q}_3}+\frac{8}{3} e^{3 \hat{q}_3}+4 e^{\hat{q}_2+\hat{q}_3} \notag 
\end{eqnarray}
  Here $\hat{\gamma}$ is the highest root $\hat{\gamma} = \hat{\alpha}_1 +2 \hat{\alpha}_2+2\hat{\alpha}_3$ in $\hat{\Delta}_{B_3}$. We notice that unlike for the $A_n$-case the number of particles already matches the rank of $B_3$ in the standard representation $\hat{\alpha}_1=(1,-1,0)$, $\hat{\alpha}_2=(0,1,-1)$ and $\hat{\alpha}_3=(0,0,1)$.
  
  \begin{figure}[h]
  	\centering         
  	\begin{minipage}[b]{0.49\textwidth}     
  		\includegraphics[width=\textwidth]{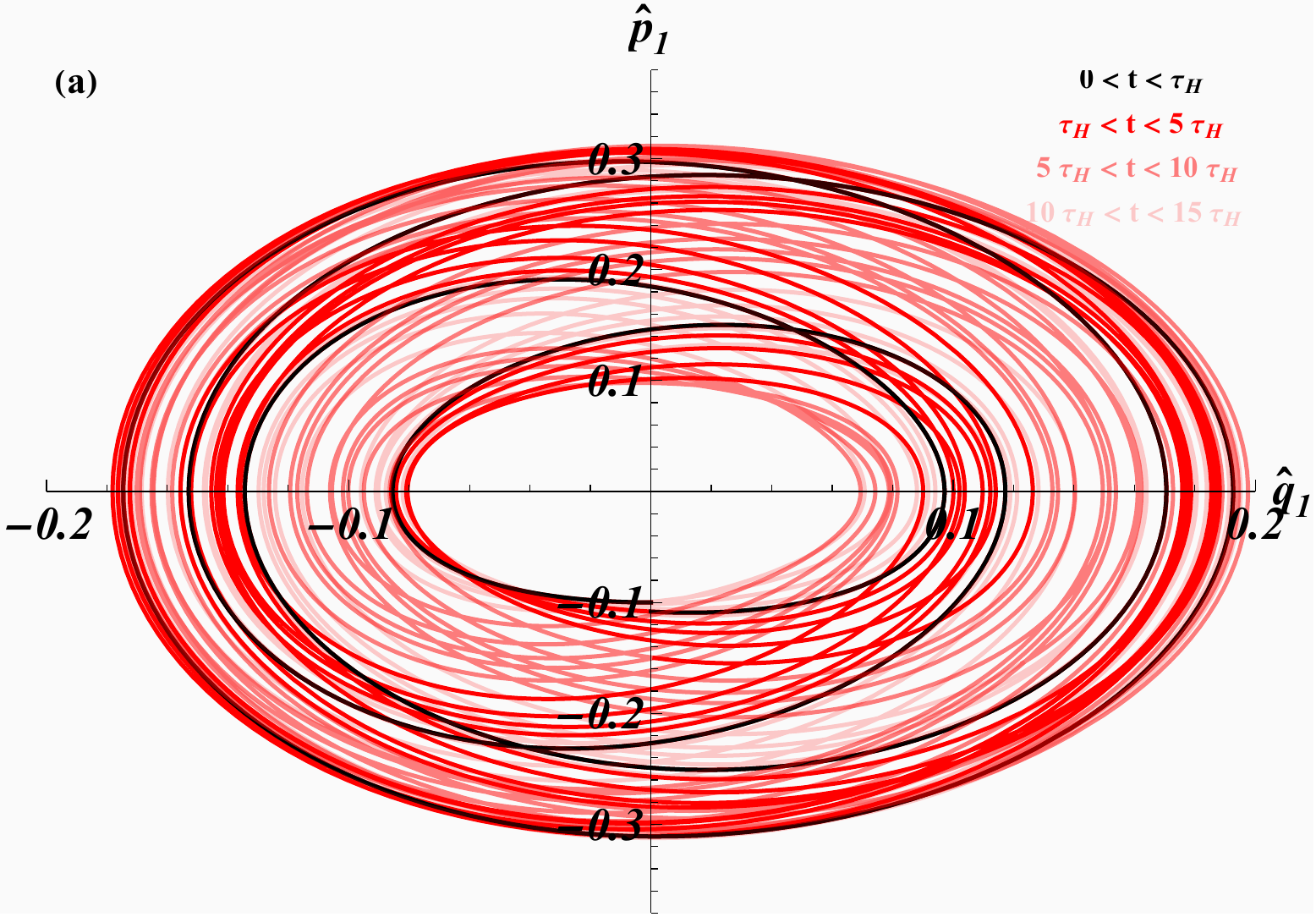}
  	\end{minipage}   
  	\begin{minipage}[b]{0.245\textwidth}        
  		\includegraphics[width=\textwidth,height=1.7cm]{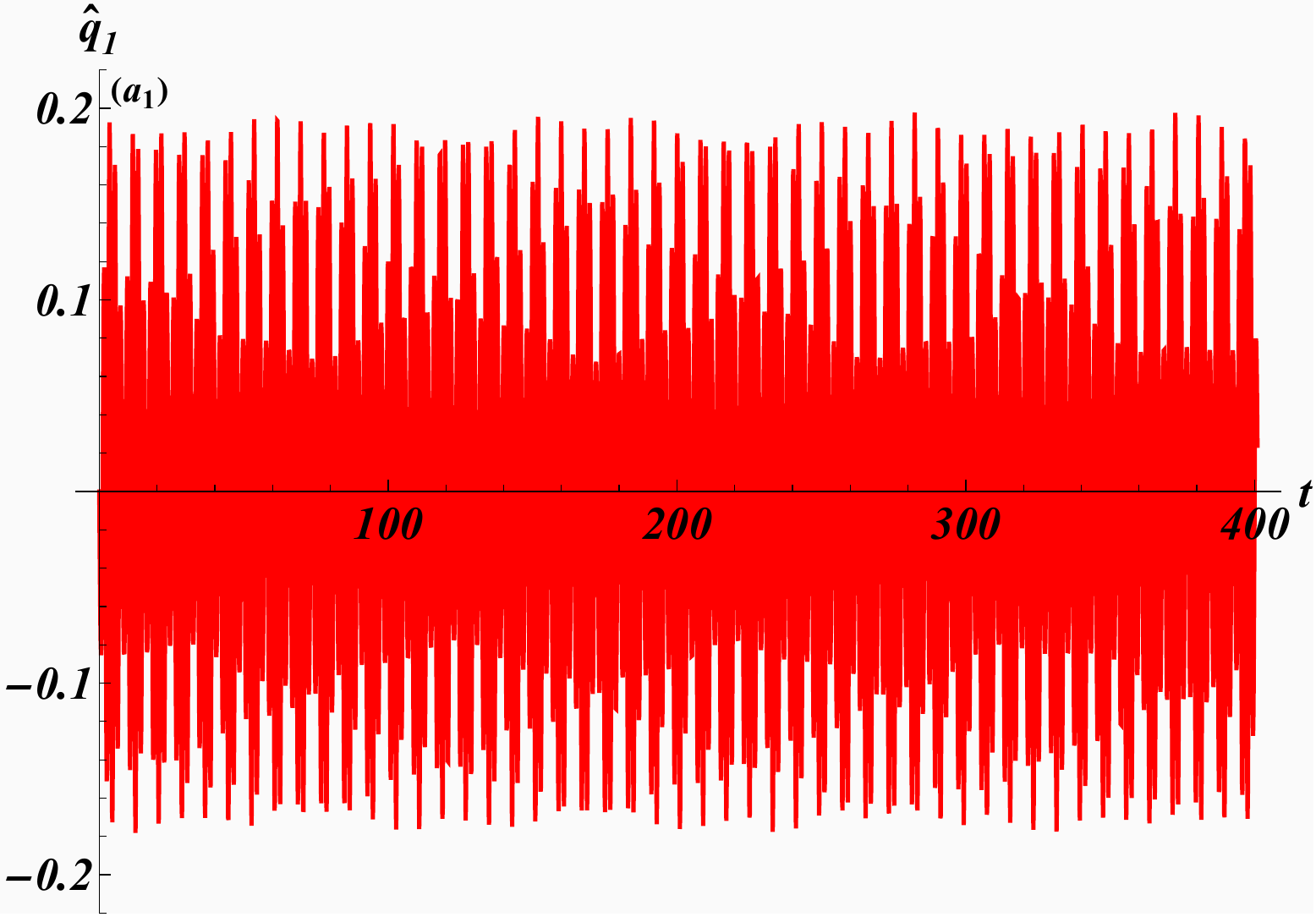}
  		\includegraphics[width=\textwidth,height=1.7cm]{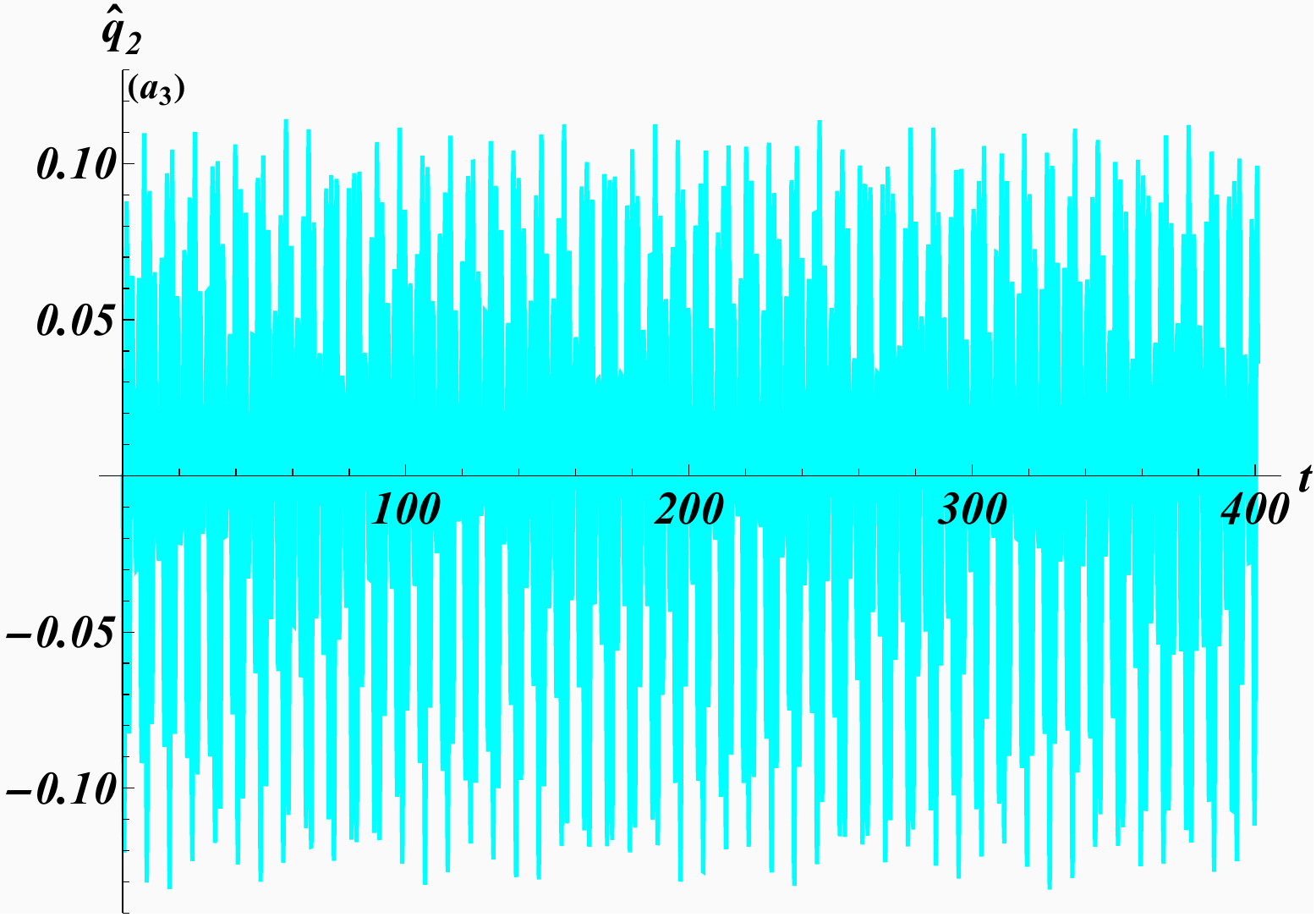}
  		\includegraphics[width=\textwidth,height=1.7cm]{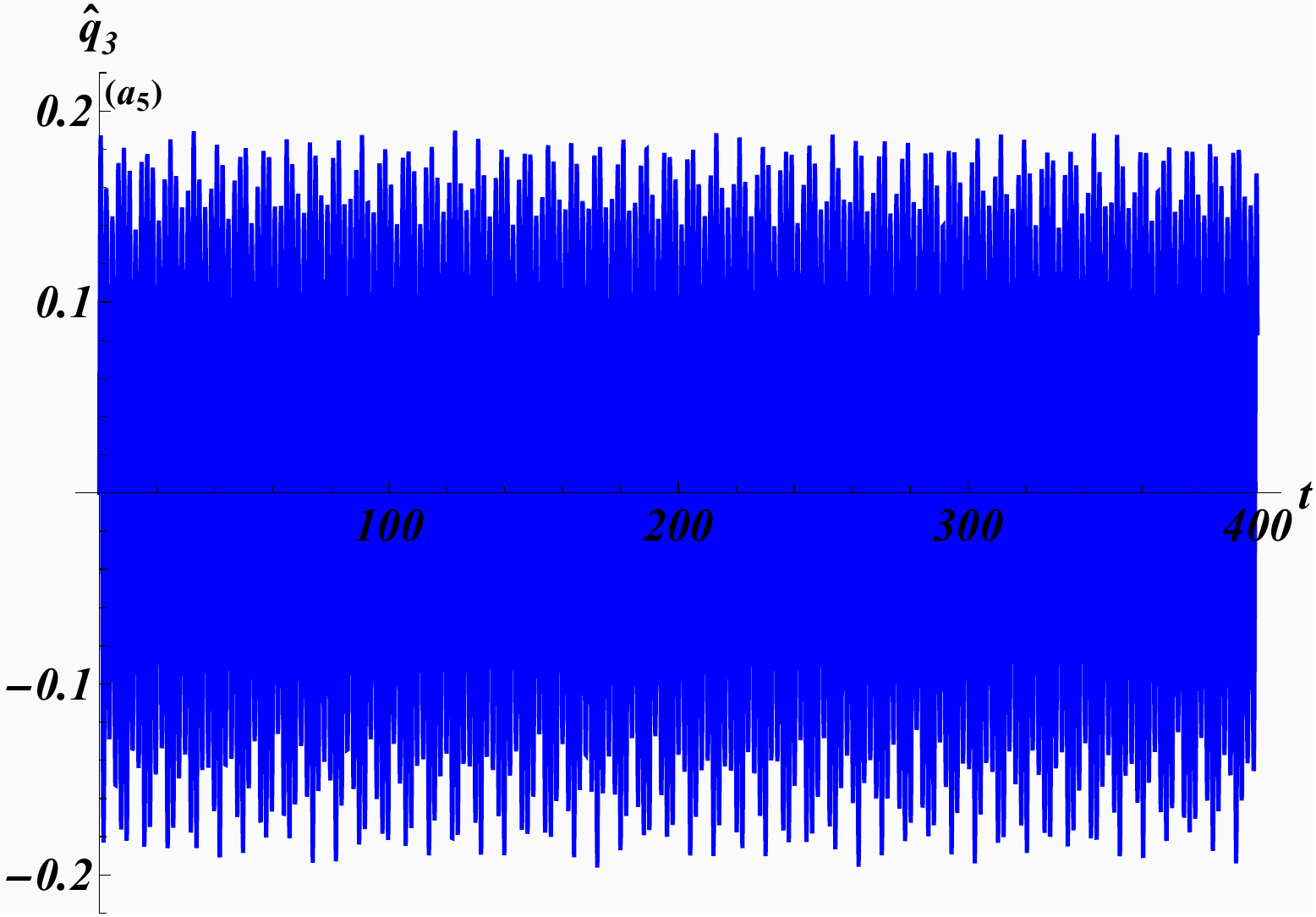}
  	\end{minipage}   
  	\begin{minipage}[b]{0.245\textwidth}        
  		\includegraphics[width=\textwidth,height=1.7cm]{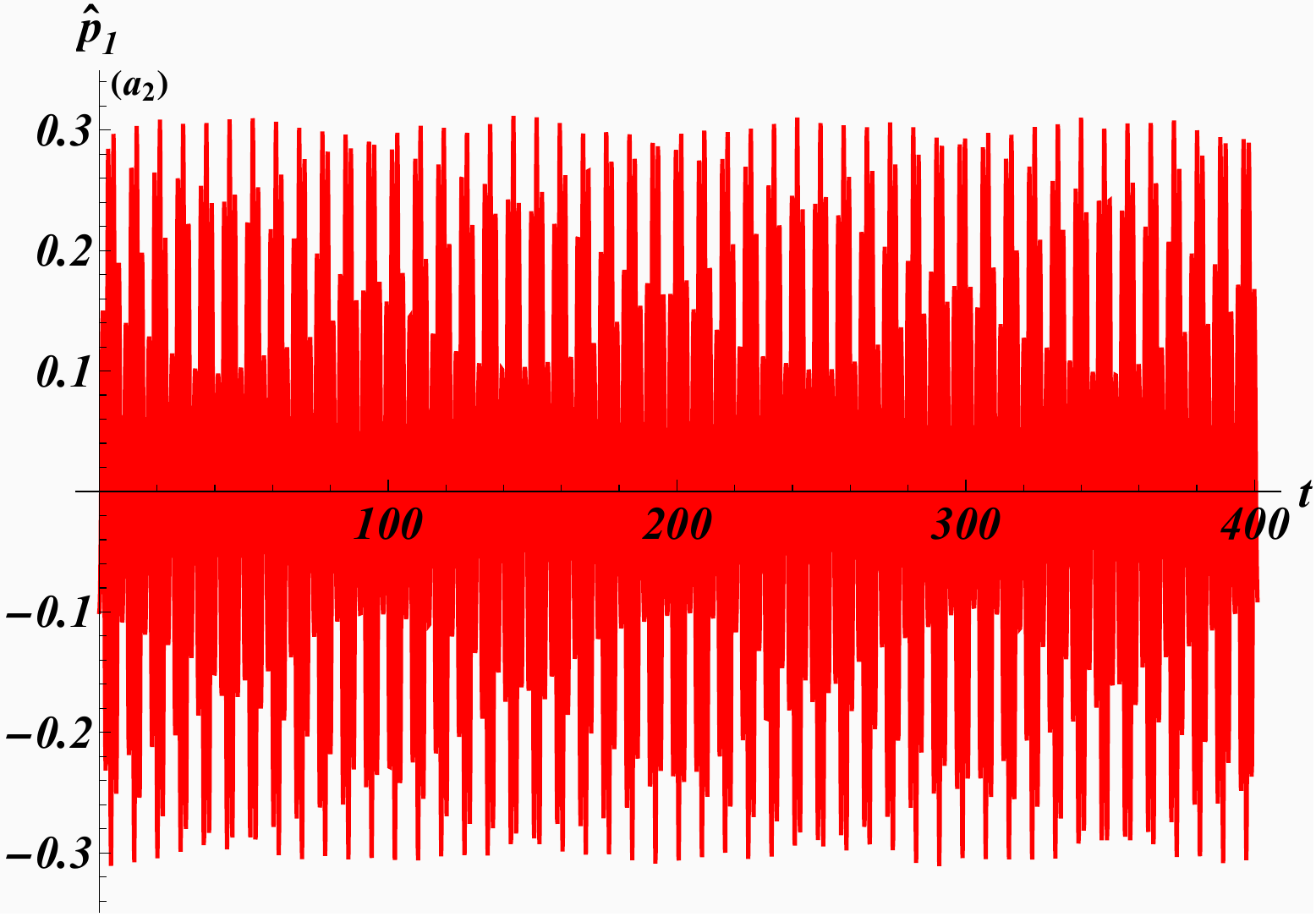}
  		\includegraphics[width=\textwidth,height=1.7cm]{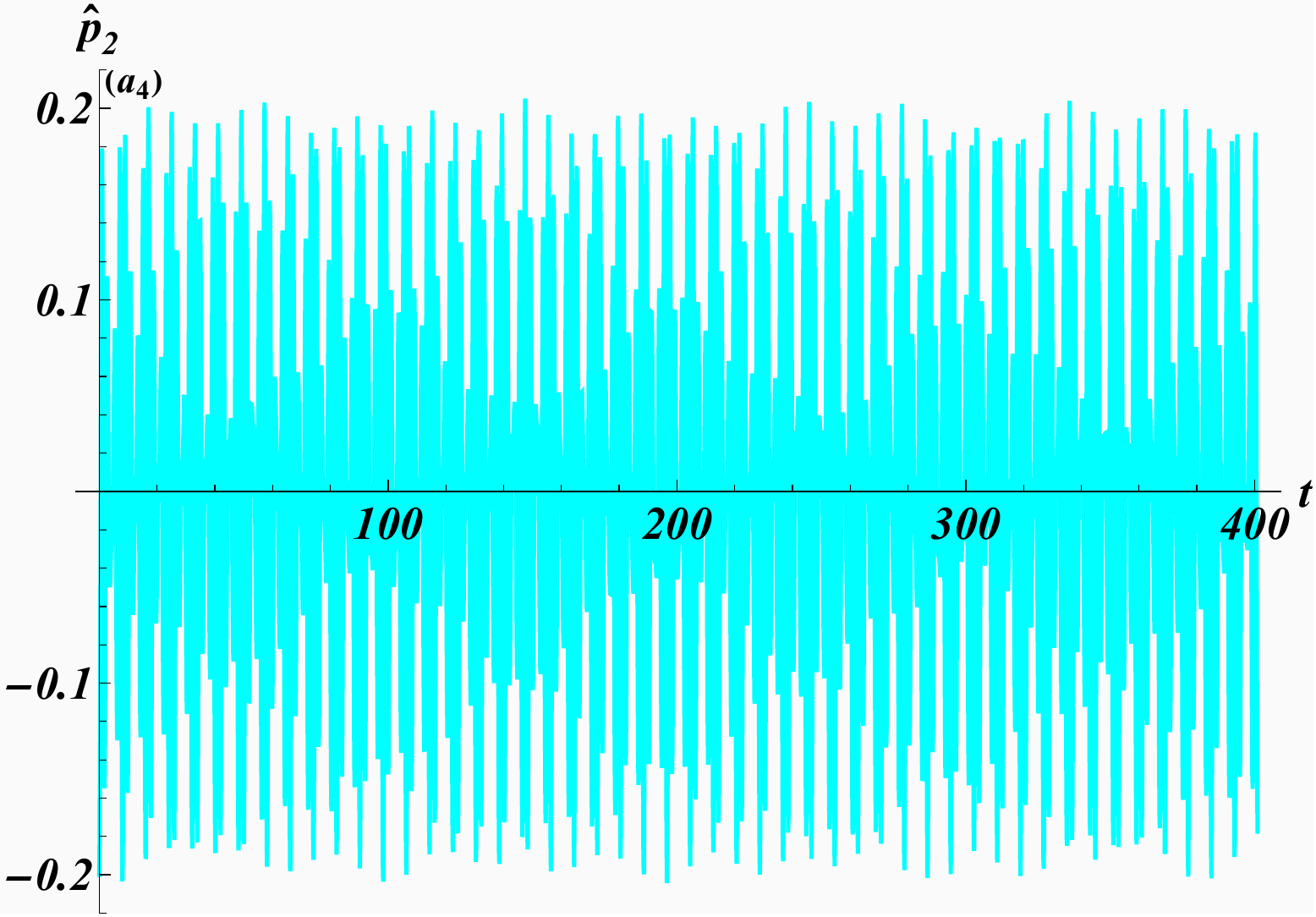}
  		\includegraphics[width=\textwidth,height=1.7cm]{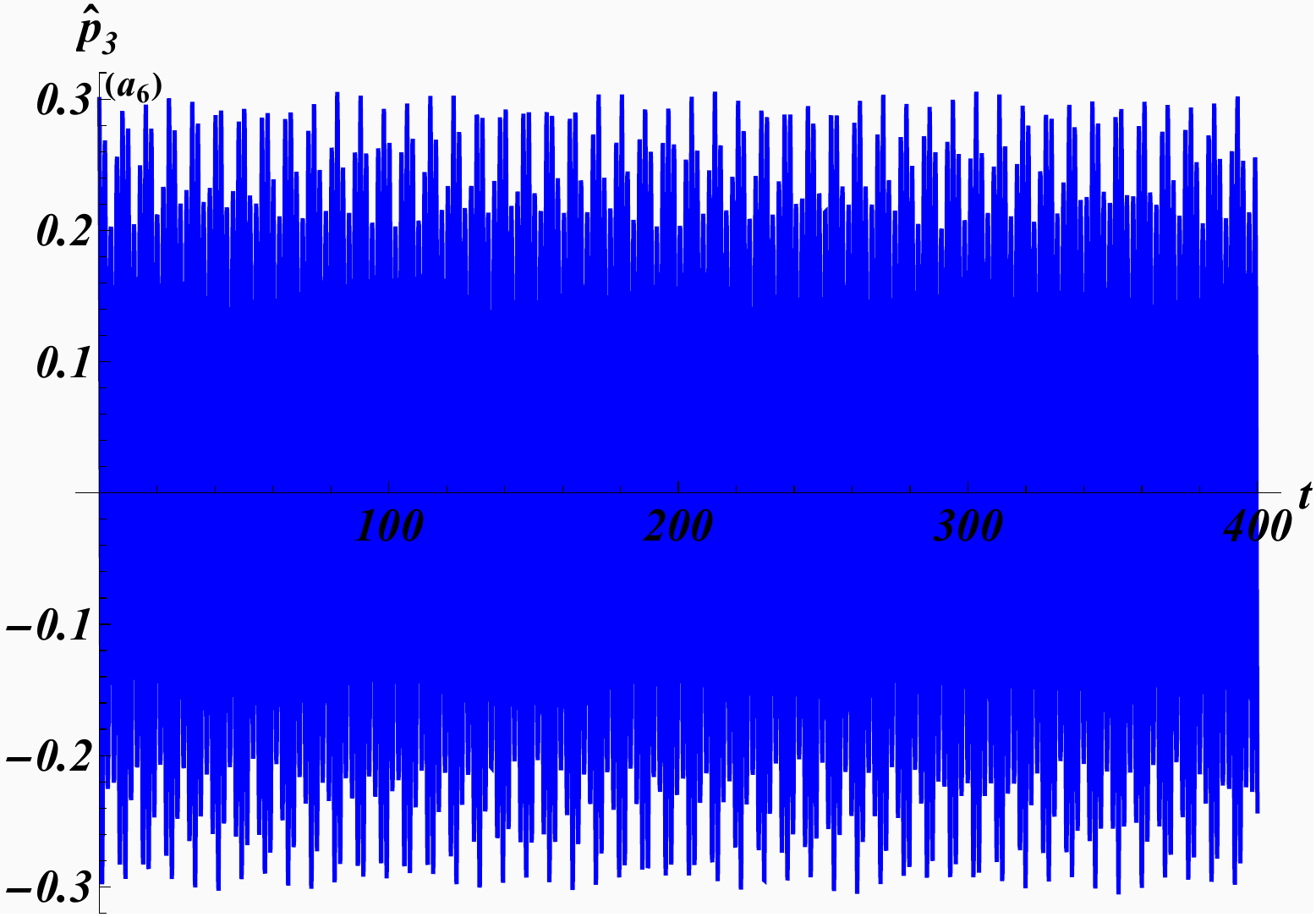}
  	\end{minipage}   
  	\begin{minipage}[b]{0.49\textwidth}           \includegraphics[width=\textwidth]{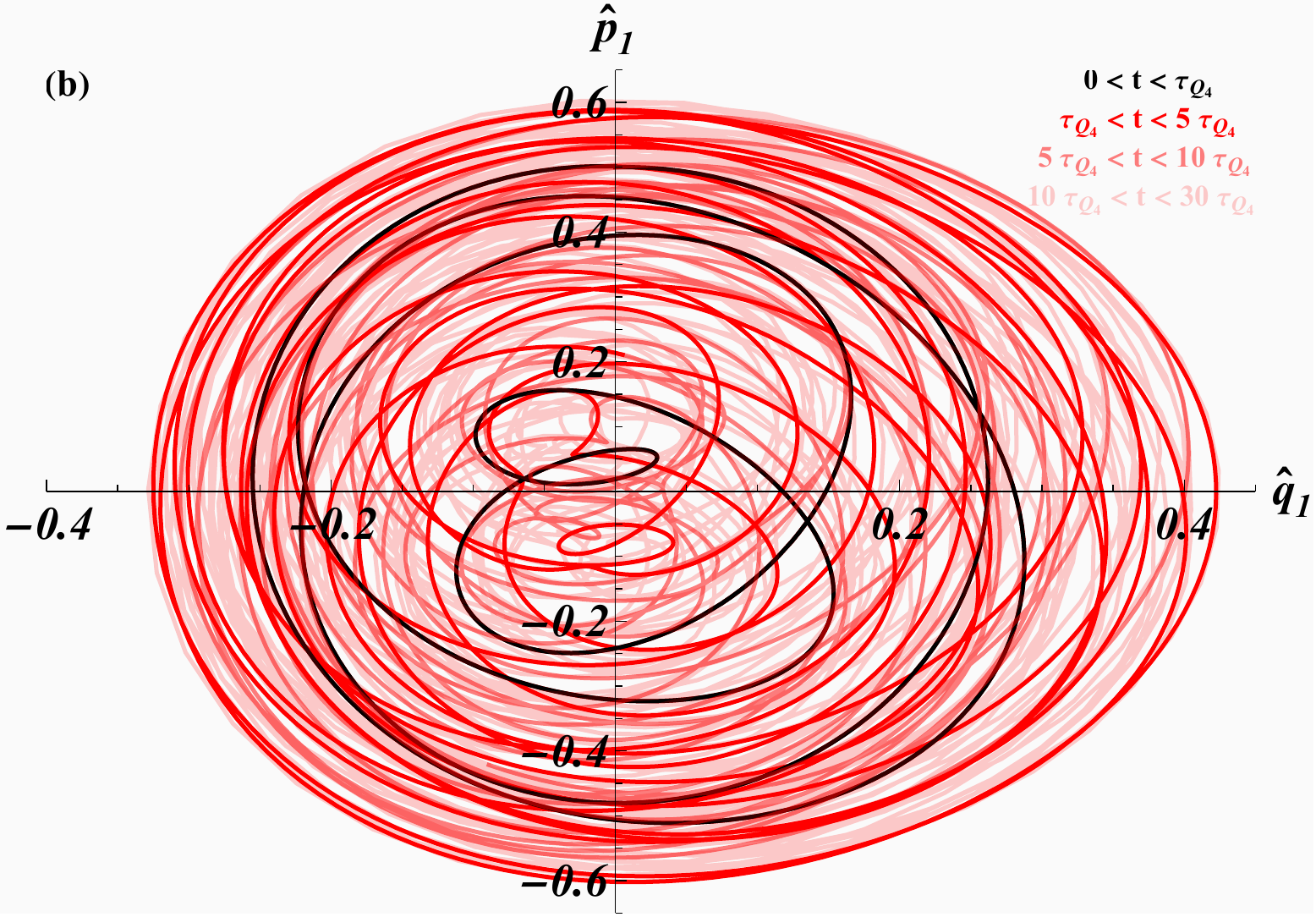}
  	\end{minipage}   
  	\begin{minipage}[b]{0.245\textwidth}        
  		\includegraphics[width=\textwidth,height=1.7cm]{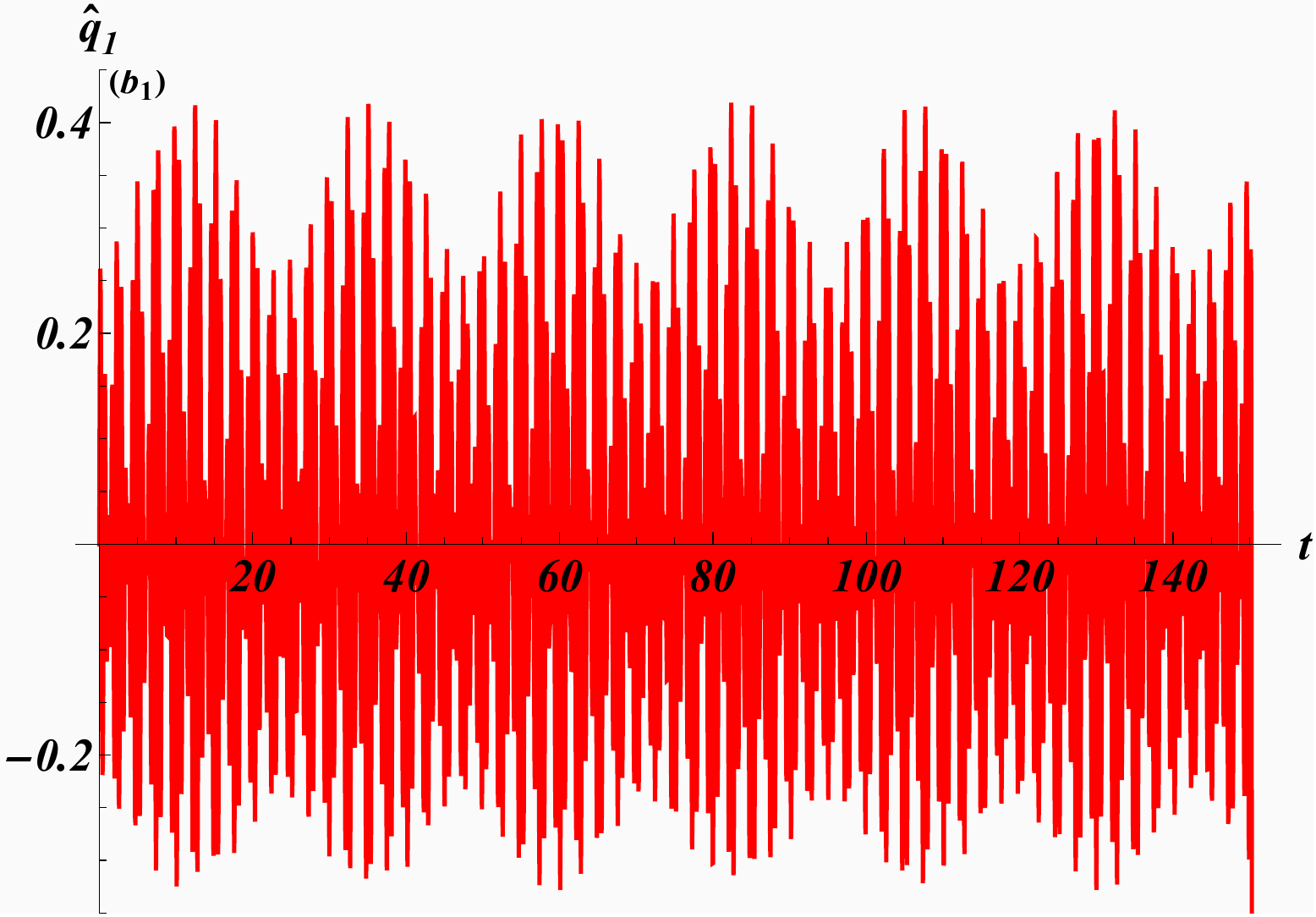}
  		\includegraphics[width=\textwidth,height=1.7cm]{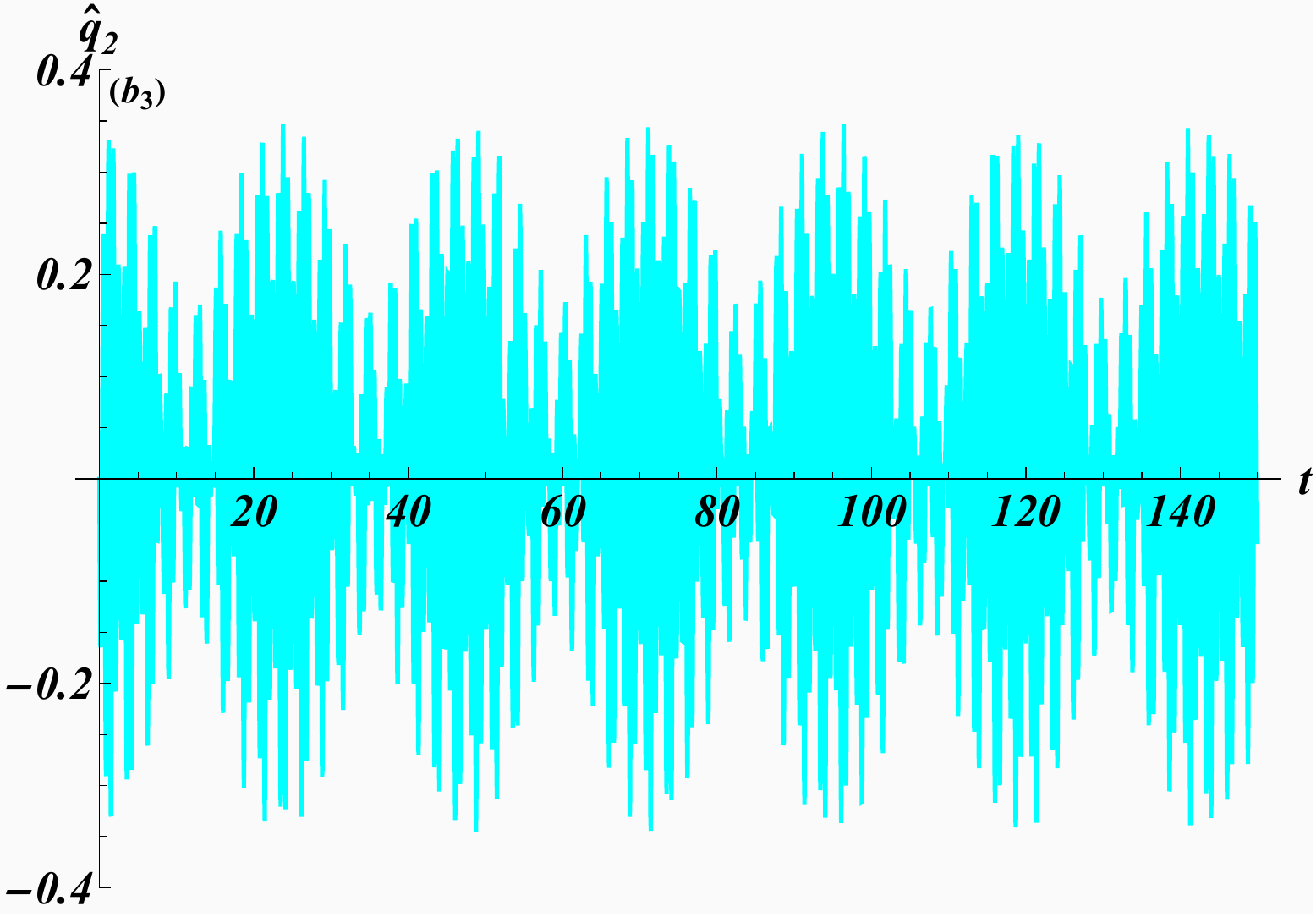}
  		\includegraphics[width=\textwidth,height=1.7cm]{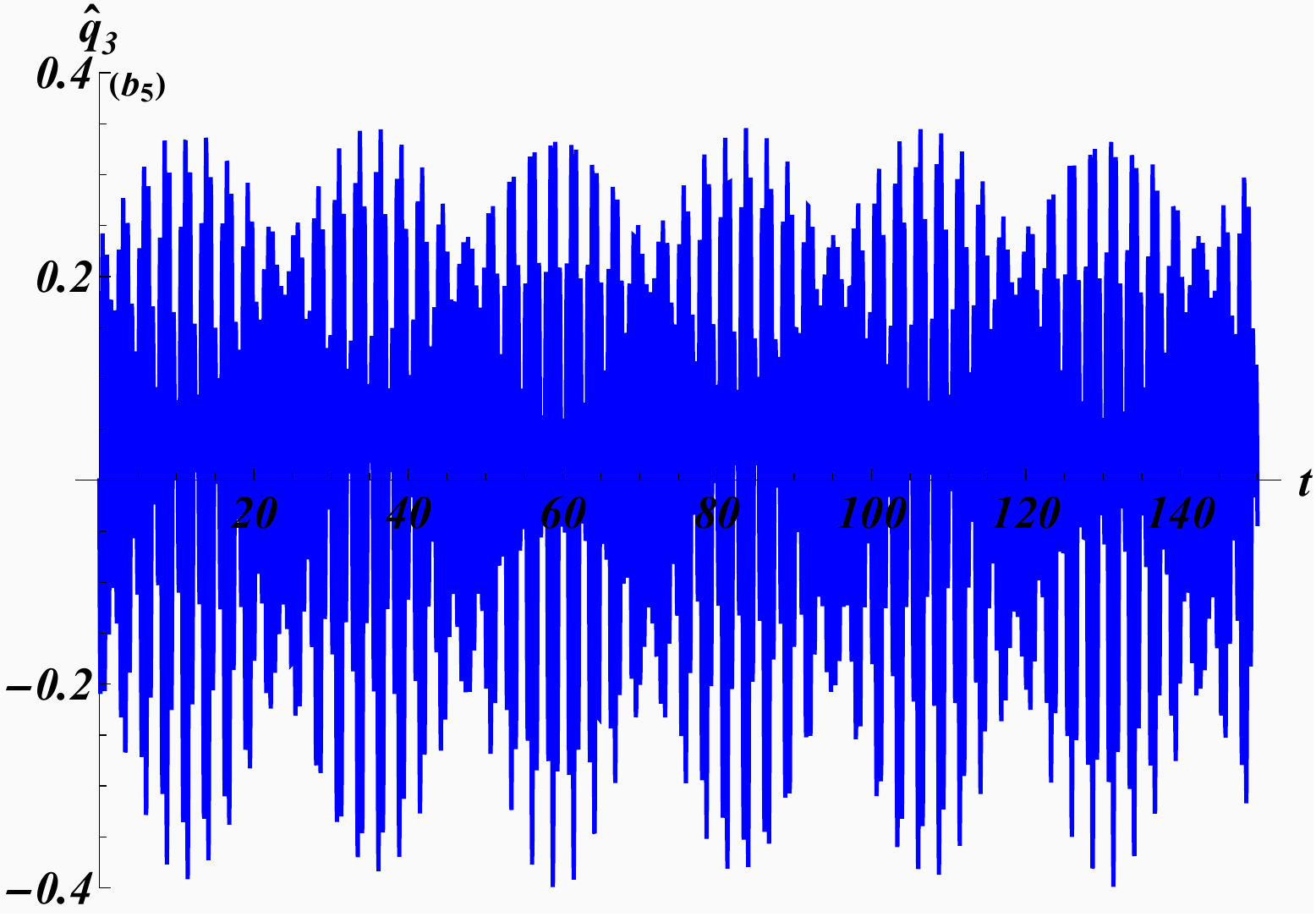}
  	\end{minipage}   
  	\begin{minipage}[b]{0.245\textwidth}        
  		\includegraphics[width=\textwidth,height=1.7cm]{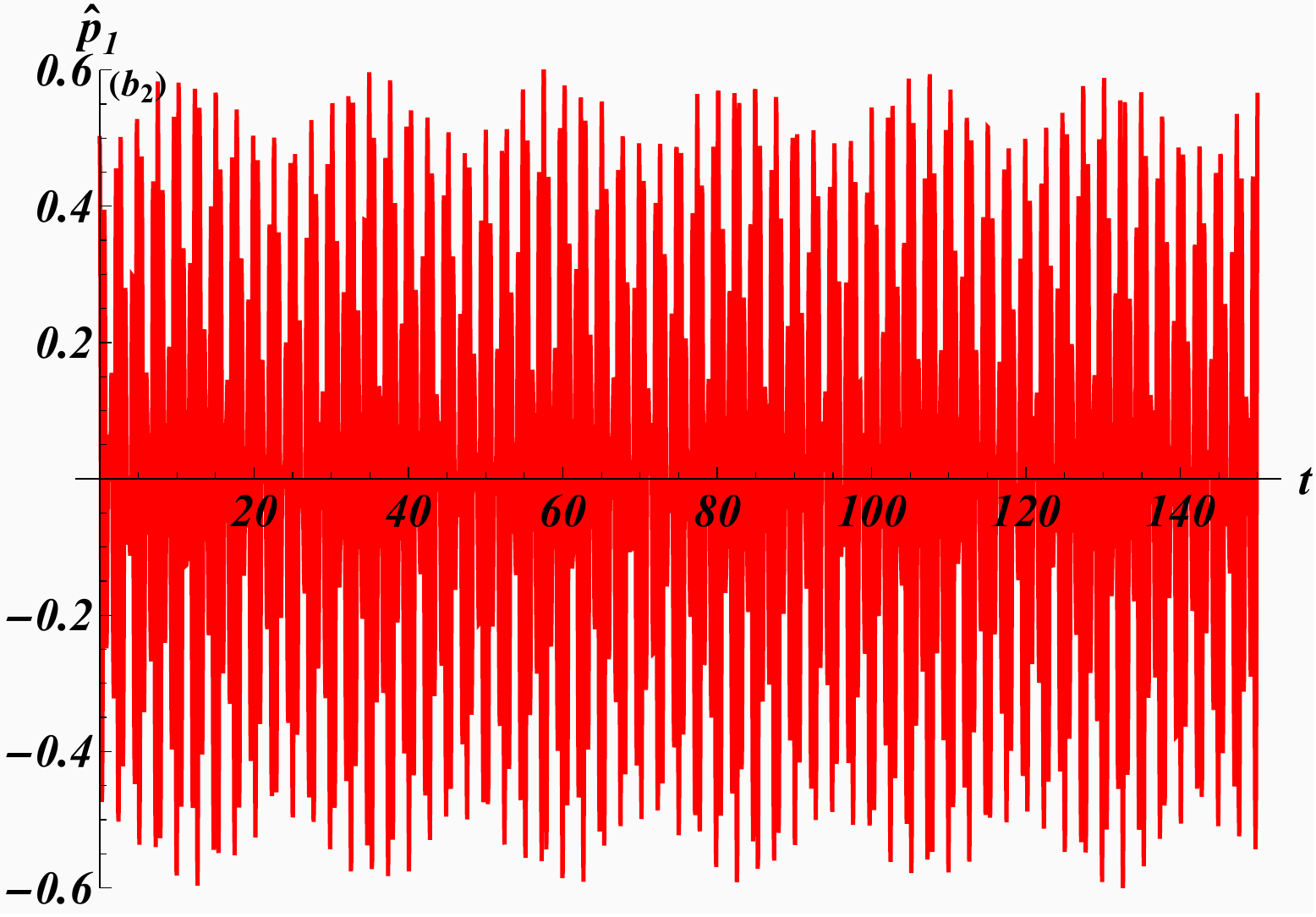}
  		\includegraphics[width=\textwidth,height=1.7cm]{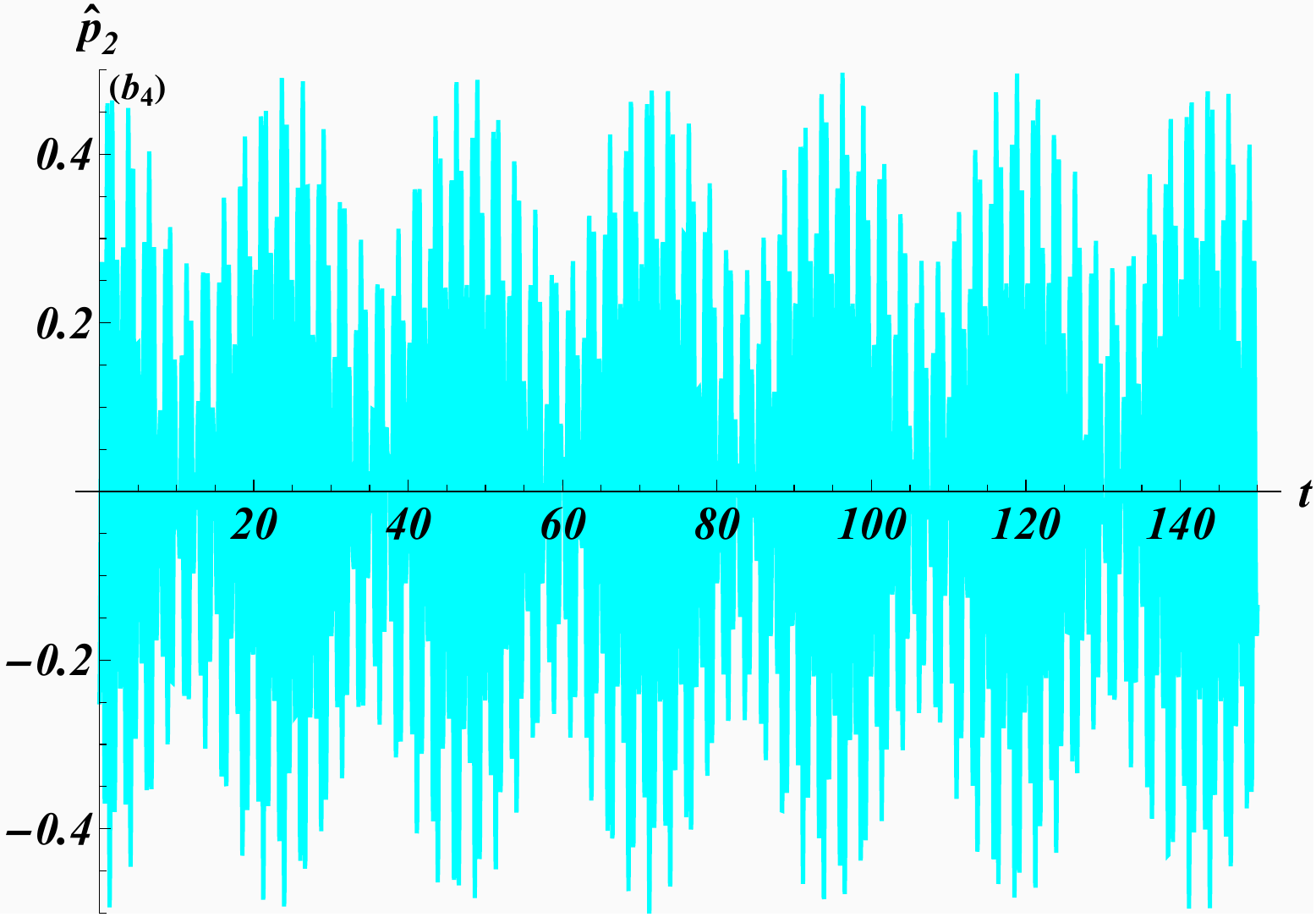}
  		\includegraphics[width=\textwidth,height=1.7cm]{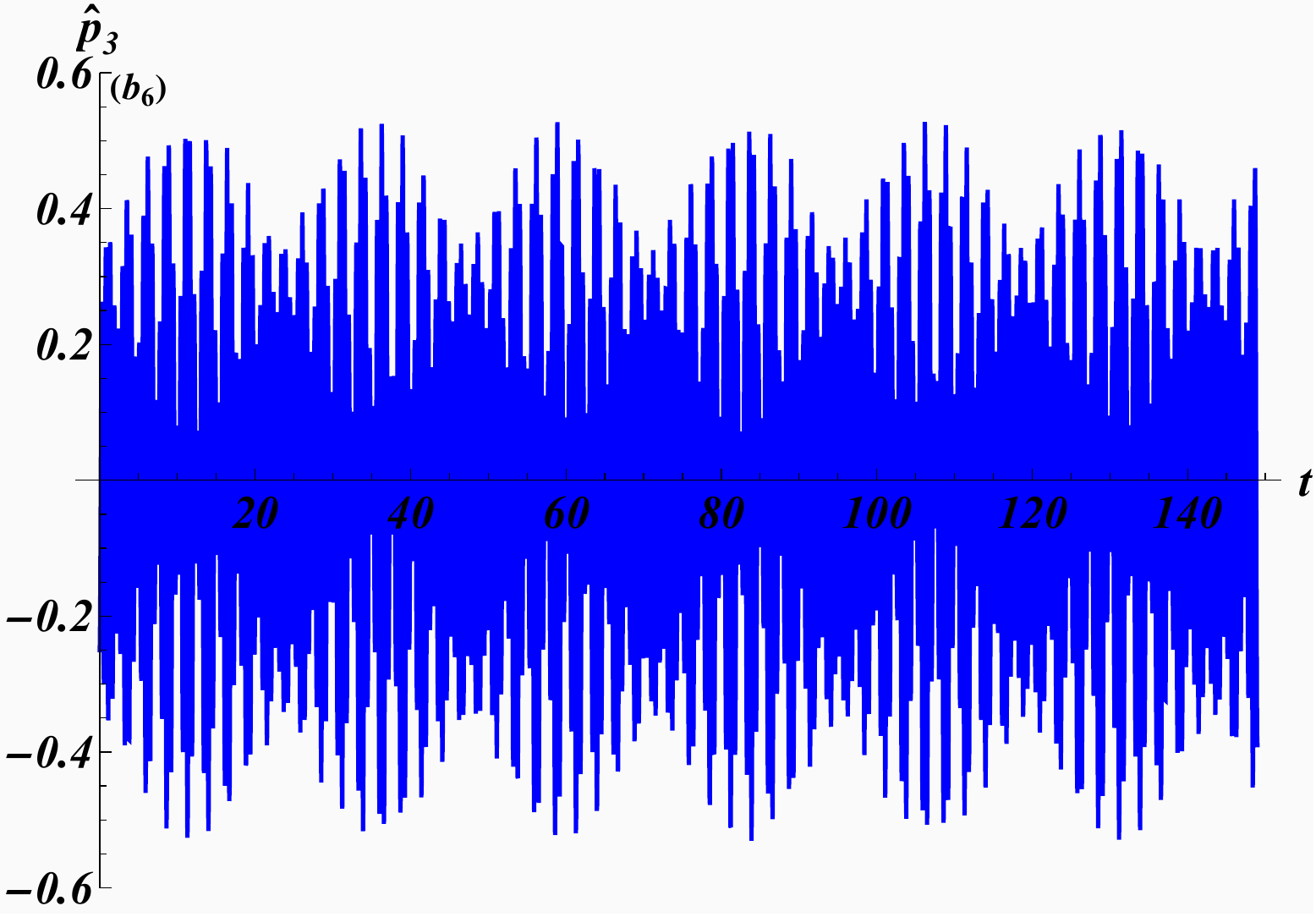}
  	\end{minipage}  
  	\begin{minipage}[b]{0.49\textwidth}           \includegraphics[width=\textwidth]{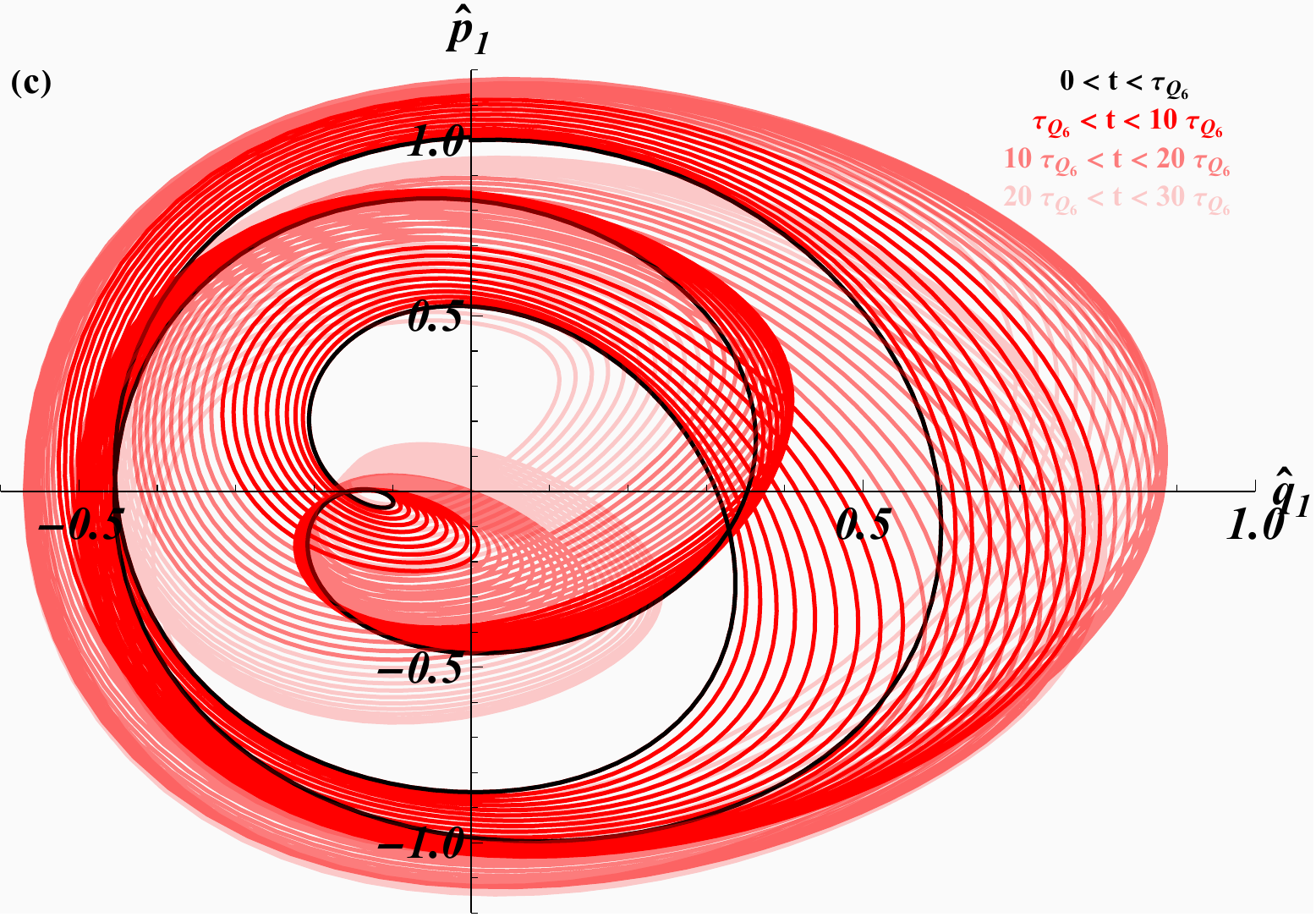}
  	\end{minipage}   
  	\begin{minipage}[b]{0.245\textwidth}        
  		\includegraphics[width=\textwidth,height=1.7cm]{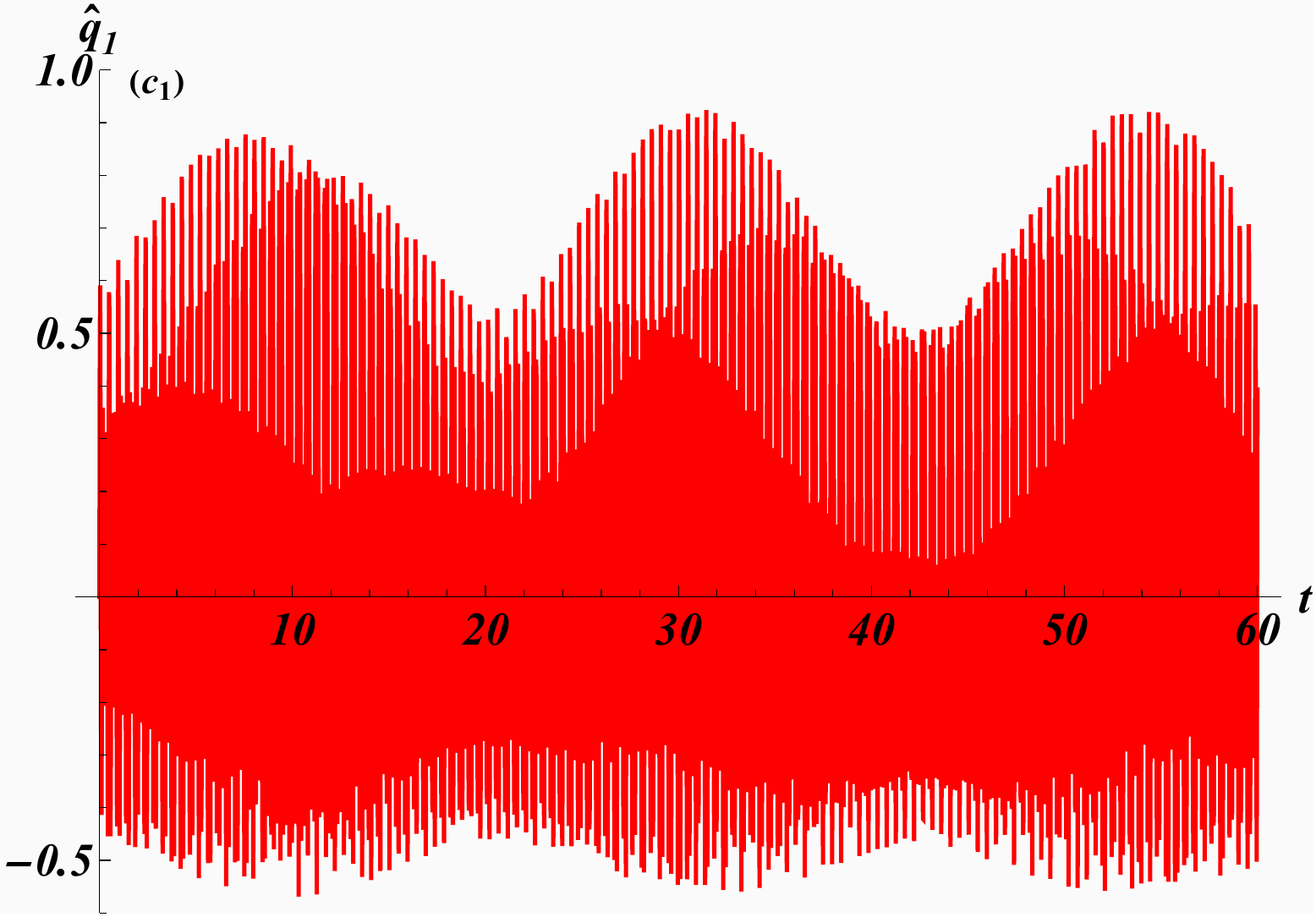}
  		\includegraphics[width=\textwidth,height=1.7cm]{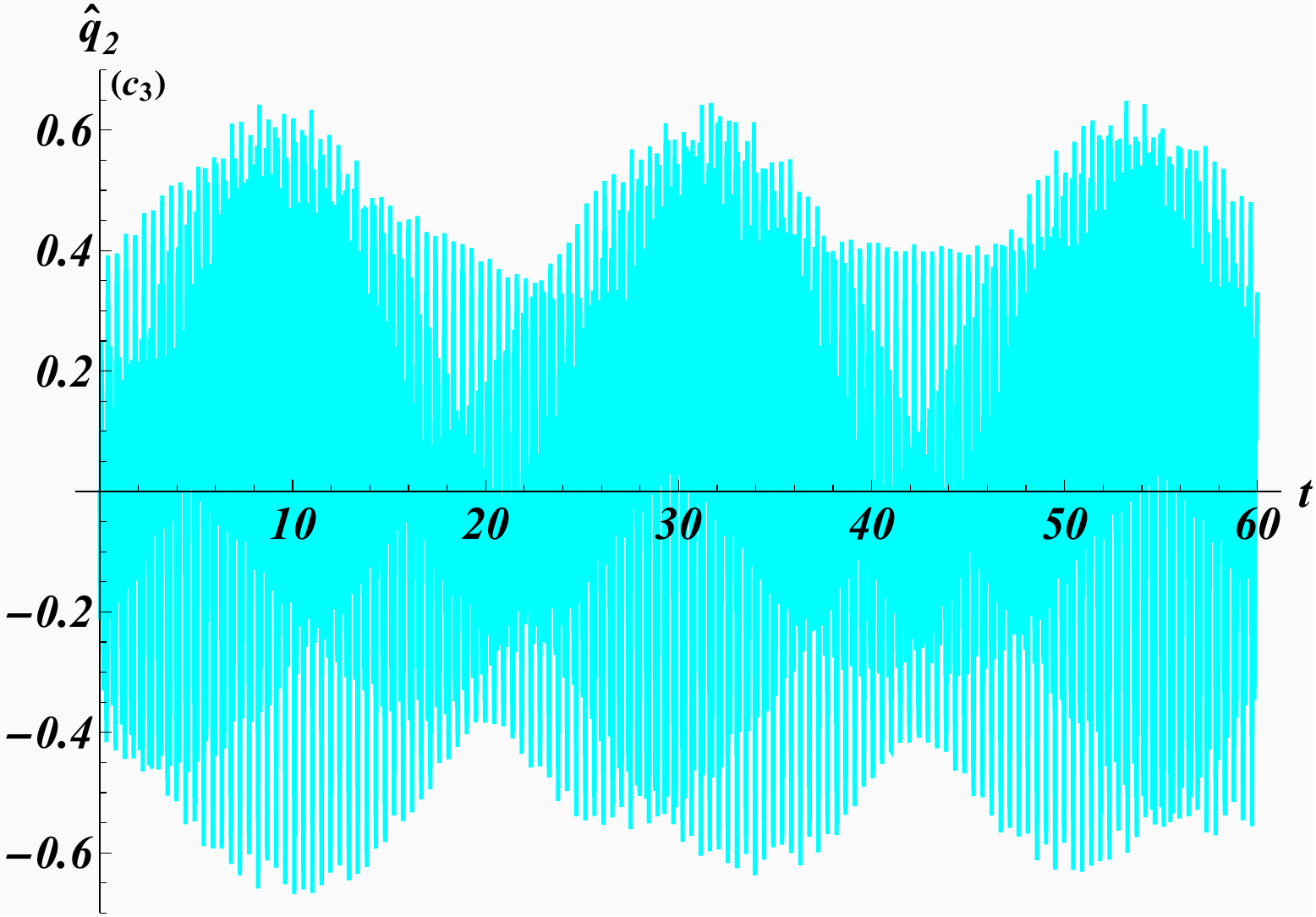}
  		\includegraphics[width=\textwidth,height=1.7cm]{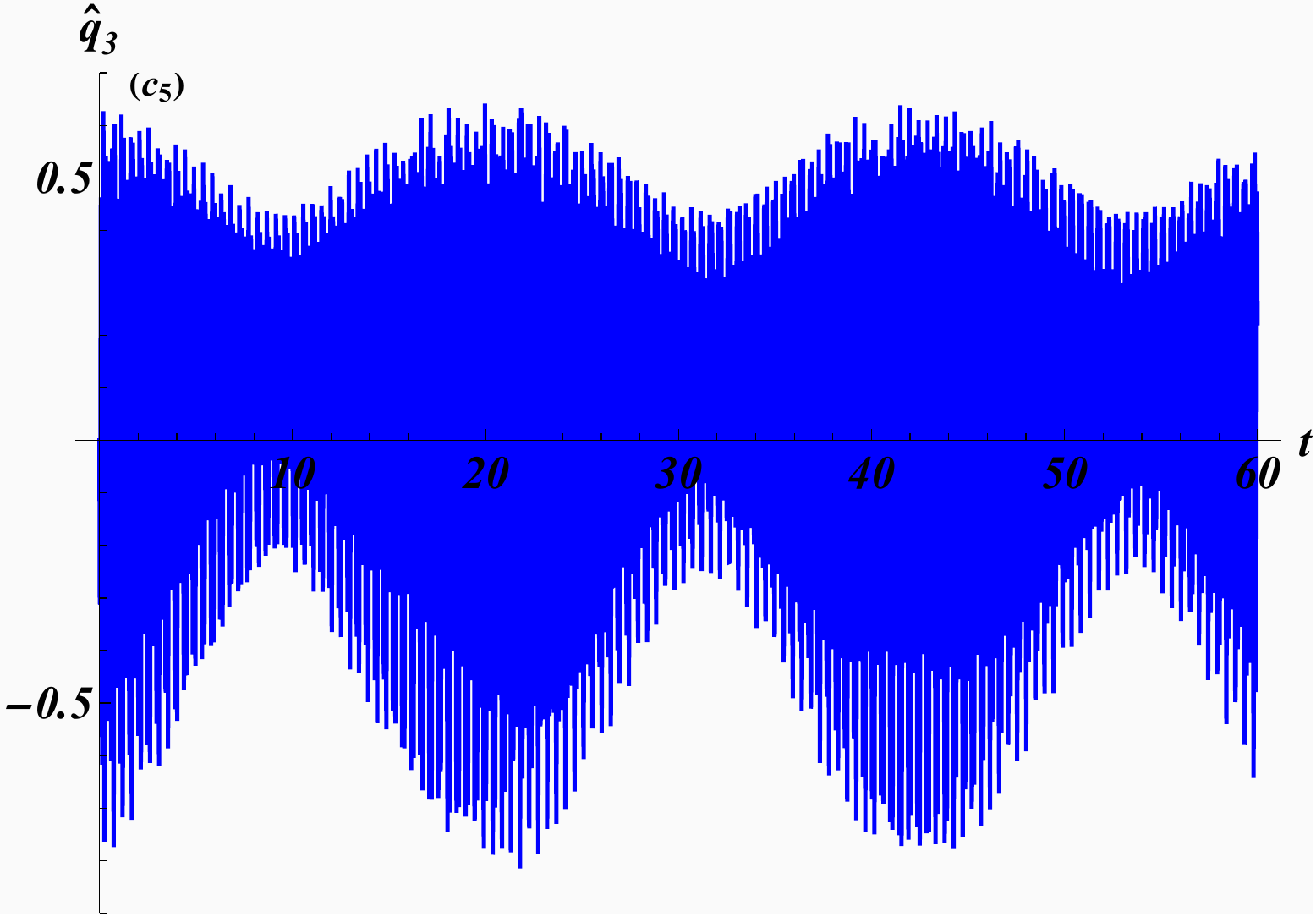}
  	\end{minipage}   
  	\begin{minipage}[b]{0.245\textwidth}        
  		\includegraphics[width=\textwidth,height=1.7cm]{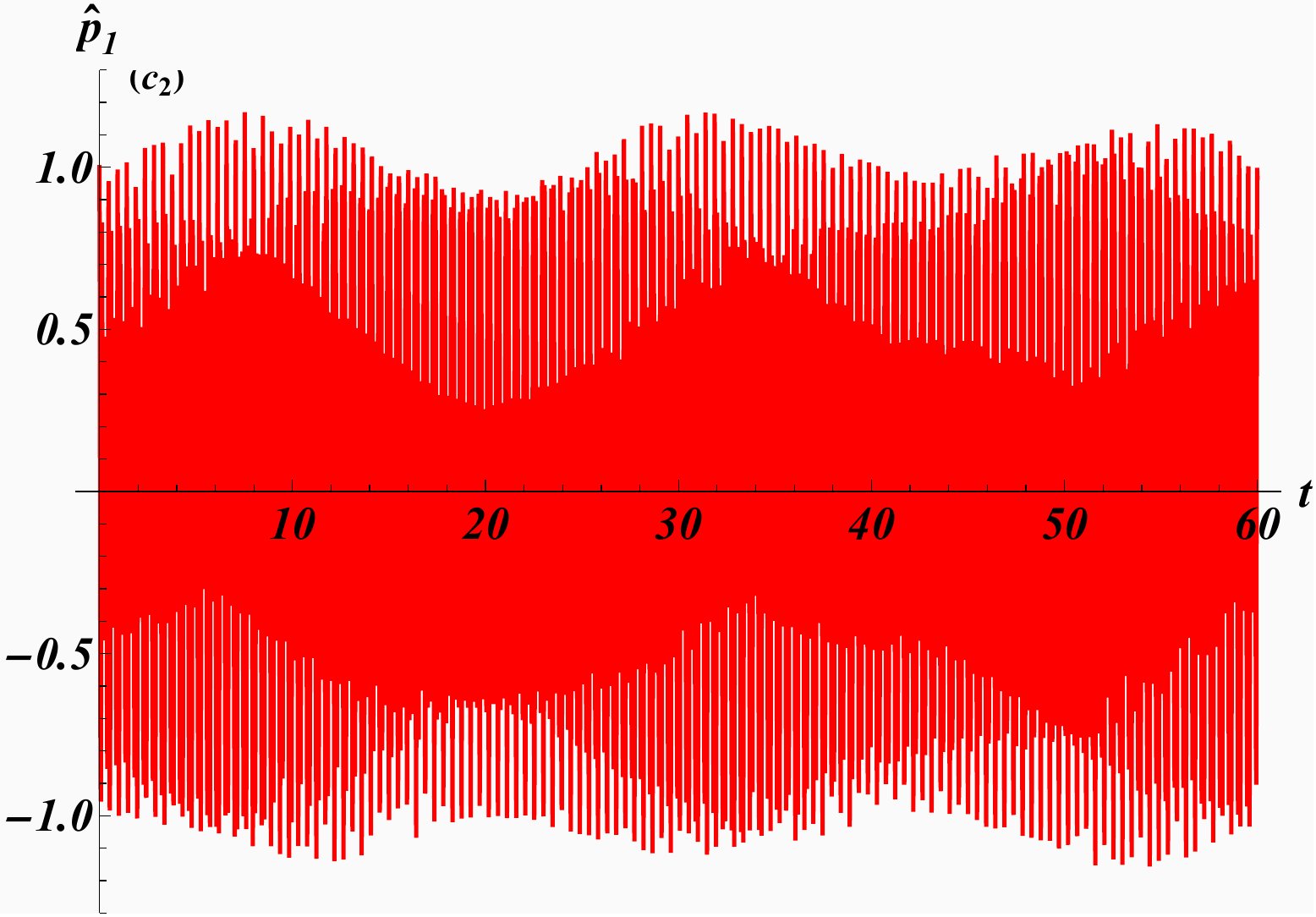}
  		\includegraphics[width=\textwidth,height=1.7cm]{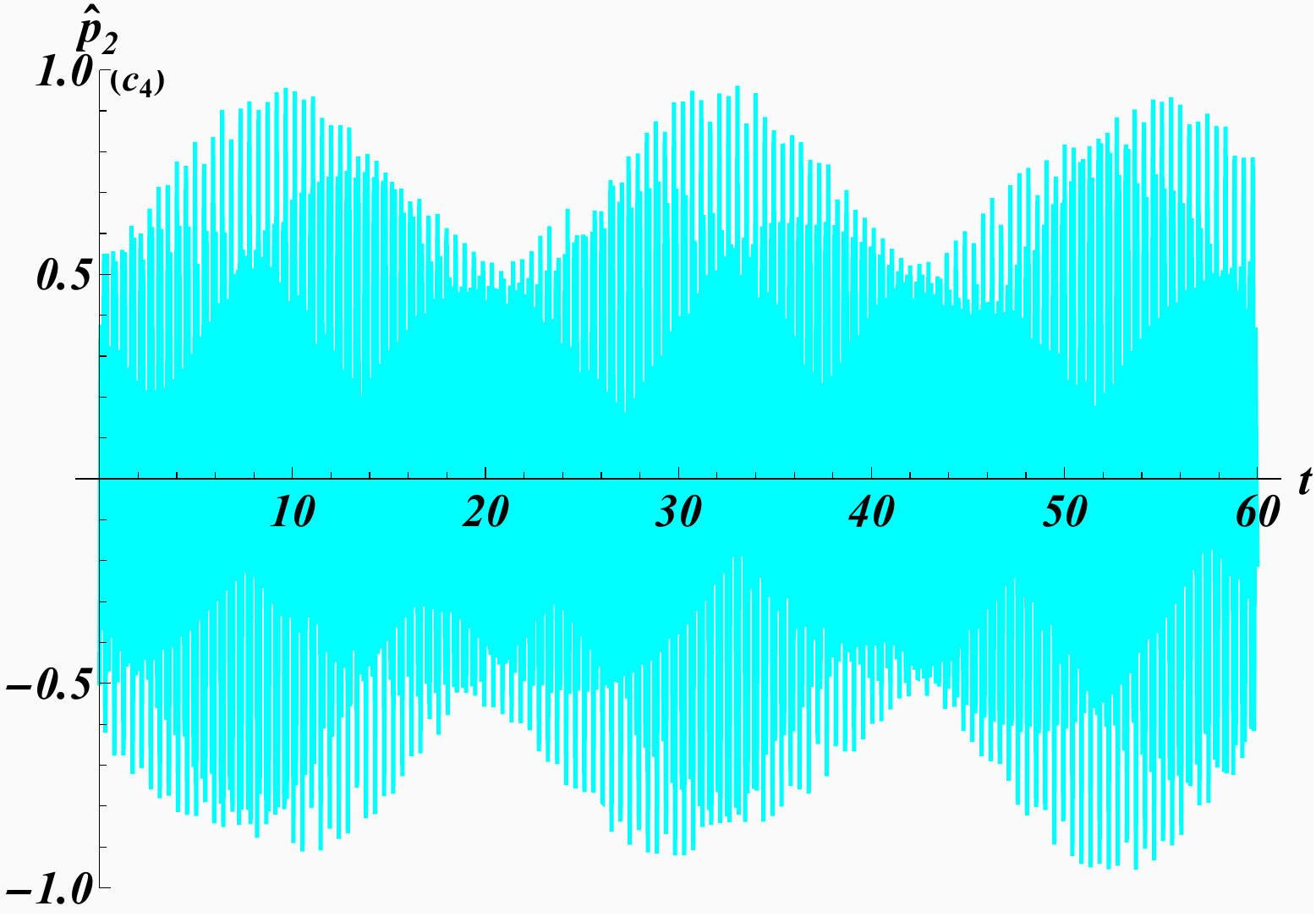}
  		\includegraphics[width=\textwidth,height=1.7cm]{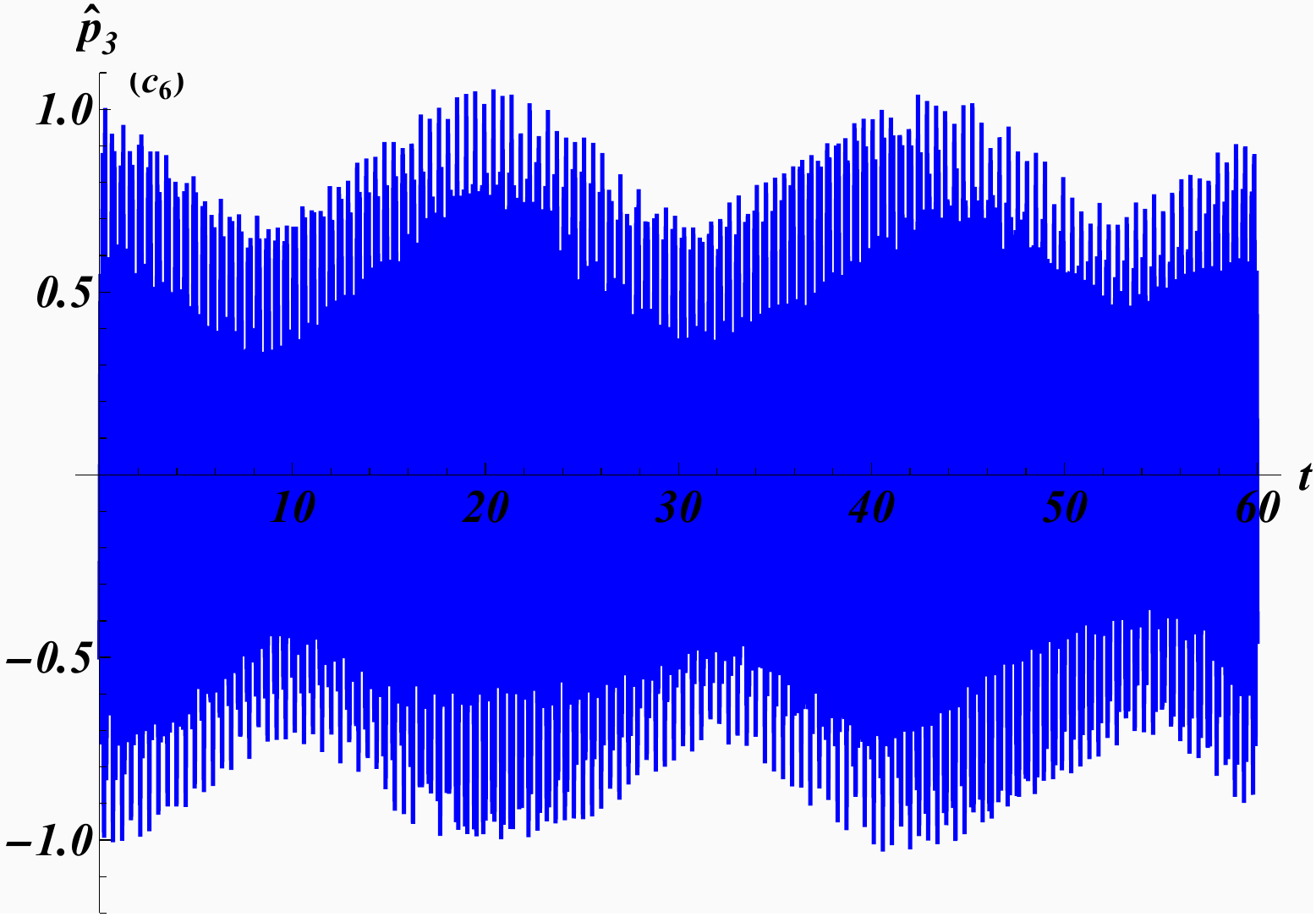}
  	\end{minipage}  
  	\caption{Affine $B_3$-Toda lattice phase space $(\hat{q}_1,\hat{p}_1)$, as function of $t$ for the standard Hamiltonian in panel (a), for $\hat{Q}_4$ taken as higher derivative Hamiltonian in panel (b) and for $\hat{Q}_6$ taken as higher derivative Hamiltonian in panel (c). The corresponding functions $x_i(t)$ and $p_i(t)$ are displayed in the respective panels $a_i,b_i,c_i$ for $i=1,\ldots,6$. For the initial condition we always chose $\hat{q}_1(0) =\hat{q}_2(0)=\hat{q}_3(0)=0$ and $\hat{p}_1(0)=-0.1$,  $\hat{p}_2(0)=-0.2$, $\hat{p}_3(0)=0.3$, in panels (a),  $\hat{p}_1(0)=0.5$,  $\hat{p}_2(0)=\hat{p}_3(0)=-0.25$, in panels (b) and $\hat{p}_1(0)=1$,  $\hat{p}_2(0)=\hat{p}_3(0)=-0.5$ in panels (c). The quasi-periods are: panel (a): $\tau_{H} \approx 7.9824 $,  panel (b): $\tau_{Q_4}\approx  2.7181 $,  panel (c): $\tau_{\hat{Q}_6} \approx 0.4719 $.} 
  	\label{PhaseB3}
  \end{figure}

The charges are once more all in involution, but there is no equivalent to relation (\ref{Poisson}) in this case. Instead we find that $\left\{ \hat{H} , \left\{ \hat{\chi} , \tilde{Q}_{4/6}  \right\}   \right\} \neq 0$, i.e. the quantities $\left\{ \hat{\chi} , \tilde{Q}_{4/6}  \right\}$ are not conserved. 
  
   Again we interpret all charges as Hamiltonians and compute their corresponding phase spaces. As depicted in figure \ref{PhaseB3}, all trajectories are of oscillatory nature and are confined in phase space. 
  
  As in this case the dimension of the standard representation already equals the rank of the algebra there was no need for a reduction or the imposition of any constraints in order to obtain trajectories of oscillatory nature for the higher charge Hamiltonians. Nonetheless, for the sake of interest we consider now the reverse scenario and construct a theory in which the roots are represented in a larger dimensional space. When drawing on the $A_2$-example one might expect divergent trajectories in this case, but as we will demonstrate this is not the case.
  
  Thus we solve once more equation (\ref{redmat}) for the orthogonal matrix $A$ and the four-dimensional roots $\hat{\beta}_i$ reproducing the $B_3$-Cartan matrix
  \begin{equation}
  	K=  \left(  \begin{array}{ccc}
  		2 & -1 & 0 \\
  		-1 & 2 & -2 \\
  		0 & -1 & 2
  	\end{array}   \right).
  \end{equation}
We find the four dimensional representation for the roots
  \begin{equation}
  	\hat{\beta}_1=(1,0,1,0)\; \; \hat{\beta}_2=\left(-\frac{1}{2},\frac{\sqrt{3}}{2},-\frac{1}{2},\frac{\sqrt{3}}{2}\right)\; \; \hat{\beta}_3=\left(0,-\frac{2+\sqrt{2}}{2\sqrt{3}},0,-\frac{2-\sqrt{2}}{2\sqrt{3}}\right), 
  \end{equation}
  together with the orthogonal matrix
  \begin{equation}
  	\hat{A}=\left(\begin{matrix}
  		\frac{1}{2} & \frac{1-\sqrt{2}}{2\sqrt{3}} & \frac{1}{2} & \frac{1+\sqrt{2}}{2\sqrt{3}} \\
  		-\frac{1}{2} &\frac{1-\sqrt{2}}{2\sqrt{3}} & -\frac{1}{2} & \frac{1+\sqrt{2}}{2\sqrt{3}} \\
  		0 & -\frac{2+\sqrt{2}}{2\sqrt{3}} & 0 &  \frac{1-\sqrt{2}}{\sqrt{6}} \\
  		-\frac{1}{\sqrt{2}} & 0 & \frac{1}{\sqrt{2}} & 0
  	\end{matrix}\right). 
  \end{equation}
  This in turn leads to the coordinate transformation
  \begin{eqnarray}
  		\hat{q} & =& (\hat{q}_1,\hat{q}_2,\hat{q}_3,0) = A (\hat{\rho_1},\hat{\rho_2},\hat{\rho_3},\hat{\rho_4}) \\
  		&=& \left(\frac{1}{2}\hat{\rho}_1+\frac{1-\sqrt{2}}{2\sqrt{3}}\hat{\rho}_2+\frac{1}{2}\hat{\rho}_3+\frac{1+\sqrt{2}}{2\sqrt{3}}\hat{\rho}_4,-\frac{1}{2}\hat{\rho}_1+\frac{1-\sqrt{2}}{2\sqrt{3}}\hat{\rho}_2-\frac{1}{2}\hat{\rho}_3+\frac{1+\sqrt{2}}{2\sqrt{3}}\hat{\rho}_4, \right. \qquad \quad \notag \\  && \left. -\frac{\sqrt{2}+1}{\sqrt{6}}\hat{\rho}_2+\frac{-\sqrt{2}+1}{\sqrt{6}}\hat{\rho}_4,\frac{1}{\sqrt{2}}( \hat{\rho}_3 - \hat{\rho}_1) \right) . \notag
  \end{eqnarray}
 Thus instead of the centre-of-mass constraint $\sum_{i=1}^4 \rho_i=0$ we have now the constraint $\hat{\rho}_1 = \hat{\rho}_3$ as we can read off from the last component in the four dimensional system. Indeed, when computing the new coordinates we find precisely this dependence in the first and third coordinate 
  \begin{eqnarray}
  		\hat{\rho} & =& (\hat{\rho}_1,\hat{\rho}_2,\hat{\rho}_3,\hat{\rho}_4)  =\hat{A}^{-1} (\hat{q}_1,\hat{q_2},\hat{q_3},0) 	\label{coords2} \\ &=&\frac{1}{2}\left(\hat{q}_1-\hat{q}_2,\frac{1-\sqrt{2}}{\sqrt{3}}(\hat{q}_1+\hat{q}_2+\hat{q}_3)+\frac{1}{\sqrt{3}}\hat{q}_3 , \hat{q}_1-\hat{q}_2, \frac{\sqrt{2}+1}{\sqrt{3}}(\hat{q}_1+\hat{q}_2+\hat{q}_3)-\frac{3}{\sqrt{3}}\hat{q}_3\right).
  \notag
  \end{eqnarray}
When setting any other component in $\hat{q}$ to zero we will obtain more complicated dependencies.

\begin{figure}[h]
	\centering 
	\begin{minipage}{.57\textwidth}
		\includegraphics[width=\textwidth]{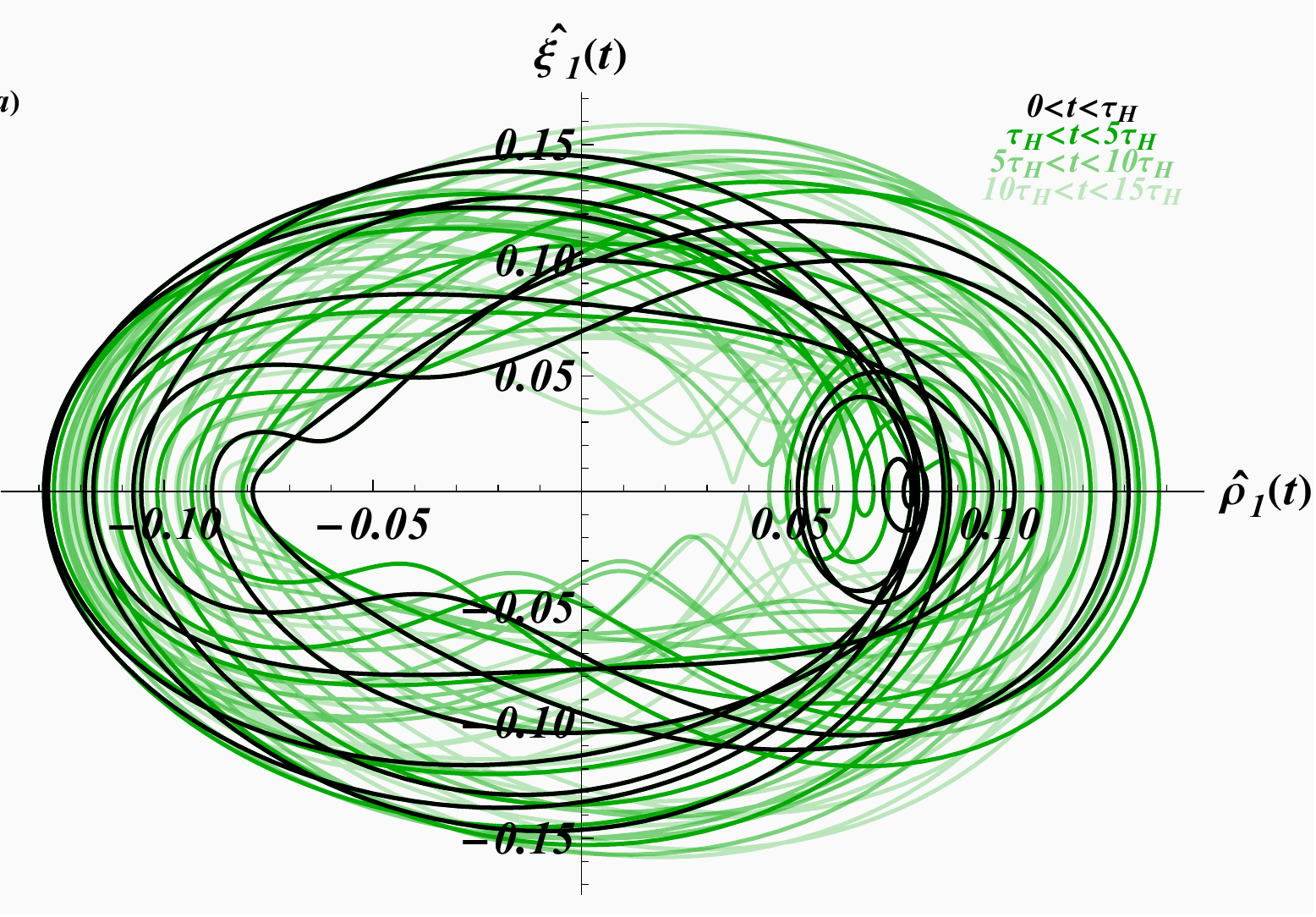}
	\end{minipage}
	\begin{minipage}{.42\textwidth}
		\includegraphics[width=\textwidth,height=3.0cm]{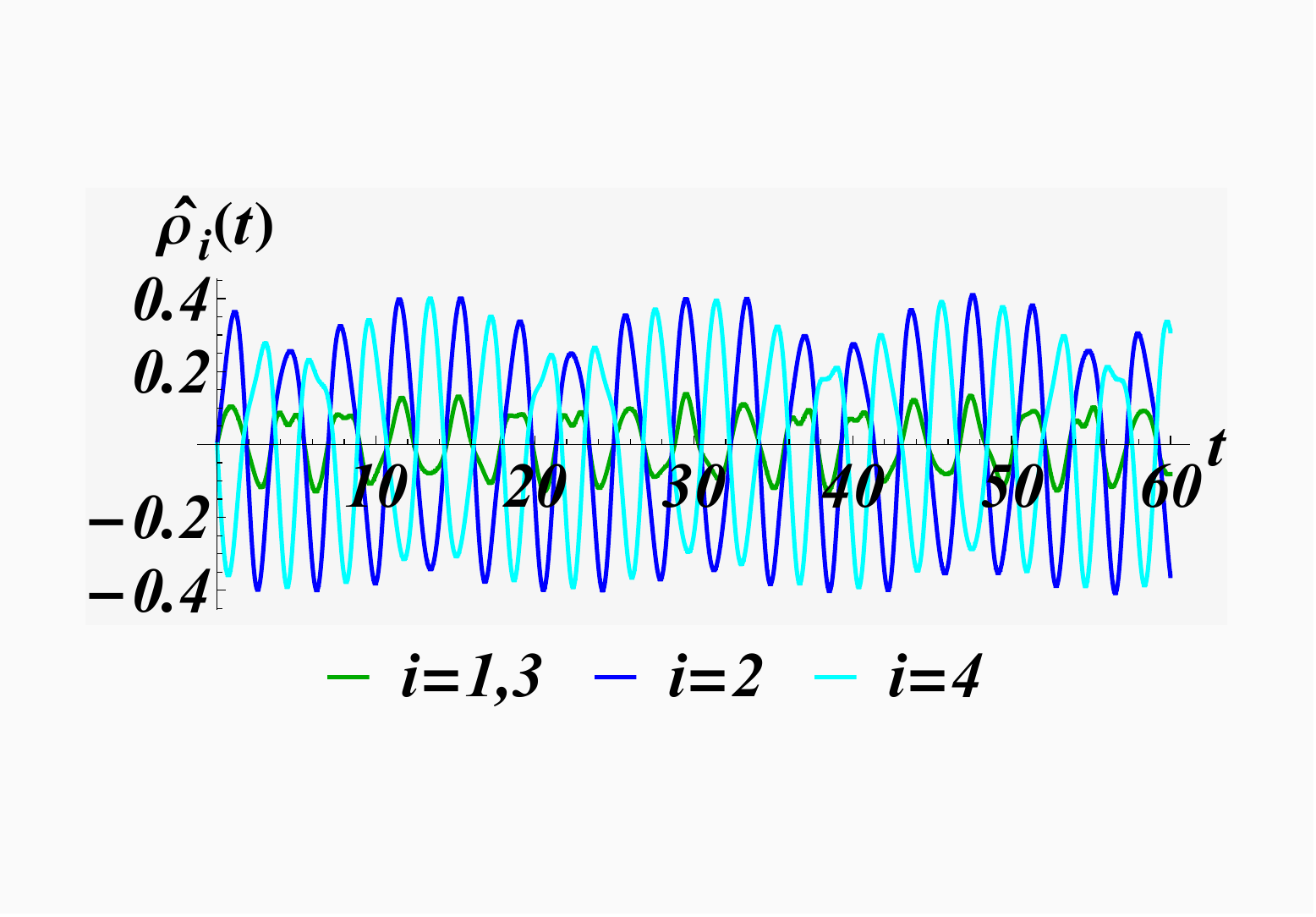}
		\includegraphics[width=\textwidth,height=3.0cm]{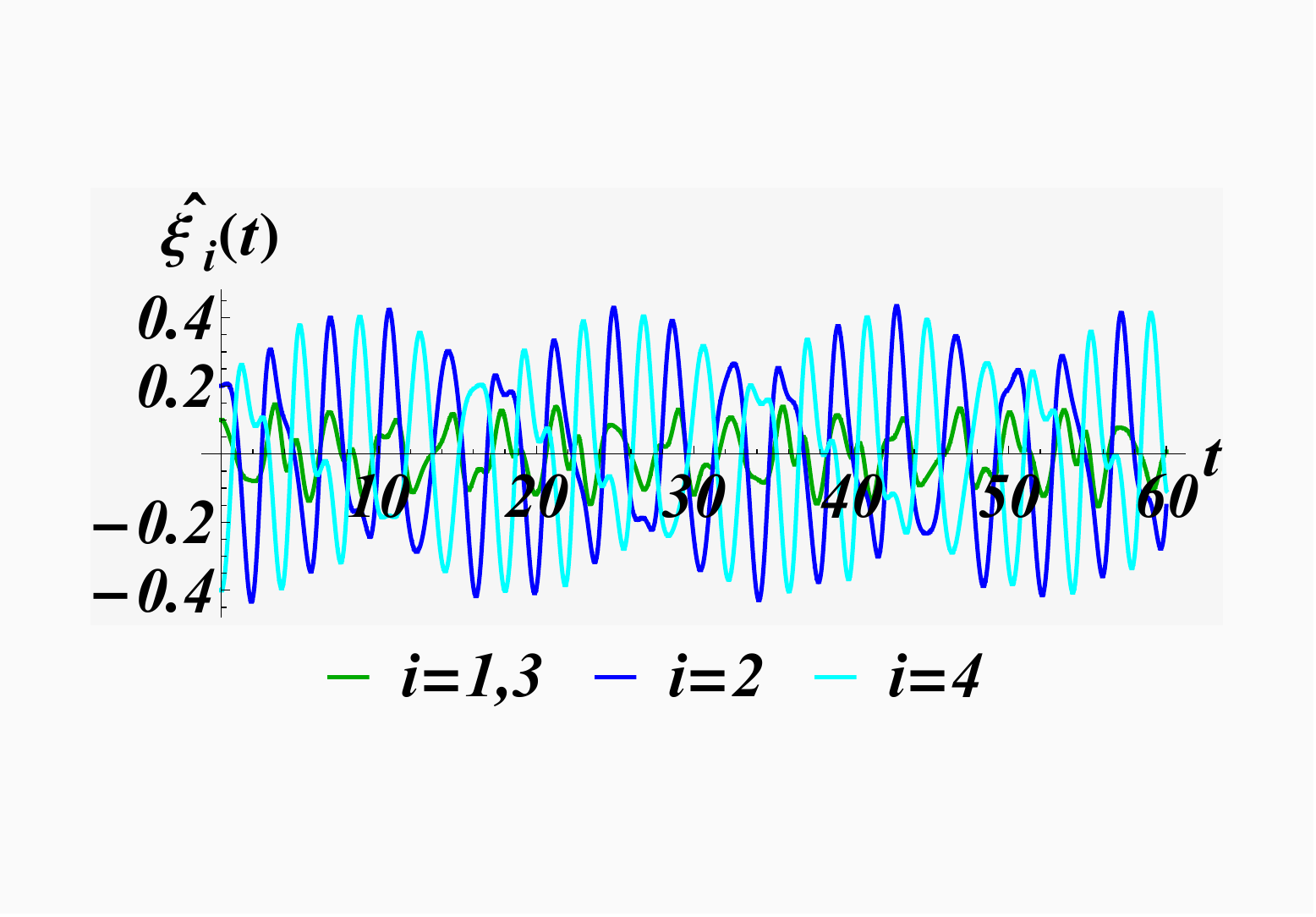}
	\end{minipage}
	\begin{minipage}{.57\textwidth}
		\includegraphics[width=\textwidth]{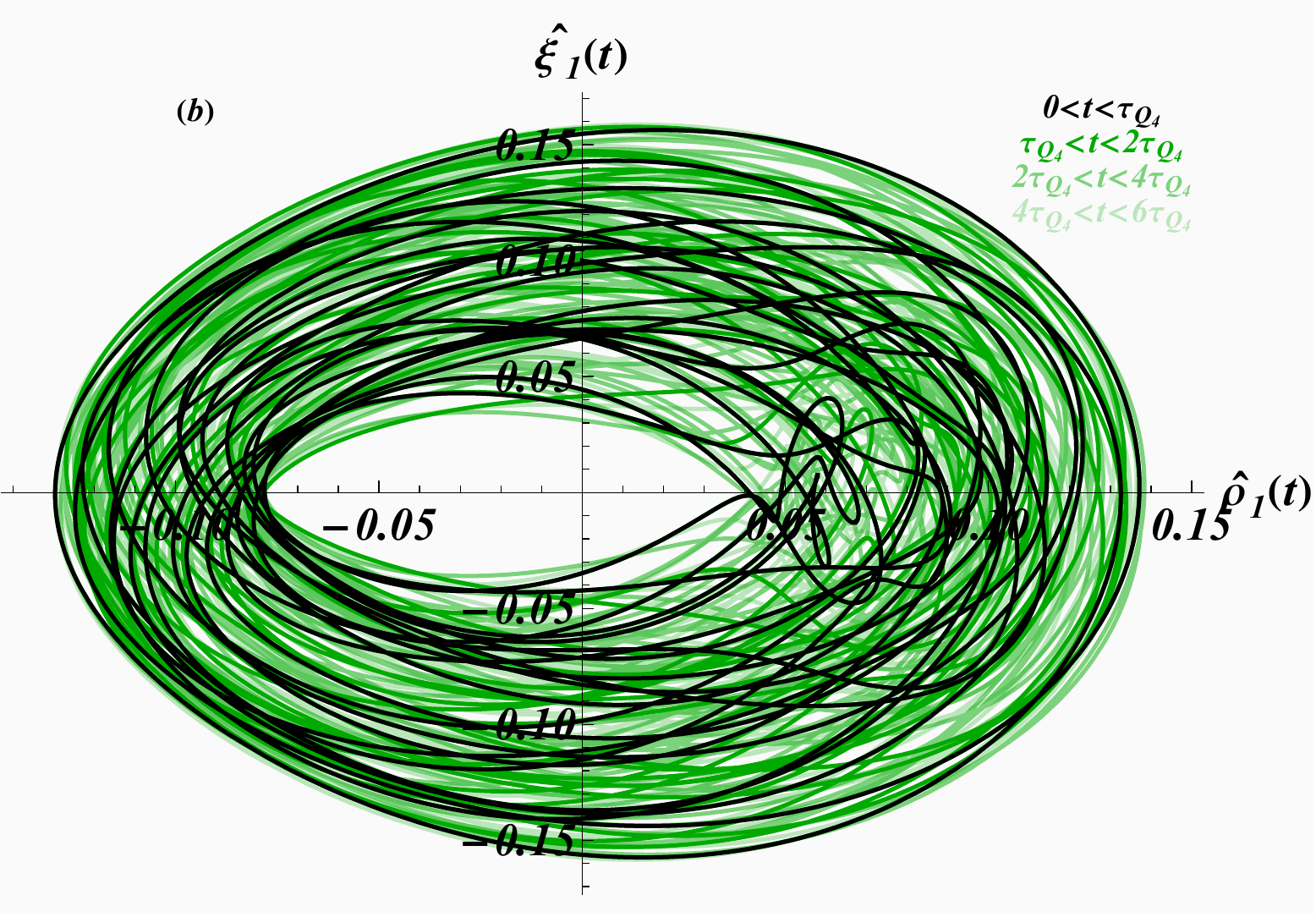}
	\end{minipage}
	\begin{minipage}{.42\textwidth}
		\includegraphics[width=\textwidth,height=3.0cm ]{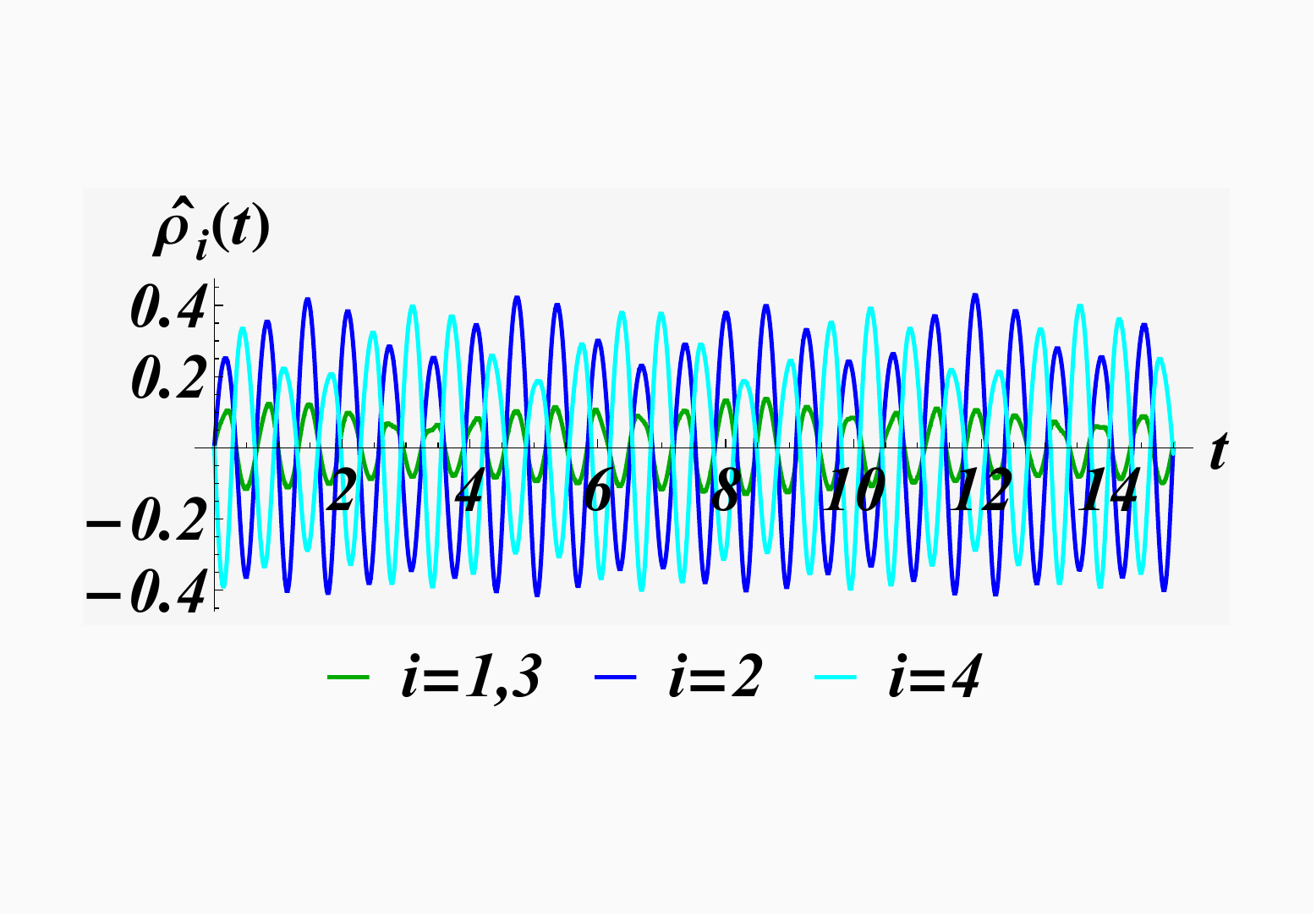}
		\includegraphics[width=\textwidth,height=3.0cm]{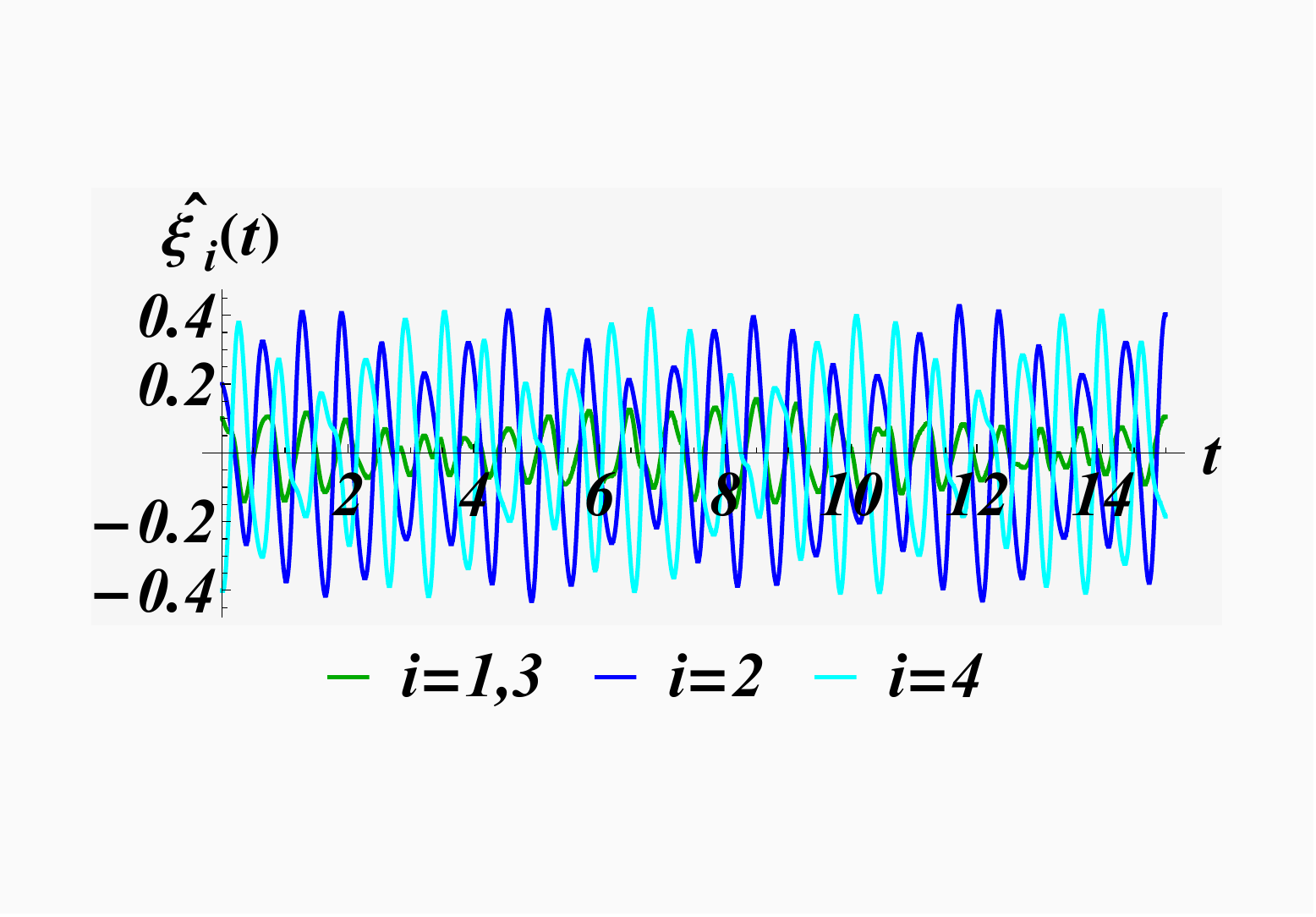}
	\end{minipage} 
	\begin{minipage}{.57\textwidth}
		\includegraphics[width=\textwidth]{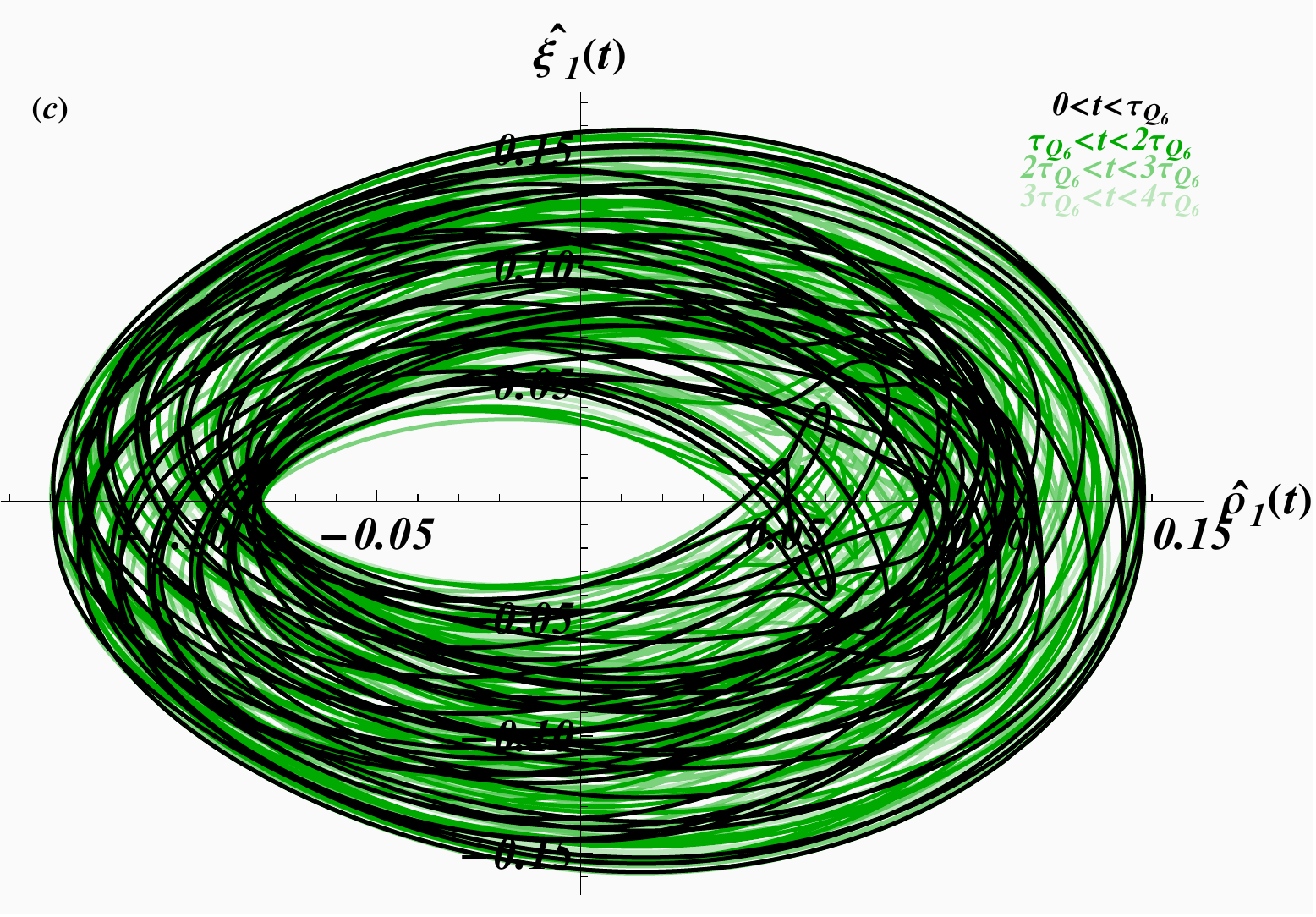}
	\end{minipage}
	\begin{minipage}{.42\textwidth}
		\includegraphics[width=\textwidth,height=3.0cm  ]{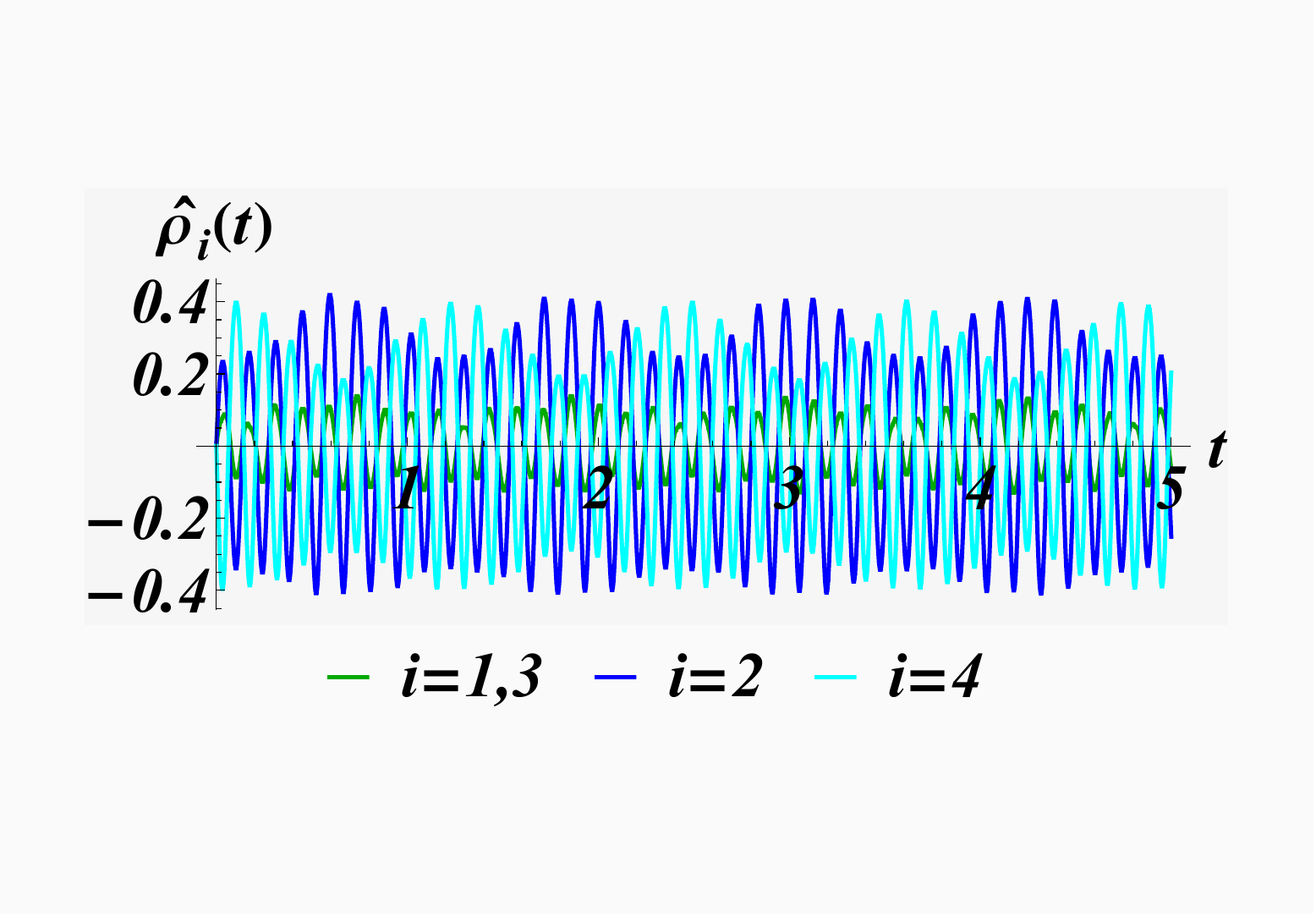}
		\includegraphics[width=\textwidth,height=3.0cm  ]{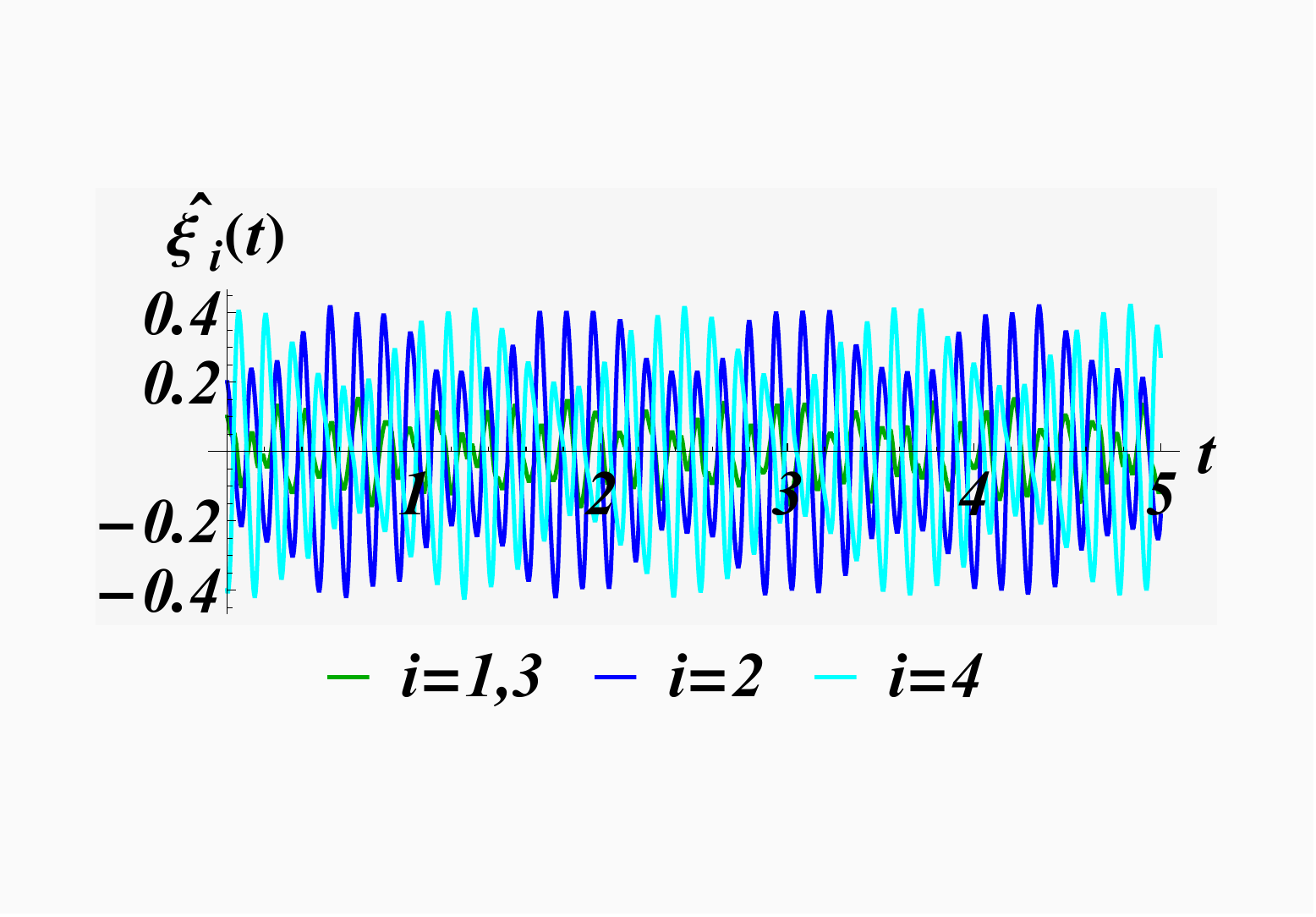}
	\end{minipage}
	\caption{Affine $B_3$-Toda lattice phase space $(\hat{\xi}_1, \hat(\rho)_1)$ as functions of $t$ for the standard Hamiltonian panel (a), for $Q_4$ taken as higher derivative Hamiltonian panel (b) and for $Q_6$  taken as higher derivative Hamiltonian panel (c). The initial conditions are taken in all cases as $ \hat{\rho}(0)=(0,0,0,0), \hat{\xi}(0)=\frac{1}{10}(1,2,1,-4)$. The quasi-periods are: $\tau_H \approx 28.80, \tau_{Q_4} \approx 14.99, \tau_{Q_6} \approx 4.103$.}
	\label{PhaseB34D}
\end{figure}

Transforming now also the $B_3$ charges into the new coordinates we proceed as previously and compute the classical trajectories numerically. Our results are shown in figure \ref{PhaseB34D}. The main observation is that all solutions found are of oscillatory nature. Thus unlike for the $A_n$ cases we do not encounter divergencies in the case where the dimension of the root representation space does not match the rank of the algebra.

	\section{Higher derivative Hamiltonians from the $G_2$-affine Toda lattice}
In order to compare and identify universal features we investigate now also the $G_2$-affine Toda lattice theory that is similar to the $A_2$-theory, in the sense that it has a natural three and two particle representation, with the former being the standard one. Its Hamiltonian in the standard form reads 
		\begin{equation}
		\tilde{H} = \frac{1}{2} \tilde{p}^2+ 3 e^{\alpha_1 \cdot \tilde{q}}+ 2 e^{\alpha_2 \cdot \tilde{q}} +  e^{\alpha_0 \cdot \tilde{q}} ,
    	\end{equation}
	where $\alpha_1$ and $\alpha_2$ are the two simple roots of $G_2$ and $\alpha_0=-3\alpha_1 -2 \alpha_2 $ is the negative of the highest $G_2$-root. In order to construct the higher charges we may use the reduction from $B_3$ or directly construct a higher order expression from a suitable Ansatz whose Poisson bracket vanishes with $\tilde{H}$. Here we do not use the folding procedure, as for it to apply in this case the Lax pair has to be slightly modified, but instead we use the latter approach. Taking initially the standard representation for the simple roots $\alpha_1=(1,-1,0)$ and $\alpha_2=(-2,1,1)$, \cite{Bou}, we find the non-trivial independent charges  
	\begin{eqnarray}
		\tilde{Q}_1 &=& \tilde{p}_1+ \tilde{p}_2+\tilde{p}_3\\
		\tilde{Q}_6 &=& \sum_{i,j=1}^3 \frac{1}{6}\Tilde{p}_i^6
		+\frac{3}{14}(\Tilde{p}_i^4\Tilde{p}_{i+1}^2+\Tilde{p}_i^2\Tilde{p}_{i+1}^4)+\frac{10}{21}\Tilde{p}_i^3\Tilde{p}_{i+1}^3+\frac{6}{7}\Tilde{p}_i\Tilde{p}_{i+1}\Tilde{p}_{i+2}+c_j\frac{j^3}{7}e^{3\Tilde{\alpha}_j.\Tilde{q}} \\ 
		&& +\frac{n_jn_{j+1}}{7}e^{(\Tilde{\alpha}j+\Tilde{\alpha}_{j+1}).\Tilde{q}}\left(c^{(1)}_jn_je^{\Tilde{\alpha}_j.\Tilde{q}}+c^{(2)}_jn_{j+1}e^{\Tilde{\alpha}_{j+1}.\Tilde{q}}+c^{(3)}_{ij}\Tilde{p}_i\Tilde{p}_{j+1}+c^{(4)}_{ij}\Tilde{p}_i^2\right) \notag \\ 
		&& +\frac{n_j}{7}e^{\Tilde{j}.\Tilde{q}}\left(c^{(5)}_{ij}\Tilde{p}_{i}^4+ c^{(6)}_{ij}\Tilde{p}_i^2\Tilde{p}_{i+1}^2 +c^{(7)}_{ij}\Tilde{p}_i^3\Tilde{p}_{i+1}+c^{(8)}_{ij}\Tilde{p}_i\Tilde{p}_{i+1}^3+c^{(9)}_{ij}\Tilde{p}_i^2\Tilde{p}_{i+1}\Tilde{p}_{i+2}\right) \notag \\ 
		&&+\frac{n_j^2}{7}e^{2\Tilde{\alpha}_j.\Tilde{q}}\left(c^{(10)}_{ij}\Tilde{p}_i^2+c^{(11)}_{ij}\Tilde{p}_i\Tilde{p}_{i+1}\right)
		 , \notag
	\end{eqnarray} 
with abbreviations $c=(2,6,6)$,
$c^{(1)}=(6,18,-18)$,
$c^{(2)}=(18,18,42)$, \\
$c^{(3)} =\left(\begin{matrix}
22 & -2 & -20 \\
46 & -2 & 40 \\
10 & 34 & 40
\end{matrix}\right)$, \,\,
$c^{(4)} =\left(\begin{matrix}
26 & 26 & 32 \\
8 & 44 & 32 \\
14 & 26 & 2
\end{matrix}\right)$,\,\,
$c^{(5)} =\left(\begin{matrix}
7 & 7 & 7 \\
7 & 7 & 7 \\
3 & 7 & 7
\end{matrix}\right)$,\,\,
$c^{(6)} =\left(\begin{matrix}
0 & 2 & 20 \\
6 & 20 & 2 \\
6 & 0 & 0
\end{matrix}\right)$, \,\,\\
$c^{(7)} =\left(\begin{matrix}
4 & 2 & -4 \\
0 & -4 & 8 \\
10 & 0 & 0
\end{matrix}\right)$, \,\,
$c^{(8)} =\left(\begin{matrix}
4 & 8 & -4 \\
10 & -4 & 2 \\
0 & 0 & 0
\end{matrix}\right)$, \,\,
$c^{(9)} =\left(\begin{matrix}
0 & 2 & -16 \\
0 & -16 & -16 \\
-30 & -16 & 2
\end{matrix}\right)$, \,\, \\
$c^{(10)} =\left(\begin{matrix}
12 & 13 & 4 \\
12 & 4 & 4 \\
21 & 4 & 13
\end{matrix}\right)$, \,\,
$c^{(11)} =\left(\begin{matrix}
18 & 8 & 26 \\
0 & 26 & 8 \\
0 & 8 & 8
\end{matrix}\right)$.\\
 We note that there is no non-trivial independent charge $\tilde{Q}_4 $, as the only quantity that one can construct at that order is proportional to $\tilde{H}^2$. The charges $\tilde{Q}_1$ and $\tilde{Q}_6$ are in involution with the Hamiltonian $\tilde{H}$ and with each other. Defining the centre-of-mass coordinate $\tilde{\chi}= \tilde{q}_1 + \tilde{q}_2  + \tilde{q}_3 $ we find that
\begin{eqnarray}
	\left\{ \tilde{\chi} , \tilde{H} \right\}  &=& \tilde{Q}_1 ,\\
	\left\{ \tilde{\chi} , \tilde{Q}_6 \right\}  &=& 6  \tilde{H}^2 \tilde{Q}_1  - \frac{10}{7} \tilde{H} \tilde{Q}_1^3 +\frac{3}{14} \tilde{Q}_1^5 . 
\end{eqnarray}
Thus we expect to find converging solution with $\tilde{H}$ and $\tilde{Q}_6$ taken as Hamiltonians for the initial conditions taken to be $\tilde{Q}_1=0$ in both cases.    

\begin{figure}[h]
		\centering    
		\begin{minipage}[b]{0.32\textwidth}              
			\includegraphics[width=\textwidth]{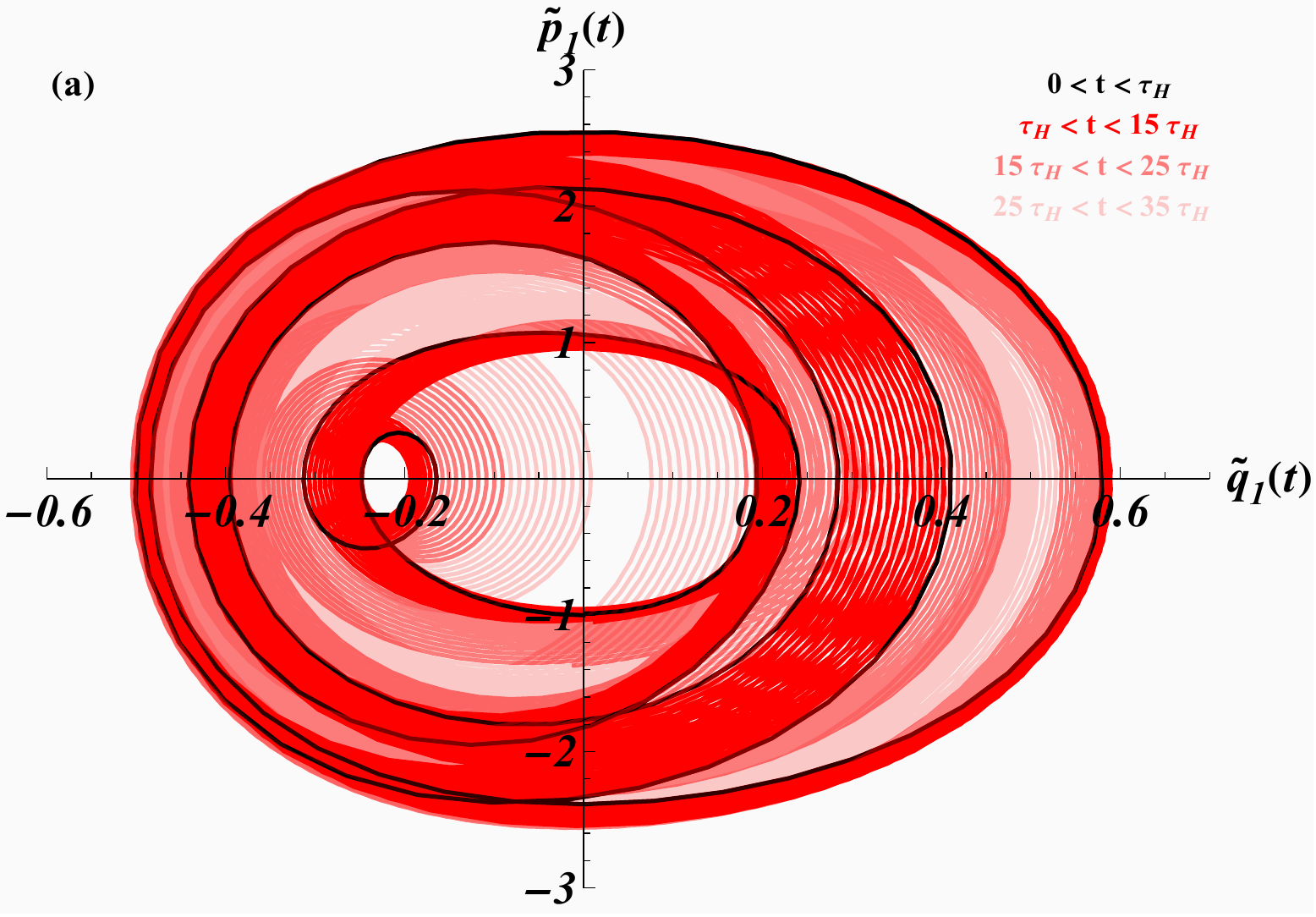}
		\end{minipage}  
		\begin{minipage}[b]{0.32\textwidth}           
			\includegraphics[width=\textwidth]{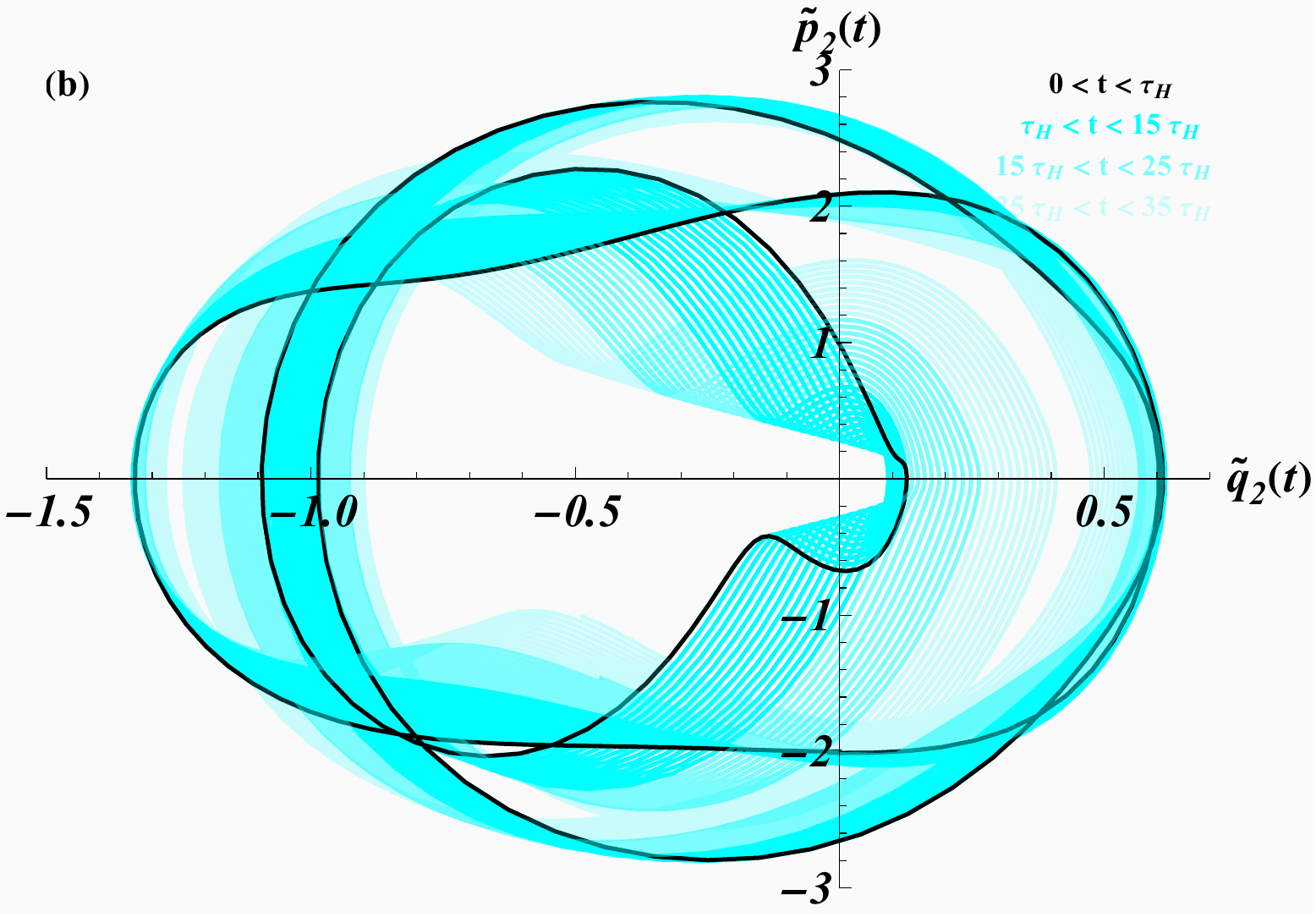}
		\end{minipage}  
		\begin{minipage}[b]{0.32\textwidth}          
			\includegraphics[width=\textwidth]{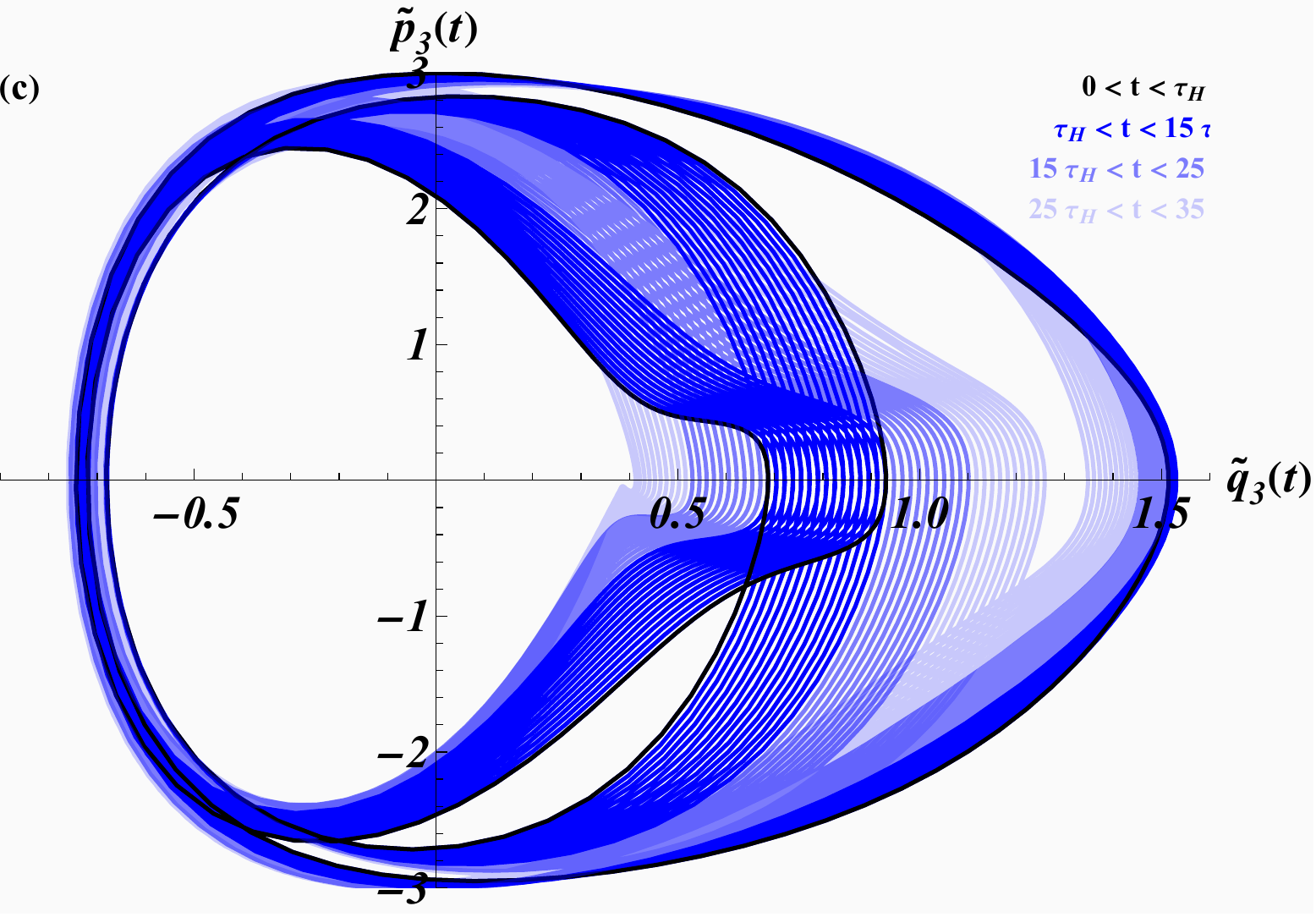}
		\end{minipage}   
		\begin{minipage}[b]{0.32\textwidth}           
			\includegraphics[width=\textwidth]{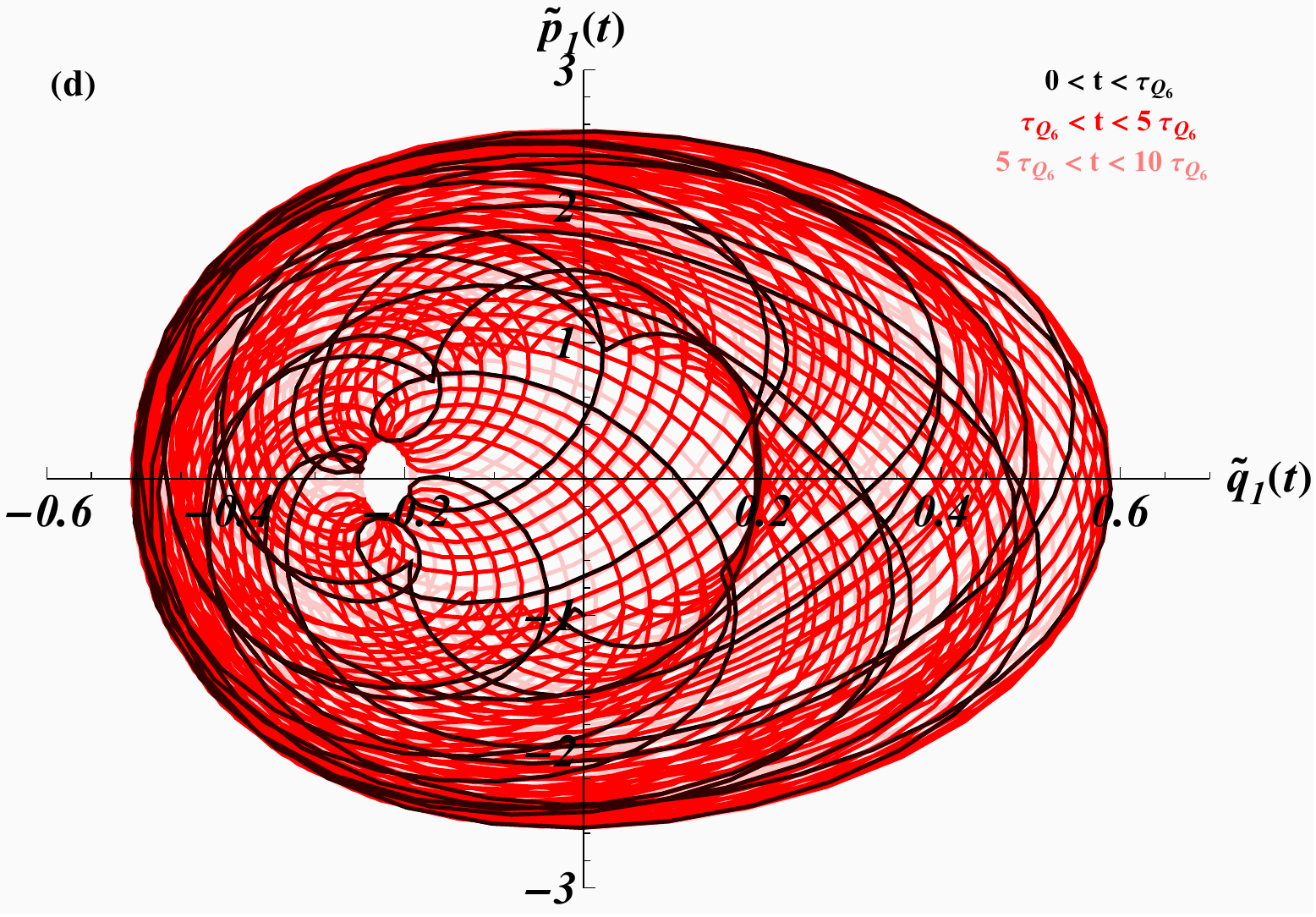}
		\end{minipage}  
		\begin{minipage}[b]{0.32\textwidth}           
			\includegraphics[width=\textwidth]{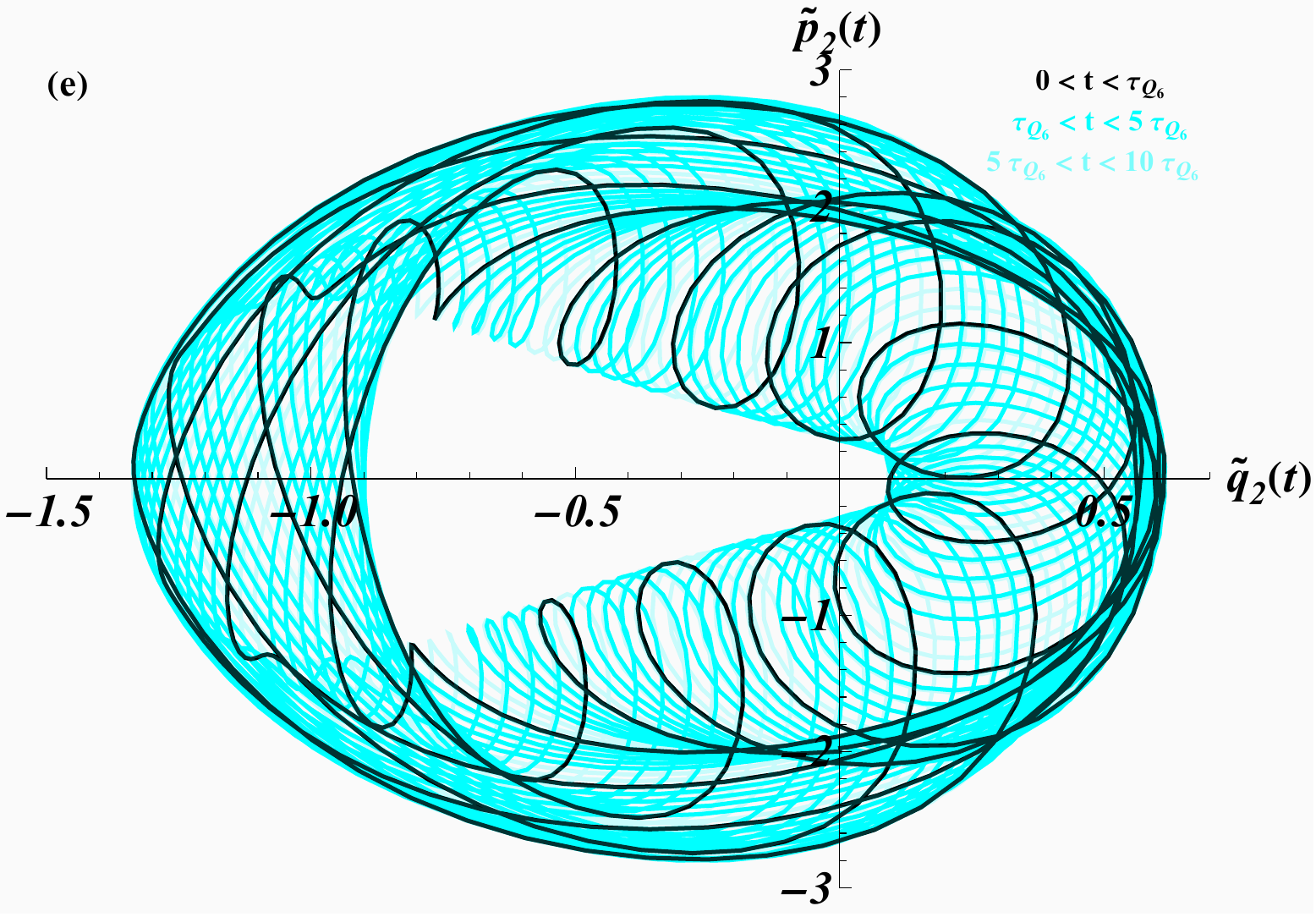}
		\end{minipage}   
		\begin{minipage}[b]{0.32\textwidth}           
			\includegraphics[width=\textwidth]{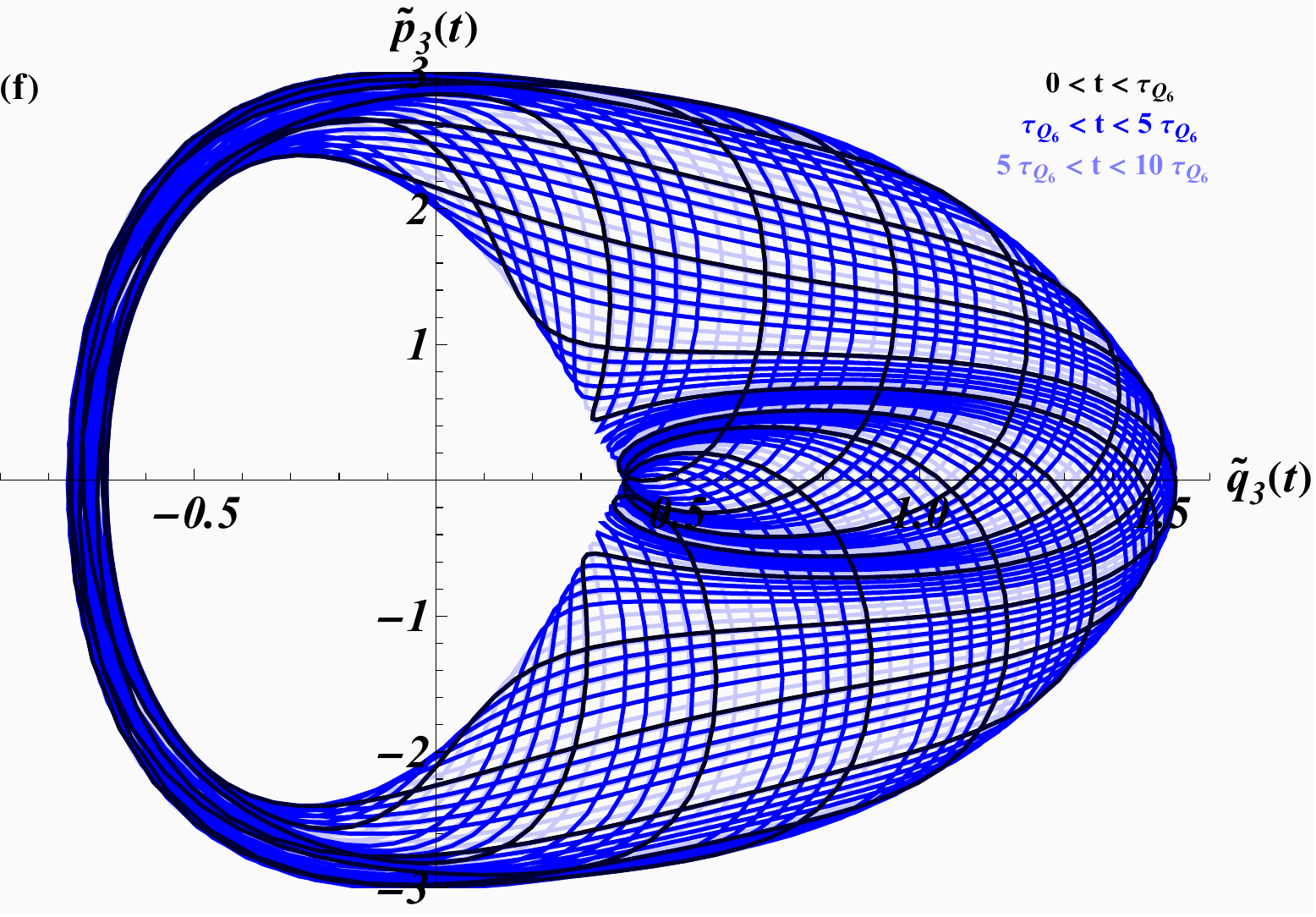}
		\end{minipage}	    
		\caption{Affine $G_2$-Toda lattice $\tilde{H}$ and $\tilde{Q}_6$ phase spaces for the variables $(\tilde{q}_i,\tilde{p}_i)$, $1,2,3$ as functions of $t$ panels (a), (b), (c) and panels (d), (e), (f), respectively.  The initial condition are taken in all cases to $\tilde{q}_1(0) =\tilde{q}_2(0)=\tilde{q}_3(0)=0$, $\tilde{p}_1(0)=-1$,  $\tilde{p}_2(0)=-2$ and  $\tilde{p}_3(0)=3$.} 
		\label{PhaseG2}
	\end{figure}

Next we solve the corresponding equations of motion for $\tilde{H}$
	\begin{eqnarray}
		\dot{\tilde{q}}_1&=&p_1, \quad \dot{\tilde{q}}_2=p_2, \quad \dot{\tilde{q}}_3=p_3, \quad \dot{\tilde{p}}_1=4 e^{-2 \tilde{q}_1+\tilde{q}_2+\tilde{q}_3}-3 e^{\tilde{q}_1-\tilde{q}_2}-e^{\tilde{q}_1+\tilde{q}_2-2 \tilde{q}_3} \label{G2Hequ1}\\
		\dot{\tilde{p}}_2&=&3 e^{\tilde{q}_1-\tilde{q}_2}-e^{\tilde{q}_1+\tilde{q}_2-2 \tilde{q}_3}-2 e^{-2 \tilde{q}_1+\tilde{q}_2+\tilde{q}_3} \quad \dot{\tilde{p}}_3 = 2 e^{\tilde{q}_1+\tilde{q}_2-2 \tilde{q}_3}-2 e^{-2 \tilde{q}_1+\tilde{q}_2+\tilde{q}_3} \notag
	\end{eqnarray}
	and $\tilde{Q}_6$, which we will not report here, numerically and depict our results in figure \ref{PhaseG2}.

	Once again we identify almost periodic motions from the trajectories in both cases with periods numerically computed as $\tau_H\approx 7.04819143$ and $\tau_{Q6} \approx 0.10294605$ for the phase spaces of $\tilde{H}$ and $\tilde{Q}_6$, respectively. In our concrete solutions for the equations of motion for $\tilde{H}$ we find $\vert \zeta_1(0)- \zeta_1(\tau_H) \vert \approx 2.3\times 10^{-9}   $,  $\vert \zeta_2(0)- \zeta_2(\tau_H) \vert =\vert \zeta_3(0)- \zeta_3(\tau_H) \vert\approx 0.0229121  $, $\vert \eta_1(0)- \eta_1(\tau_H) \vert \approx 0.0122752   $, $\vert \eta_2(0)- \eta_2(\tau_H) \vert \approx 0.009117525   $ and $\vert \eta_3(0)- \eta_3(\tau_H) \vert \approx 0.0031   $. As previously, in our depiction we distinguish between the first almost period and some further periods that illustrate how the phase space is gradually filled inward and outwardly. Comparing the  $\tilde{H}$ and $\tilde{Q}_6$ phase spaces we notice that for large time the same confined region in phase space will be filled out.

	 Furthermore we notice that, unlike as in the $A_2$-case, even for the case when the particle number does not match the rank of the algebra the motion is of oscillatory nature. 
	
	Thus in principal there is no need for a dimensional reduction to the centre-of-mass frame from the point of view to obtain finite trajectories, but for completeness we also analyse that case. For this purpose we need to solve once more equation (\ref{redmat}) for the orthogonal matrix $A$ and the two-dimensional roots $\beta_i$, but now involving the $G_2$-Cartan matrix
	\begin{equation}
	K=  \left(  \begin{array}{cc}
		2 & -1  \\
		-3 & 2 
	\end{array}   \right).
	\end{equation}
In this case we find the solutions
\begin{equation}
	A= \left(
	\begin{array}{ccc}
		\frac{1}{\sqrt{2}} & -\frac{1}{\sqrt{6}} & \frac{1}{\sqrt{3}} \\
		-\frac{1}{\sqrt{2}} & -\frac{1}{\sqrt{6}} & \frac{1}{\sqrt{3}} \\
		0 & \sqrt{\frac{2}{3}} & \frac{1}{\sqrt{3}} \\
	\end{array}
	\right), \quad \beta_1= \left( \sqrt{2},0,0\right),  \quad \beta_2= \left(-\frac{3}{\sqrt{2}},\sqrt{\frac{3}{2}},0\right), \label{orthog2}
\end{equation}
for the orthogonal matrix and the three dimensional roots.

\begin{figure}[h]
	\centering    
	\begin{minipage}[b]{0.49\textwidth}              
		\includegraphics[width=\textwidth]{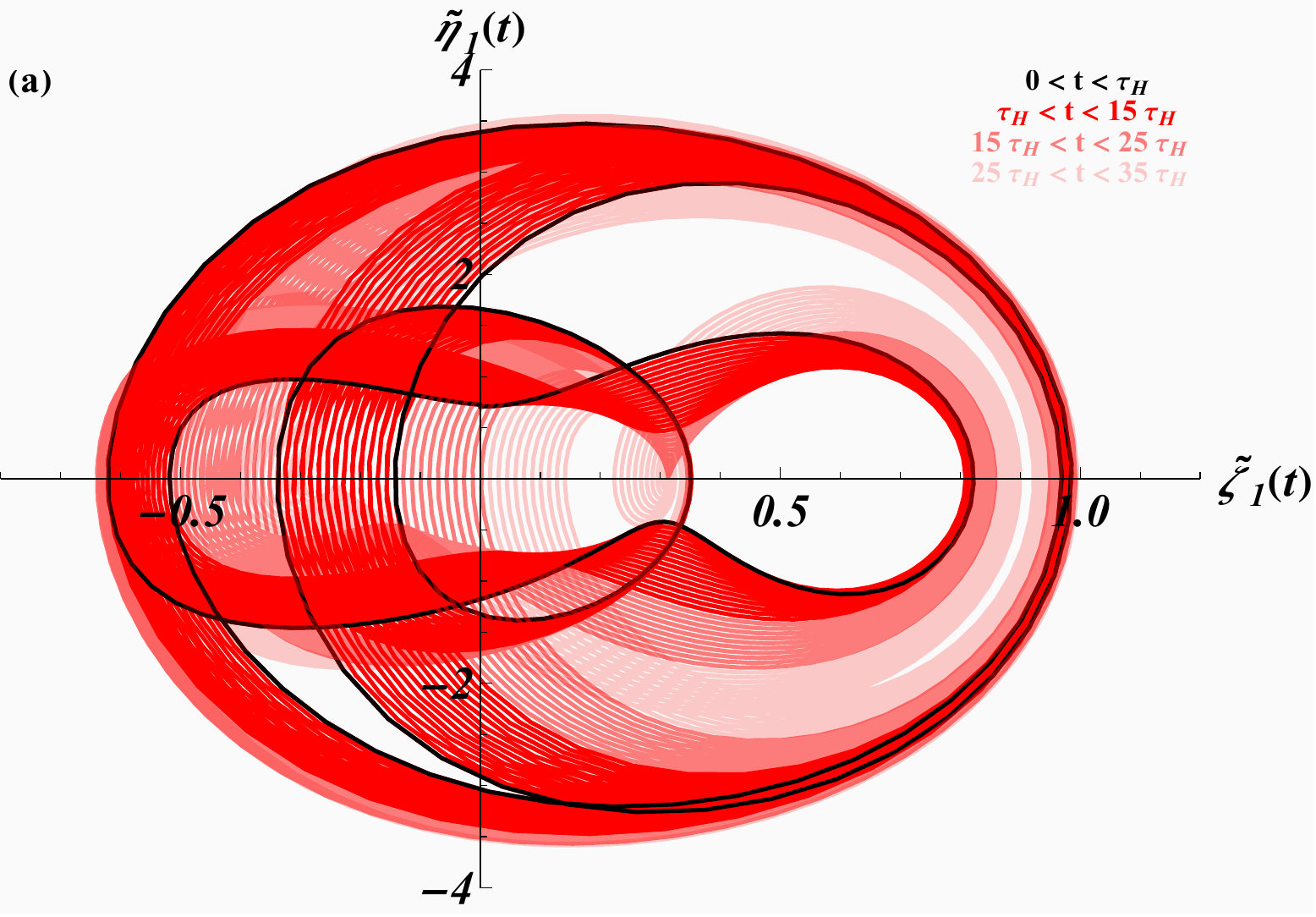}
	\end{minipage}  
	\begin{minipage}[b]{0.49\textwidth}           
		\includegraphics[width=\textwidth]{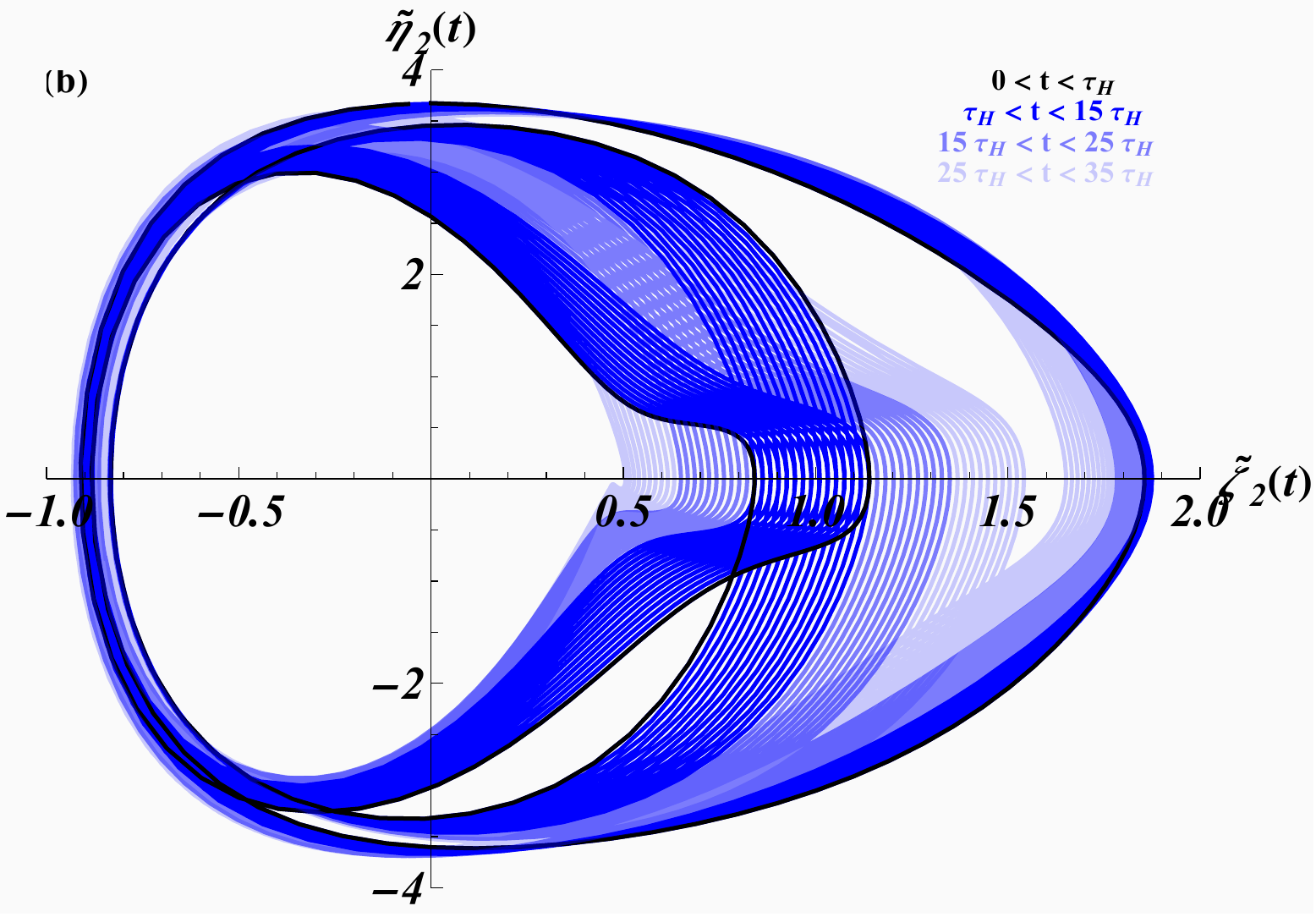}
	\end{minipage}  
	\begin{minipage}[b]{0.49\textwidth}              
		\includegraphics[width=\textwidth]{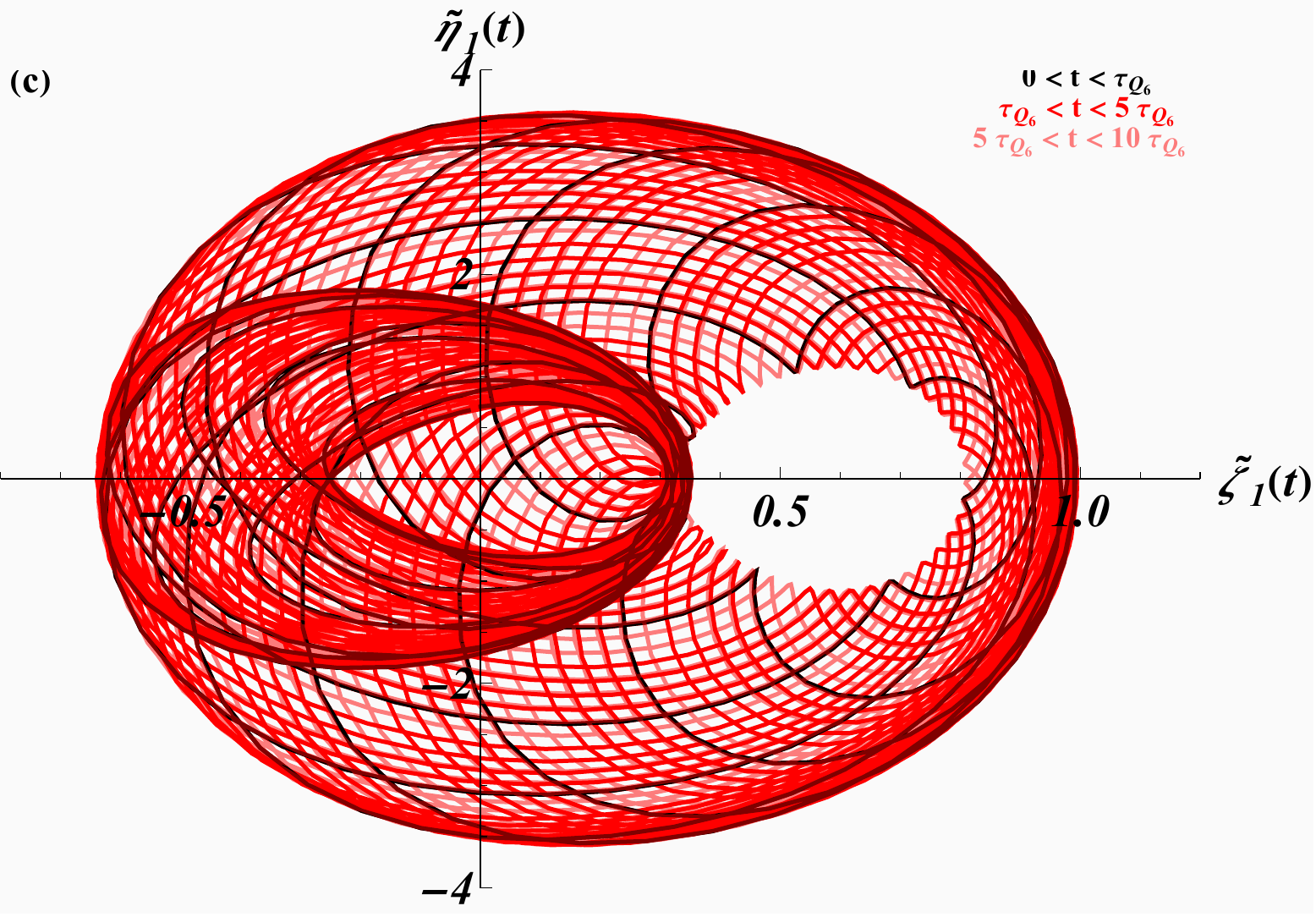}
	\end{minipage}  
	\begin{minipage}[b]{0.49\textwidth}           
		\includegraphics[width=\textwidth]{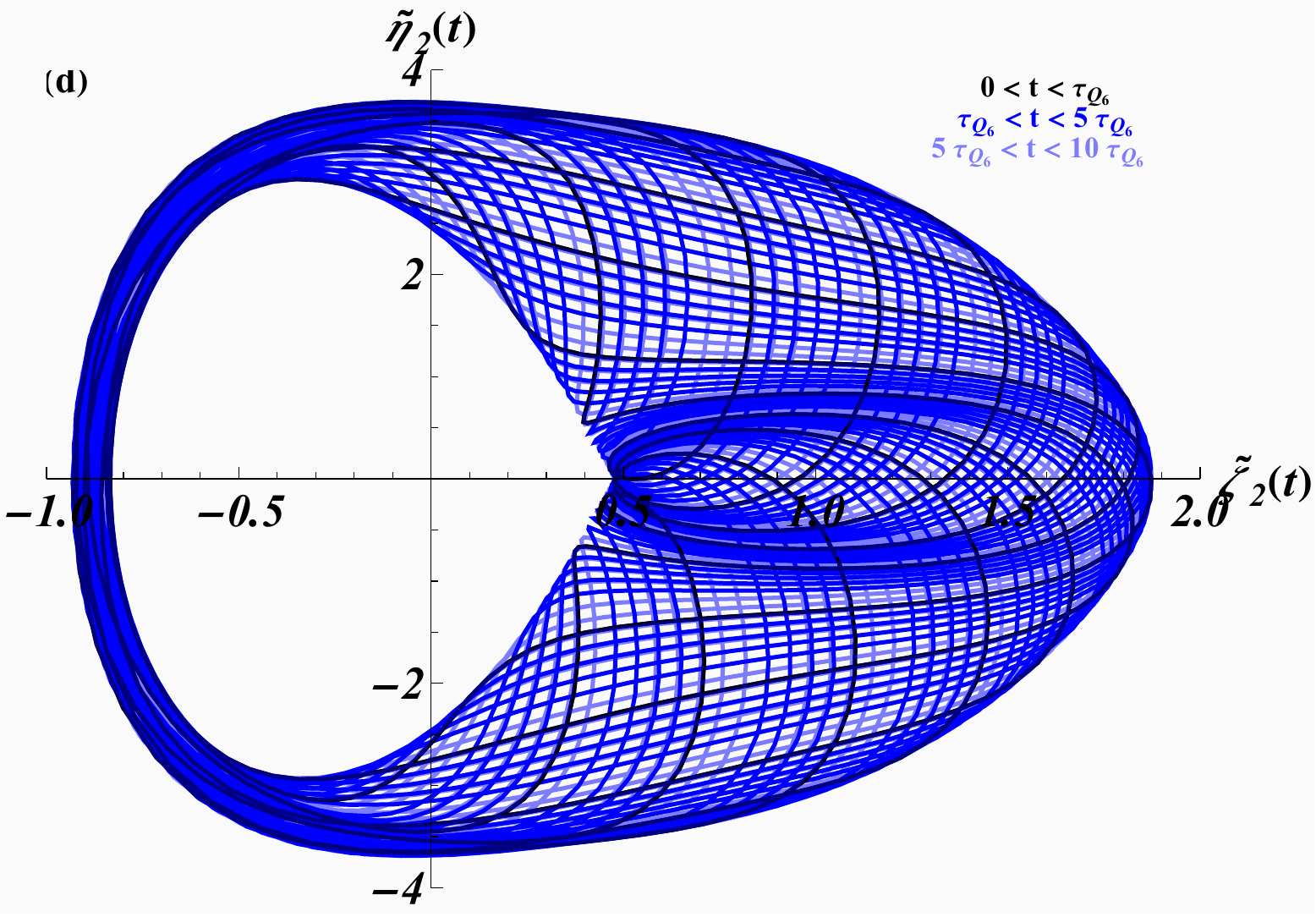}
	\end{minipage}         
	\caption{Affine $G_2$-Toda lattice 
		$\tilde{H}$ and $\tilde{Q}_6$ phase spaces for the variables $(\tilde{\zeta}_i,\tilde{\eta}_i)$, $1,2$ as functions of $t$ panels (a), (b) and panels (c), (d), respectively. The initial condition are taken in both cases as $\tilde{\zeta}_1(0) =\tilde{\zeta}_2(0)=0$, $\tilde{\eta}_1(0)=3/2\sqrt{2}$ and  $\tilde{\eta}_2(0)=-\sqrt{3}/2\sqrt{2}$.} 
	\label{PhaseG2red}
\end{figure}
Then, according to (\ref{222}) and (\ref{223}) the coordinates and momenta transform as 
\begin{eqnarray}
	\tilde{q}&=&\left(\frac{\tilde{\zeta} _1}{\sqrt{2}}-\frac{\tilde{\zeta} _2}{\sqrt{6}},-\frac{\tilde{\zeta} _1}{\sqrt{2}}-\frac{\tilde{\zeta} _2}{\sqrt{6}},\sqrt{\frac{2}{3}} \tilde{\zeta} _2\right)=(\tilde{q}_1,\tilde{q}_2,\tilde{q}_3), \label{newg2q} \\ 
	\tilde{p}&=& \left(\frac{\tilde{\eta} _1}{\sqrt{2}}-\frac{\tilde{\eta} _2}{\sqrt{6}},-\frac{\tilde{\eta} _1}{\sqrt{2}}-\frac{\tilde{\eta} _2}{\sqrt{6}},\sqrt{\frac{2}{3}} \tilde{\eta}_2\right)=(\tilde{p}_1,\tilde{p}_2,\tilde{p}_3),
\end{eqnarray}
and in reverse as
\begin{eqnarray}
	\tilde{\zeta}&=&\left(\frac{\tilde{q}_1-\tilde{q}_2}{\sqrt{2}},-\frac{\tilde{q}_1+\tilde{q}_2-2 \tilde{q}_3}{\sqrt{6}},\frac{\tilde{q}_1+\tilde{q}_2+\tilde{q}_3}{\sqrt{3}}\right)=(\tilde{\zeta}_1,\tilde{\zeta}_2,0), \label{newvg2} \\ 
	\tilde{\eta}&=& \left(\frac{\tilde{p}_1-\tilde{p}_2}{\sqrt{2}},-\frac{\tilde{p}_1+\tilde{p}_2-2 \tilde{p}_3}{\sqrt{6}},\frac{\tilde{p}_1+\tilde{p}_2+\tilde{p}_3}{\sqrt{3}}\right)=(\tilde{\eta}_1,\tilde{\eta}_2,0). \label{newv2g2}
\end{eqnarray}
Transforming the charges according to (\ref{newvg2}) and (\ref{newv2g2}) the equations of motion for the Hamiltonian $ \tilde{H}(\tilde{\zeta},\tilde{\eta})$ become
\begin{eqnarray}
	\dot{\tilde{\zeta}}_1& =& \tilde{\eta}_1, \quad 	\dot{\tilde{\zeta}}_2 = \tilde{\eta}_2, \label{eq12g2}\\ 
	\dot{\tilde{\eta}}_1 &=& 3 \sqrt{2} e^{\sqrt{\frac{3}{2}} \tilde{\zeta}_2-\frac{3 }{\sqrt{2}}  \tilde{\zeta}_1 }-3 \sqrt{2} e^{\sqrt{2} \tilde{\zeta}_1}, \quad \dot{\tilde{\eta}}_2 =\sqrt{6} e^{-\sqrt{6} \tilde{\zeta}_2}-\sqrt{6} e^{\sqrt{\frac{3}{2}} \tilde{\zeta}_2-\frac{3 }{\sqrt{2}} \tilde{\zeta}_1 } , \notag
\end{eqnarray}
and similarly for the $\tilde{Q}_6$-charge Hamiltonian that we do not report. We solve these equations numerically with the results depicted in figure \ref{PhaseG2red}. Once more we find almost periodic solutions with the same period as in the three dimensional case for all charges.

\begin{figure}[h]
	\centering    
	\begin{minipage}[b]{0.49\textwidth}              
		\includegraphics[width=\textwidth]{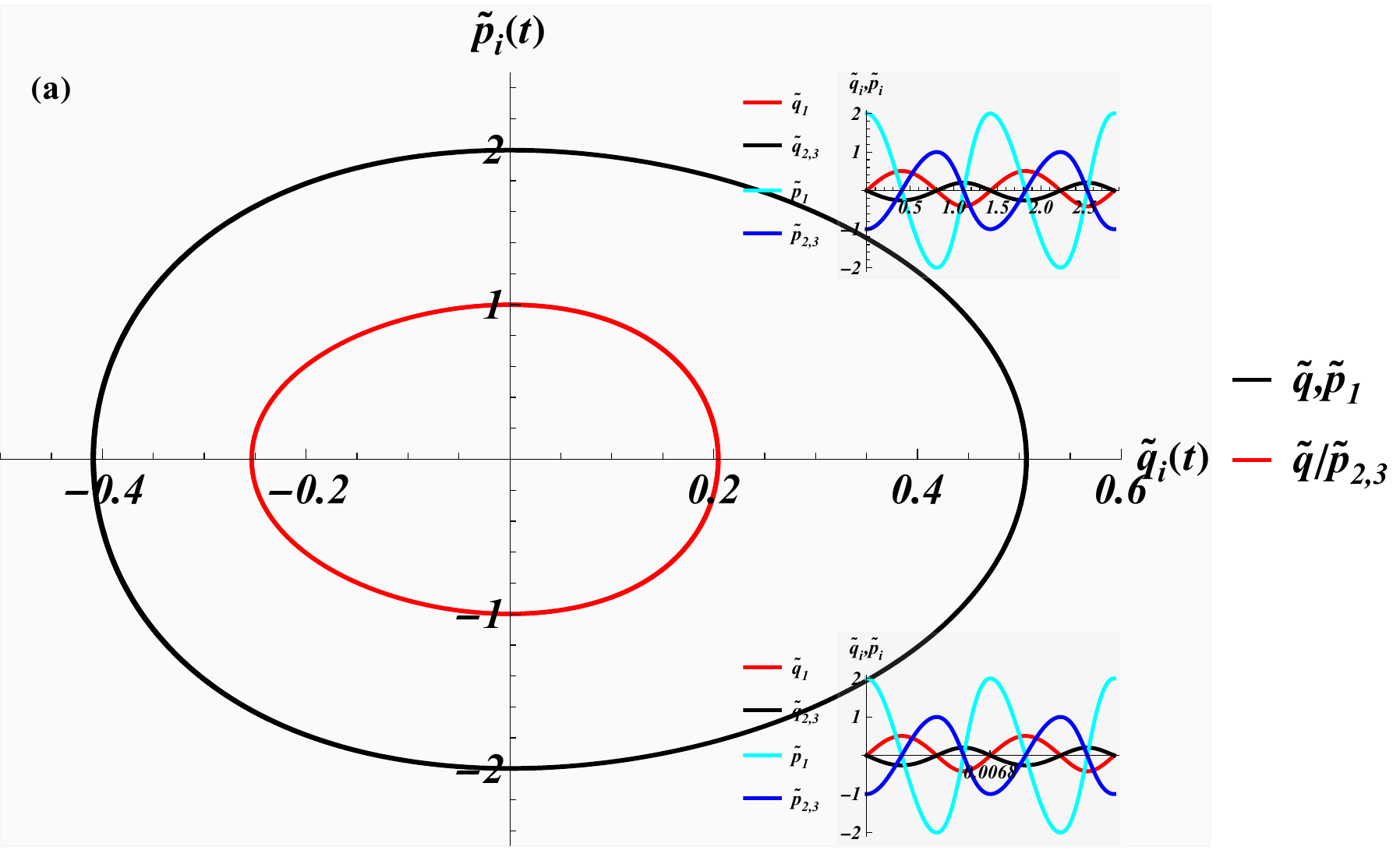}
	\end{minipage}  
	\begin{minipage}[b]{0.49\textwidth}           
		\includegraphics[width=\textwidth]{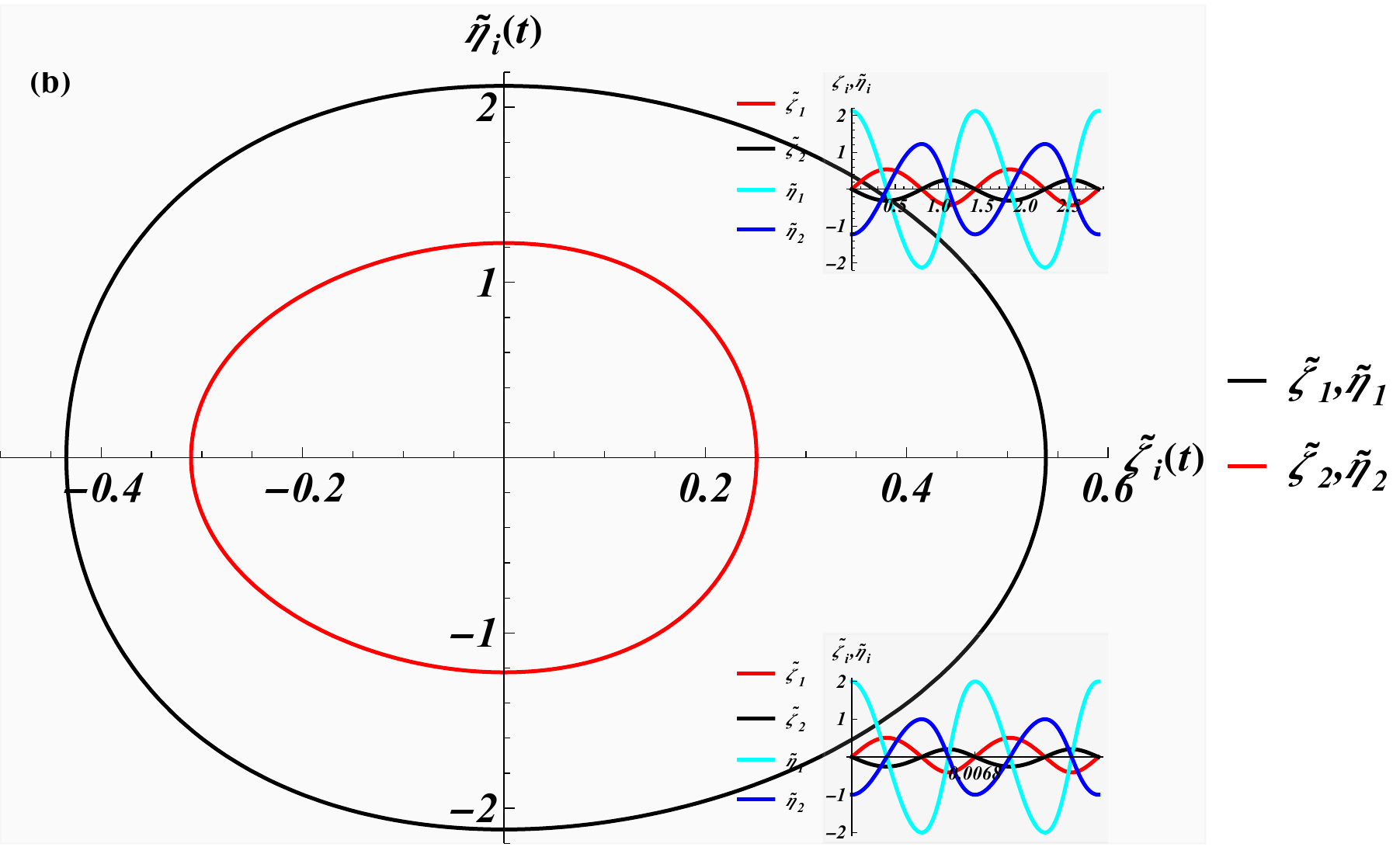}
	\end{minipage}     
	\caption{Affine $G_2$-Toda lattice 
		$\tilde{H}$ and $\tilde{Q}_6$ phase spaces for the variables $(\tilde{q}_i,\tilde{p}_i)$, $1,2,3$ and $(\tilde{\zeta}_i,\tilde{\eta}_i)$, $1,2$ as functions of $t$ panel (a) and panel (b), respectively.  The initial condition are taken in both cases as $\tilde{q}_1(0)=\tilde{q}_2(0)=\tilde{q}_3(0)=0$, $\tilde{p}_1(0)=2 $, $\tilde{p}_2(0)=\tilde{p}_3(0)=-1$ and  $\tilde{\zeta}_1(0) =\tilde{\zeta}_2(0)=0$, $\tilde{\eta}_1(0)=3/\sqrt{2}$ and  $\tilde{\eta}_2(0)=-\sqrt{3}/\sqrt{2}$.} 
	\label{PhaseG2redspecial}   
\end{figure}

Remarkably, for some special initial conditions we can also identify full periodic solutions. The numerical solutions for the three dimensional case with the special initial conditions $\tilde{p}_2(0)=\tilde{p}_3(0)$ are depicted in figure \ref{PhaseG2redspecial} panel (a). We observe that the periodic solutions for $i=2$ and $i=3$ becoming identical. Moreover, the trajectories for the $\tilde{H}$ and $\tilde{Q}_6$ phase spaces coincide. However, as seen in the inlets the periods differ by orders of magnitude with $\tau_H \approx 1.4210841$ and $\tau_{Q_6} \approx 0.0068228$.
Similar features are observed for the dimensionally reduced case depicted in panel (b). The condition $\tilde{p}_2(0) \rightarrow \tilde{p}_3(0)$ translates into $\tilde{\eta}_2(0) \rightarrow - 1/\sqrt{3} \tilde{\eta}_2(0) $.

\section{Non-integrable perturbations}	

\subsection{Sensitivity of the initial conditions}	

\begin{figure}[h]
	\centering    
	\begin{minipage}[b]{0.36\textwidth}              
		\includegraphics[width=\textwidth]{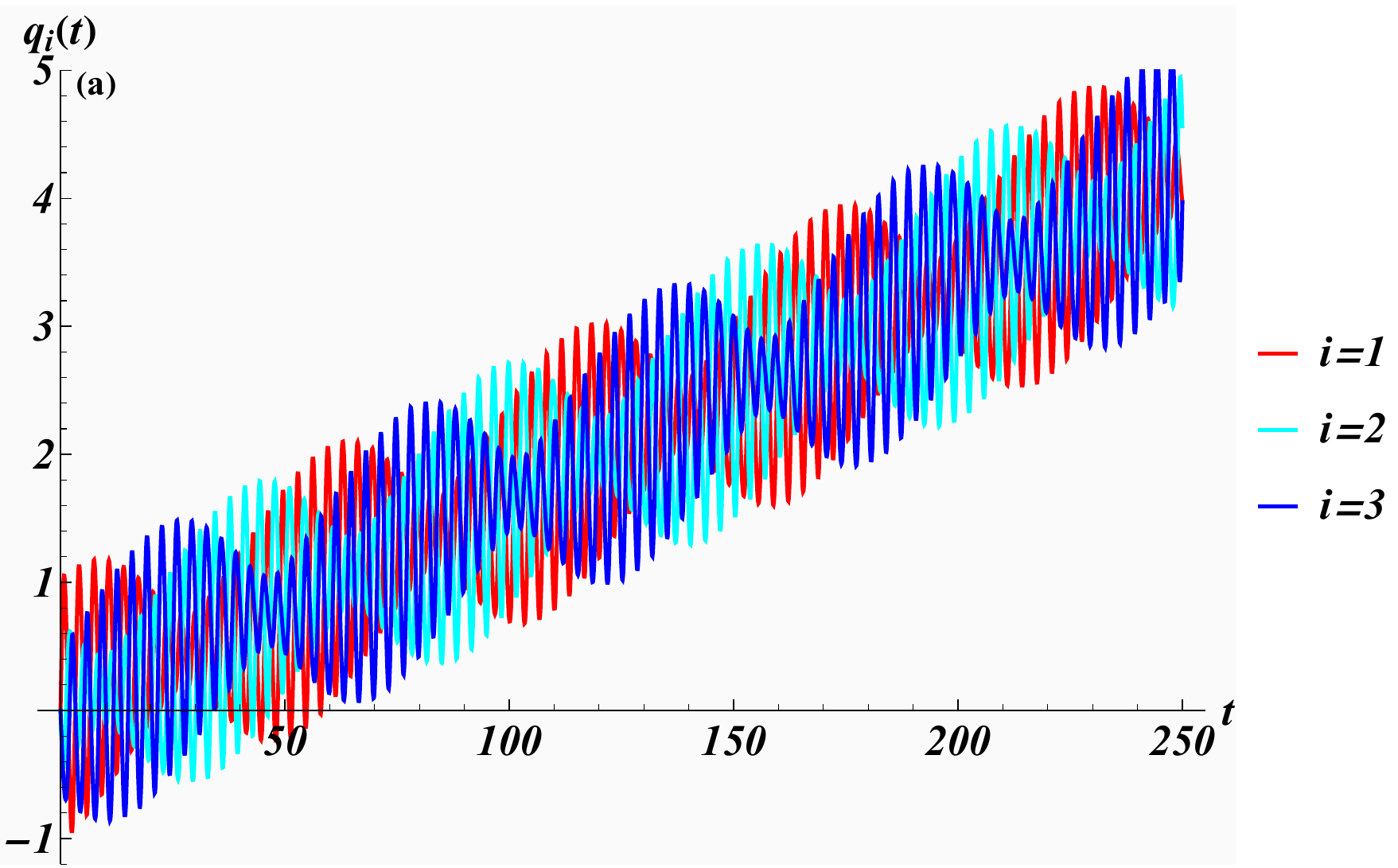}
	\end{minipage}  
\begin{minipage}[b]{0.31\textwidth}              
	\includegraphics[width=\textwidth]{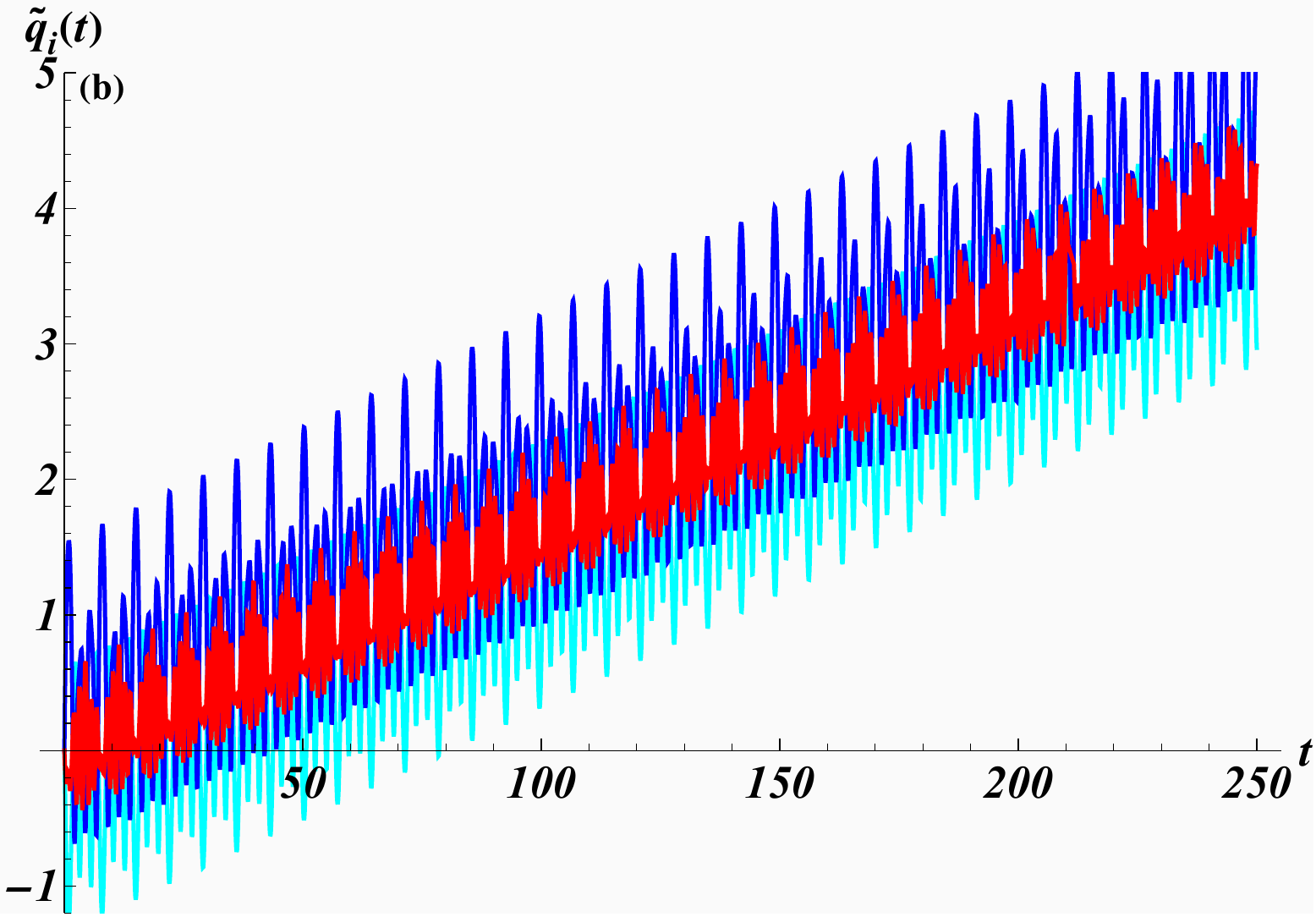}
\end{minipage}  
\begin{minipage}[b]{0.31\textwidth}              
	\includegraphics[width=\textwidth]{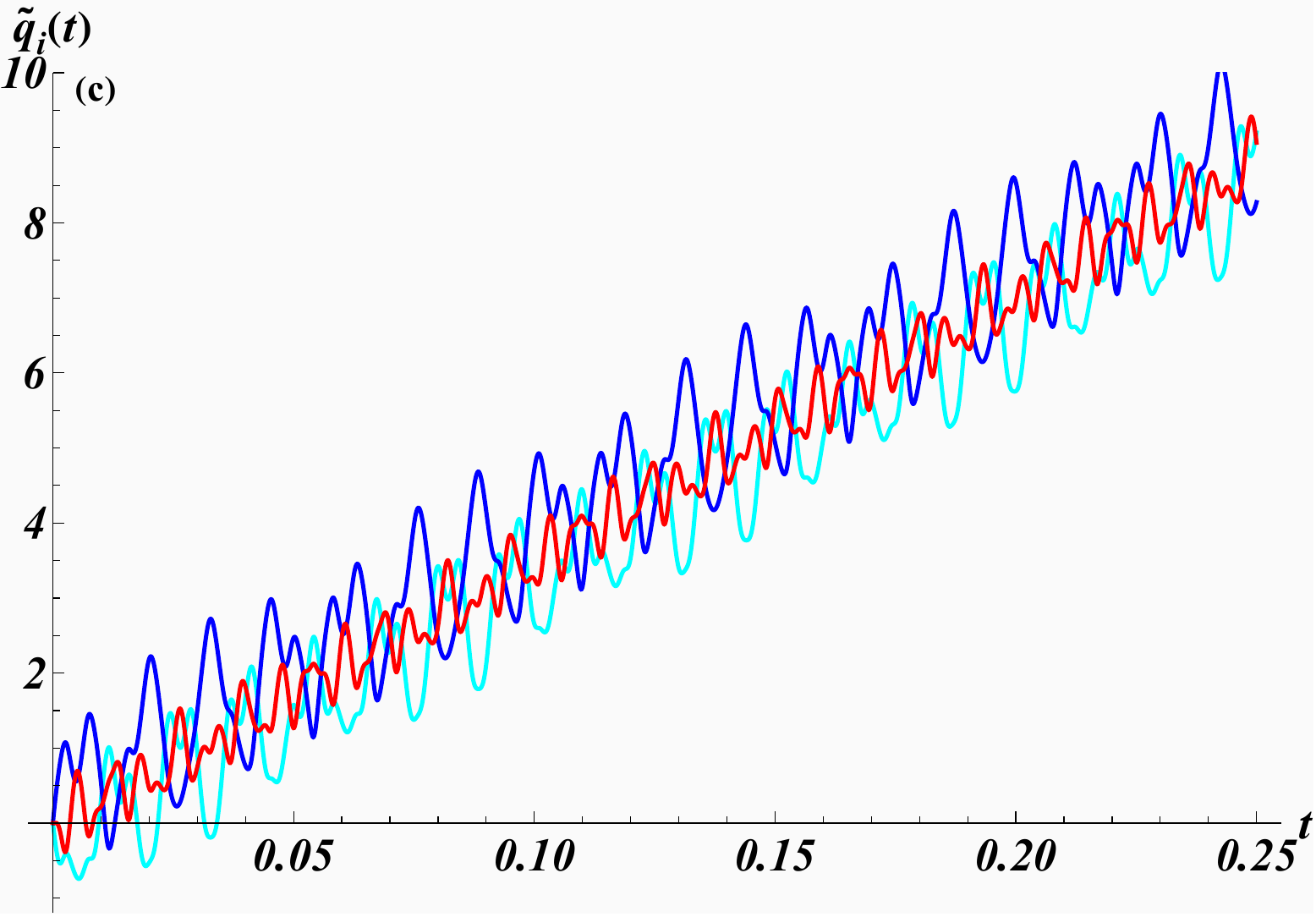}
\end{minipage}  
	\begin{minipage}[b]{0.36\textwidth}           
		\includegraphics[width=\textwidth]{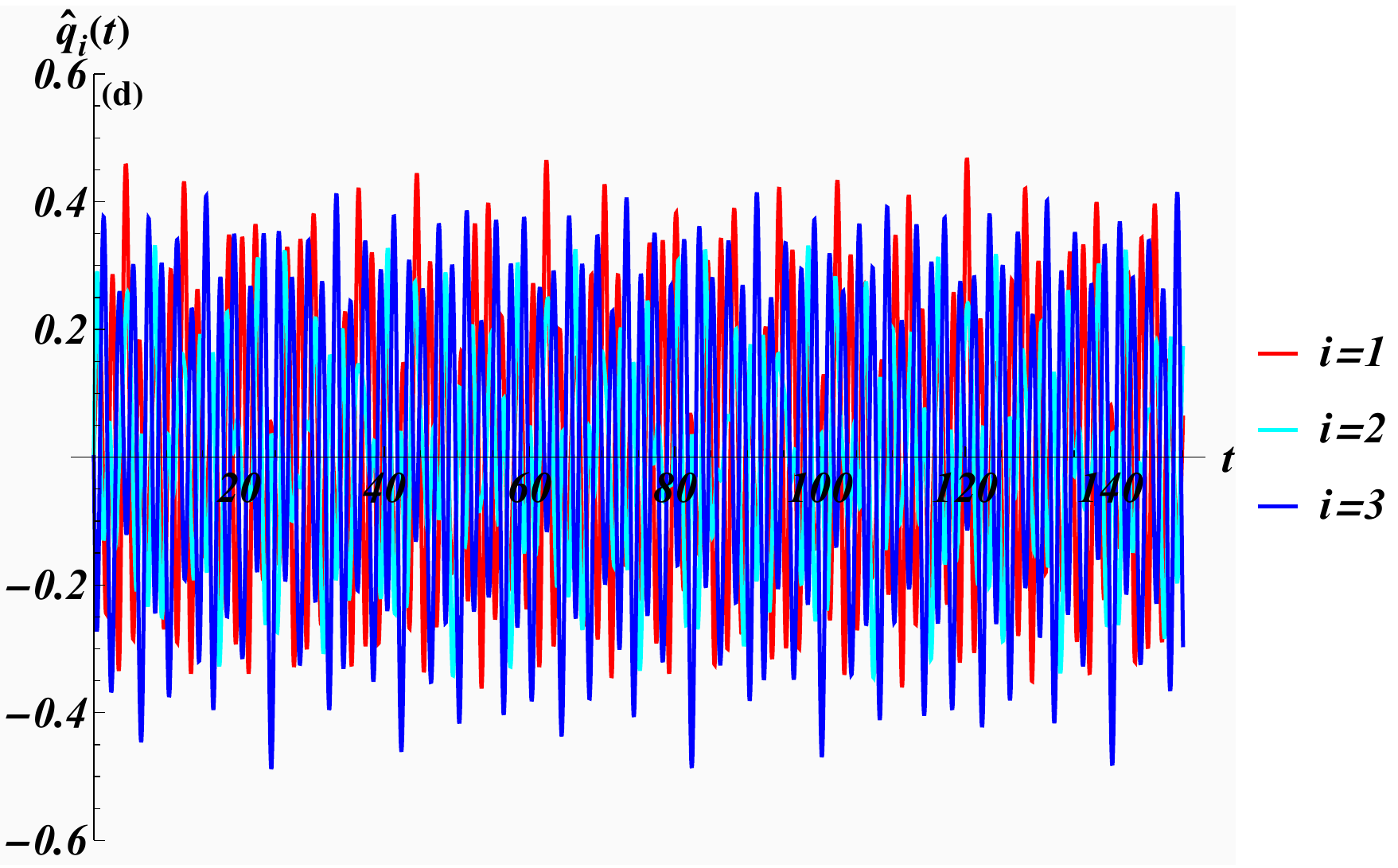}
	\end{minipage}   
\begin{minipage}[b]{0.31\textwidth}           
	\includegraphics[width=\textwidth]{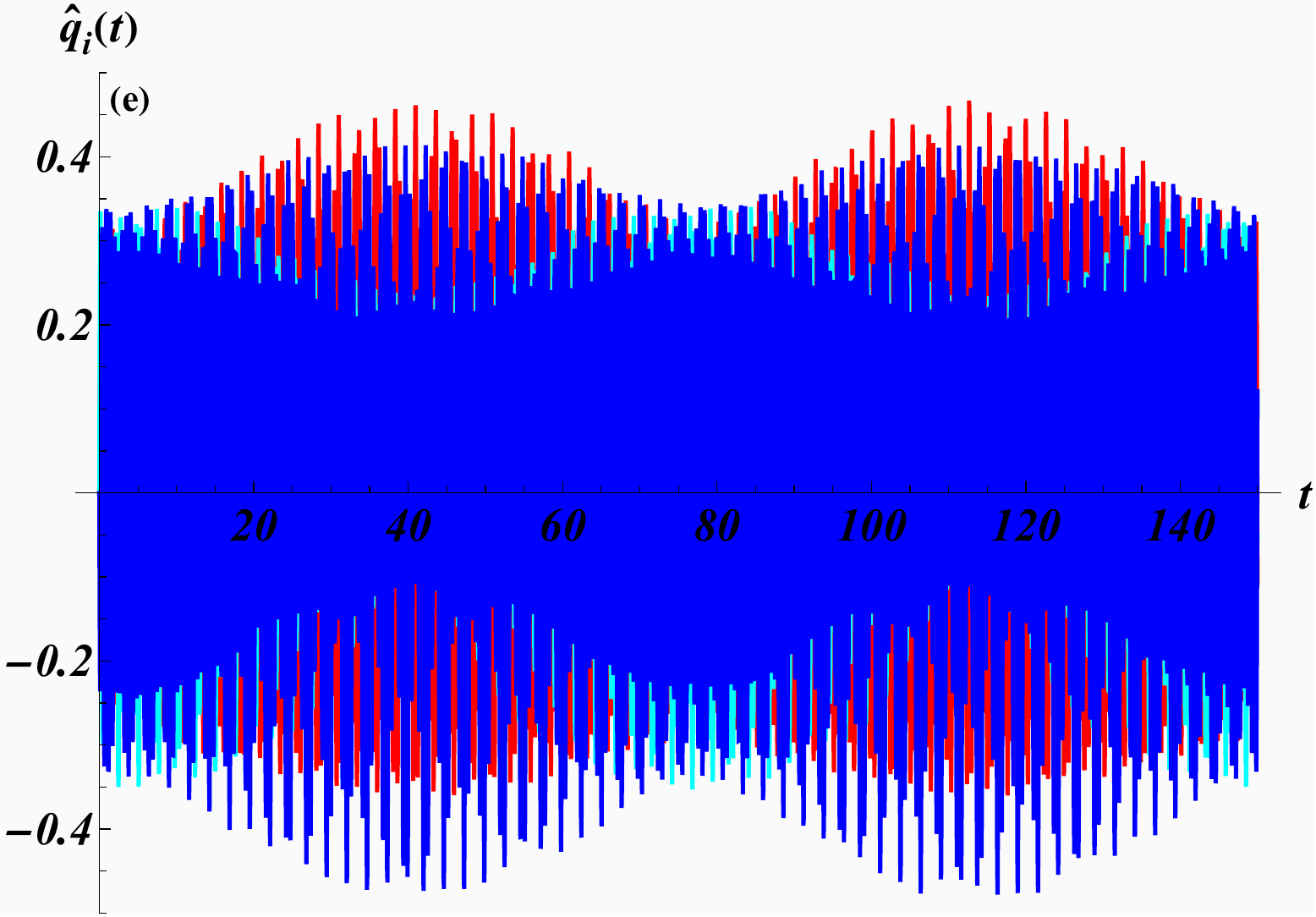}
\end{minipage}  
\begin{minipage}[b]{0.31\textwidth}           
	\includegraphics[width=\textwidth]{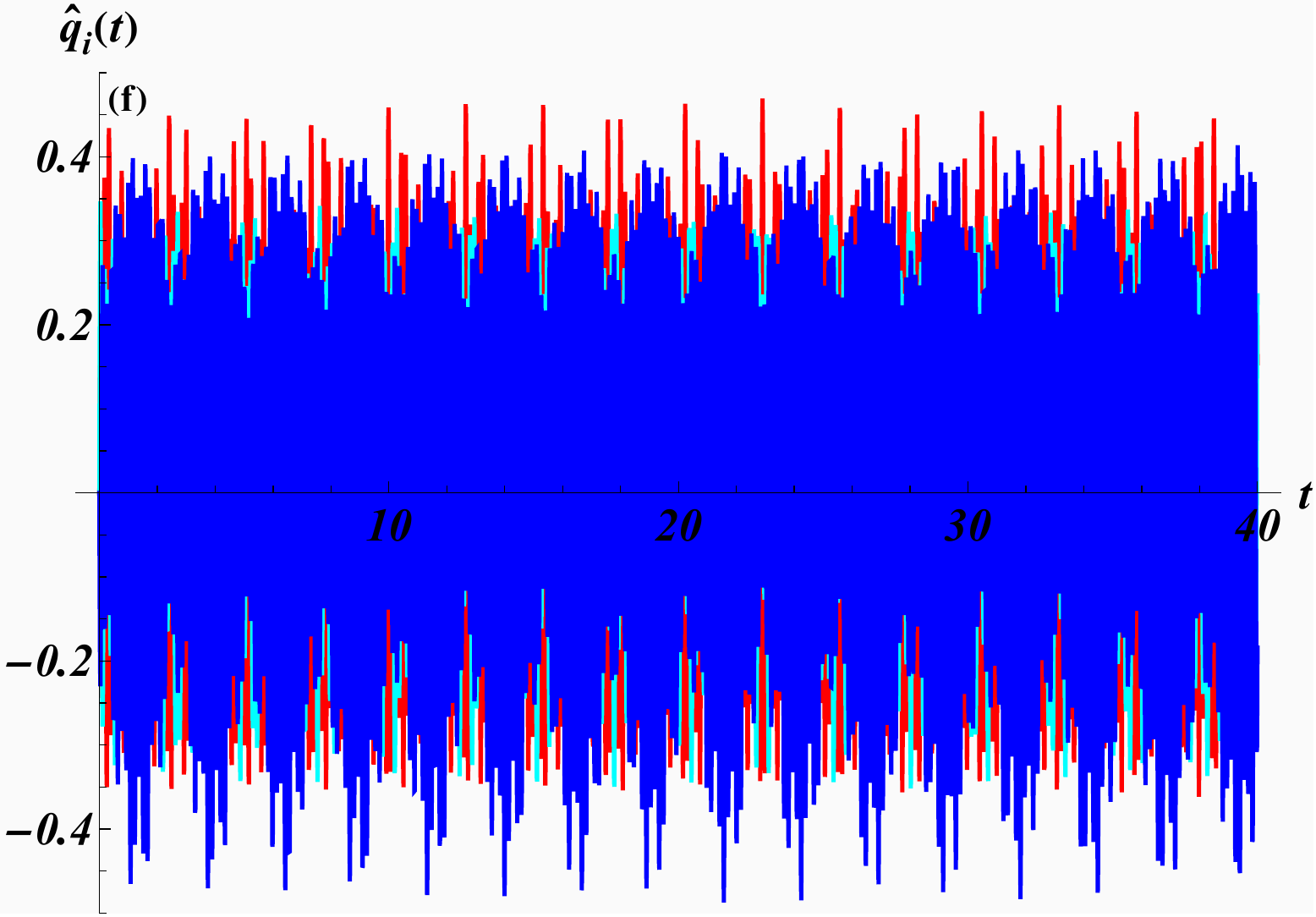}
\end{minipage}        
	\caption{Three dimensional phase space solutions for the coordinates as functions of $t$ with initial conditions $\sum_{i=1}^3 p_i \neq 0$. Panels (a), (b), (c): Manevolent $A_2$-solutions for the Hamiltonian $H$, $G_2$-solutions for the Hamiltonian $\tilde{H}$ and the charge $\tilde{Q}_6$, respectively, all with initial conditions $\tilde{q}_1(0)=\tilde{q}_2(0)=\tilde{q}_3(0)=0$, $\tilde{p}_1(0)=-1 $, $\tilde{p}_2(0)=-2  $   $\tilde{p}_3(0)=3+0.05$. Panels (d), (e), (f): Benign $B_3$-solutions for the Hamiltonian $\hat{H}$ and the charge $\hat{Q}_4$, $\hat{Q}_6$ respectively, all initial conditions are taken to $\hat{q}_1(0)=\hat{q}_2(0)=\hat{q}_3(0)=0$, $\hat{p}_1(0)=0.2 $, $\hat{p}_2(0)=-1/2  $   $\hat{p}_3(0)=1/2$.} 
	\label{PhaseABGpertini}
\end{figure}

There are various possibilities to investigate the stability of the above divergent and benign ghost solutions. The most delicate way to perturb them is to just vary the initial conditions. We have seen that the Poisson bracket relation between the centre of mass coordinate and particular charges govern in the standard representation for the $A_6$-algebra whether the solutions converge or not. It turns out that in the representations for which the dimensions do not match up with the rank of the algebra the trajectories are rather sensitive towards these impositions. In figure \ref{PhaseABGpertini} panels (a) to (c) we see that even a very small violation of the condition $\sum_{i=1}^3 p_i =0$ leads to the divergence of the trajectories in the $x_i$-directions in the phase spaces even for the Hamiltonians $H$ and $\tilde{H}$ for $A_2$ and $G_2$-theories as well as for the $\tilde{Q}_6$-charge of $G_2$. In panels (d) to (f) we observe that the trajectories for all independent $B_3$-charges remain oscillatory when $\sum_{i} p_i \neq 0$. The violation of $\sum_{i} q_i =0$ does not produce this effect. 

\begin{figure}[h]
	\centering     
	\begin{minipage}[b]{0.52\textwidth}              
		\includegraphics[width=\textwidth]{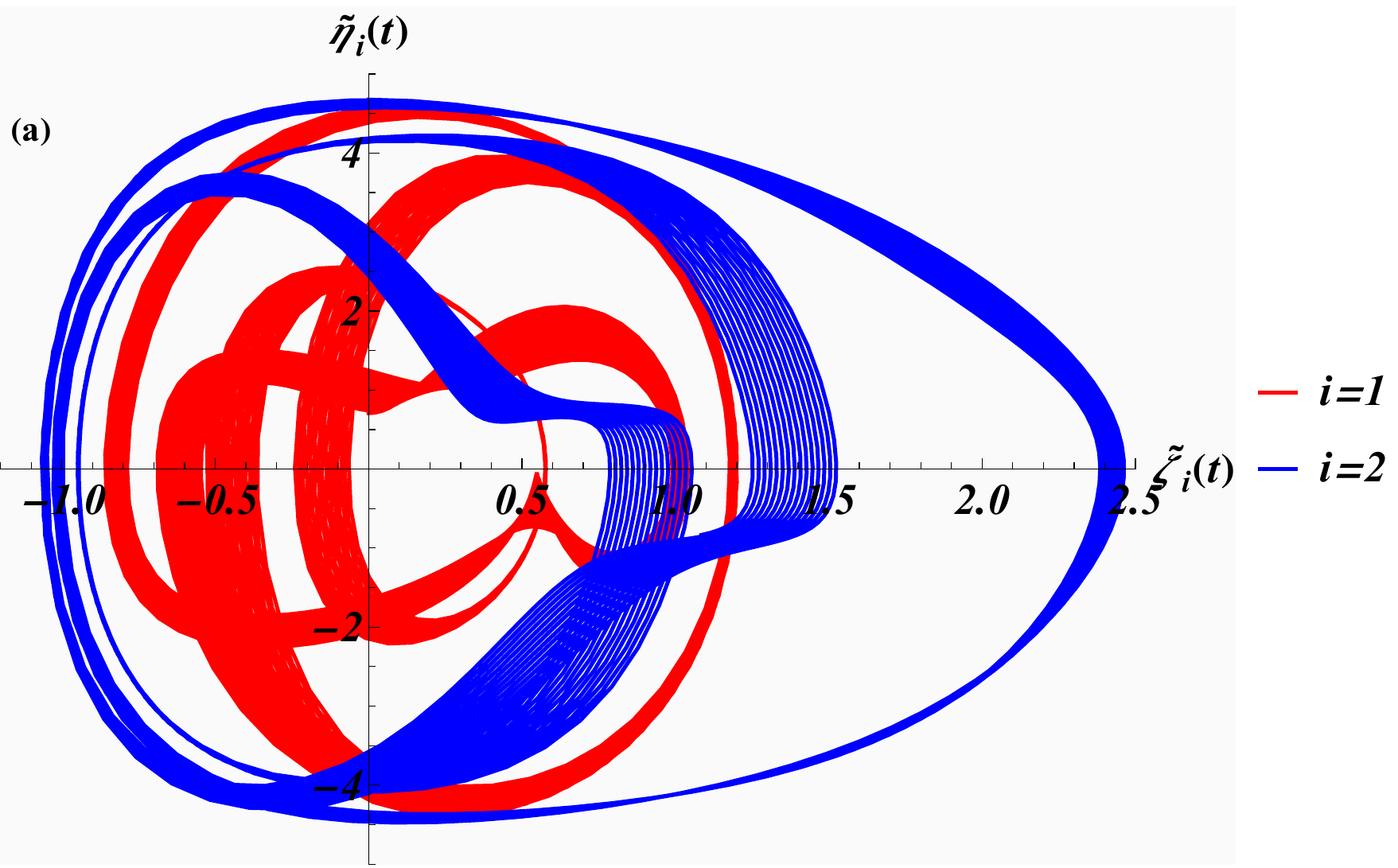}
	\end{minipage}  
	\begin{minipage}[b]{0.46\textwidth}           
		\includegraphics[width=\textwidth]{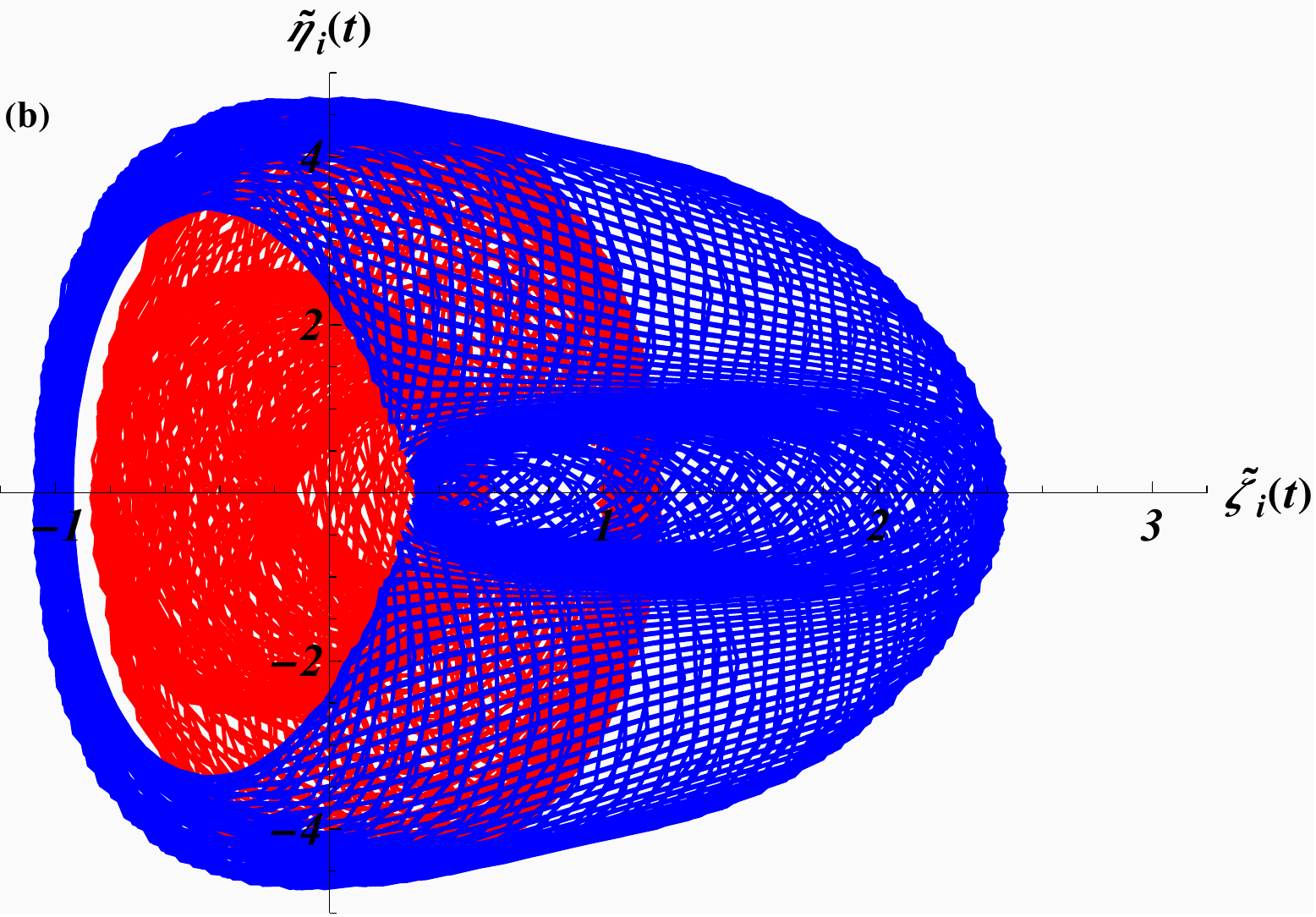}
	\end{minipage}         
	\caption{Two-dimensional affine $G_2$-Toda lattice 
		$\tilde{H}$ and $\tilde{Q}_6$ phase spaces $(\tilde{\zeta}_i,\tilde{\eta}_i)$, $1,2$, panels (a) and (b) respectively, for perturbed initial conditions as functions of $t$. The initial conditions are taken in both cases as $\tilde{\zeta}_1(0) =\tilde{\zeta}_2(0)=0$, $\tilde{\eta}_1(0)=1/\sqrt{2}$ and  $\tilde{\eta}_2(0)=3\sqrt{3}/\sqrt{2} +1$.} 
	\label{PhaseG2pertini}
\end{figure}

In contrast the trajectories in the two dimensional phase spaces are rather robust against very large perturbations of $Q_1(0) \neq 0$ as seen in figure \ref{PhaseG2pertini} panels (a) and (b), where we present the $G_2$-case. In this case the divergence would occur in the third component that has already been set to zero in the construction. The trajectories may be compared to the unperturbed case presented in figure \ref{PhaseG2red}.   

\subsection{Breaking of the integrability}	

\begin{figure}[h]
	\centering    
	\begin{minipage}[b]{0.19\textwidth}              
		\includegraphics[width=\textwidth]{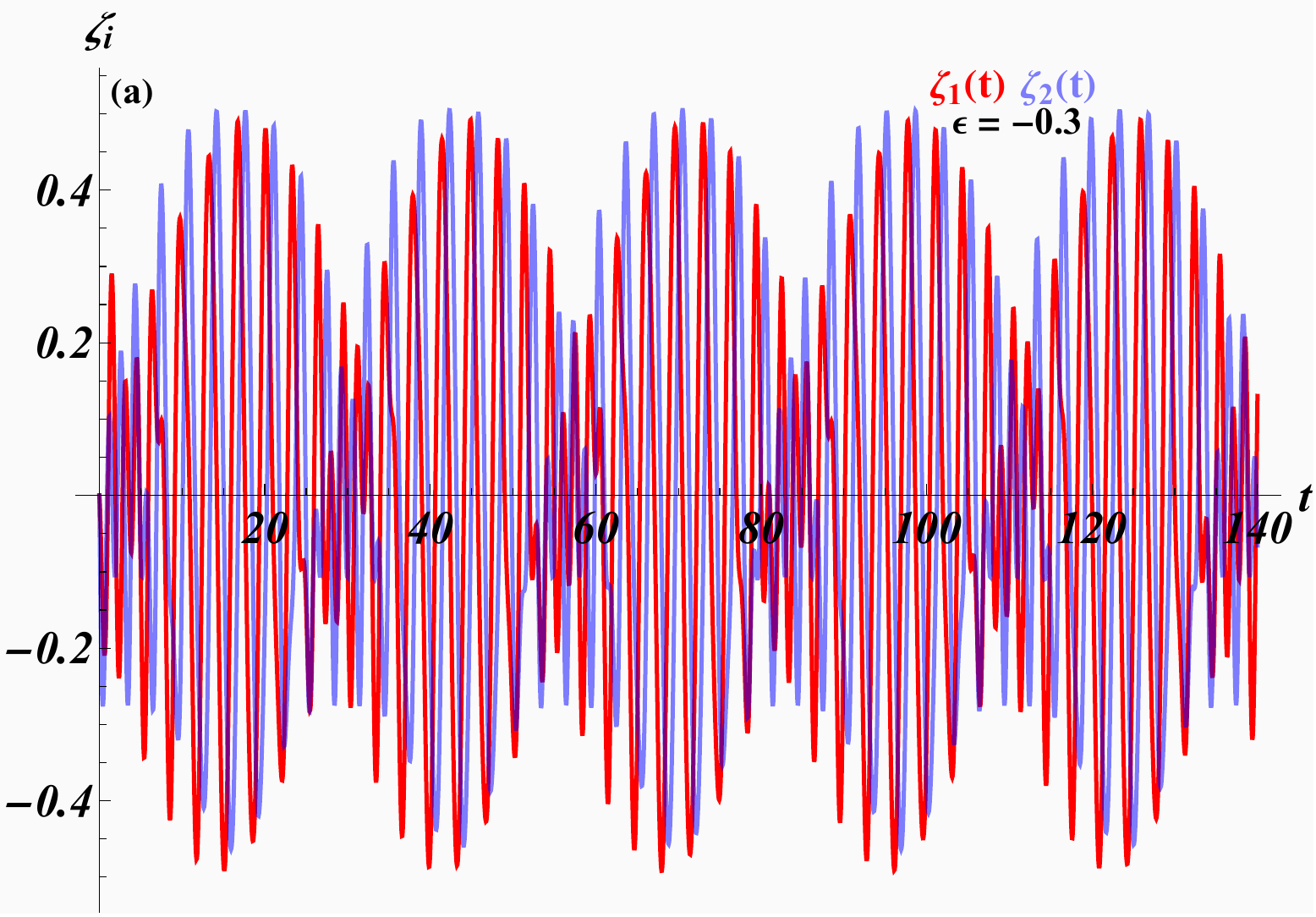}
	\end{minipage}  
	\begin{minipage}[b]{0.19\textwidth}              
		\includegraphics[width=\textwidth]{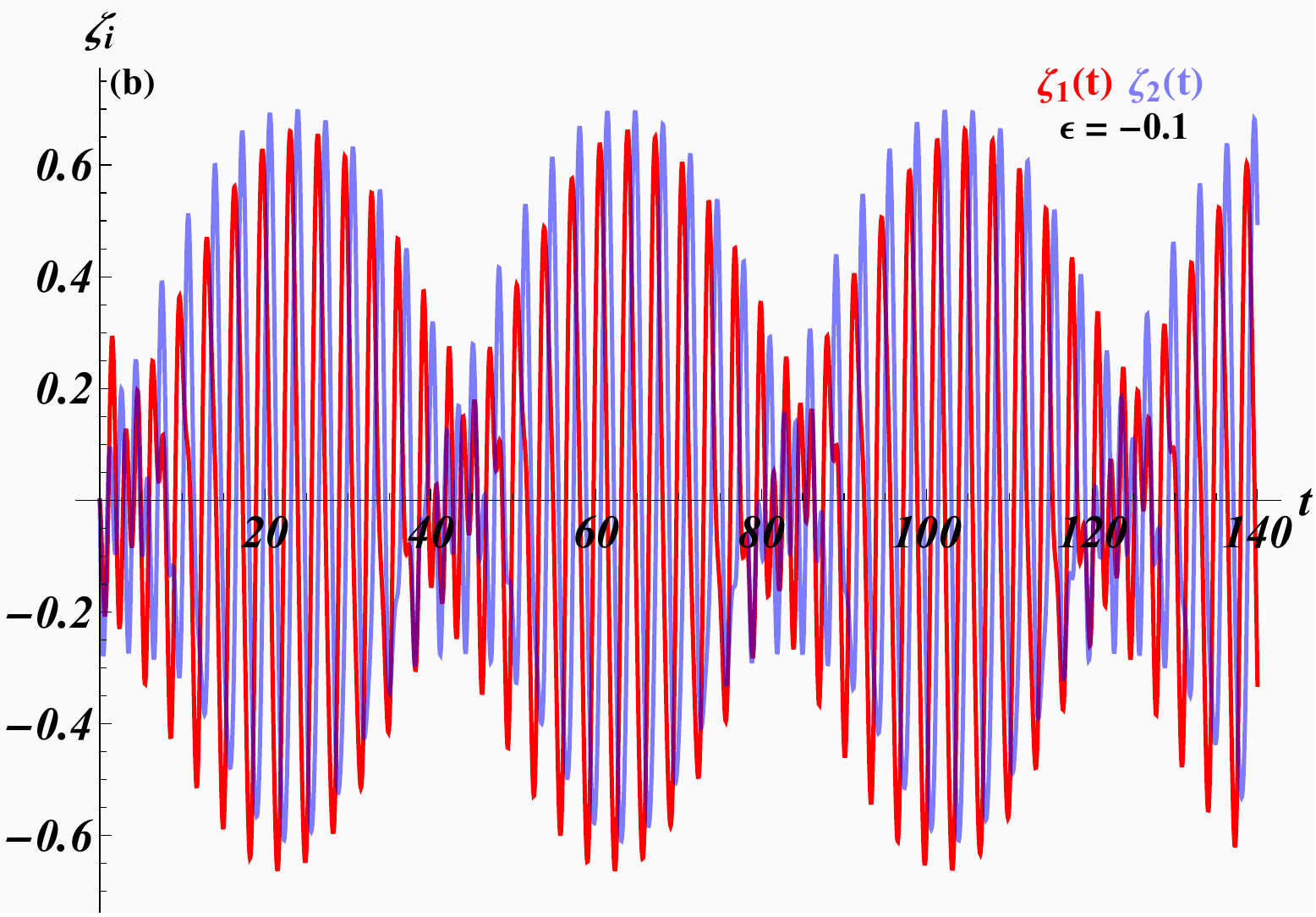}
	\end{minipage}  
\begin{minipage}[b]{0.19\textwidth}              
	\includegraphics[width=\textwidth]{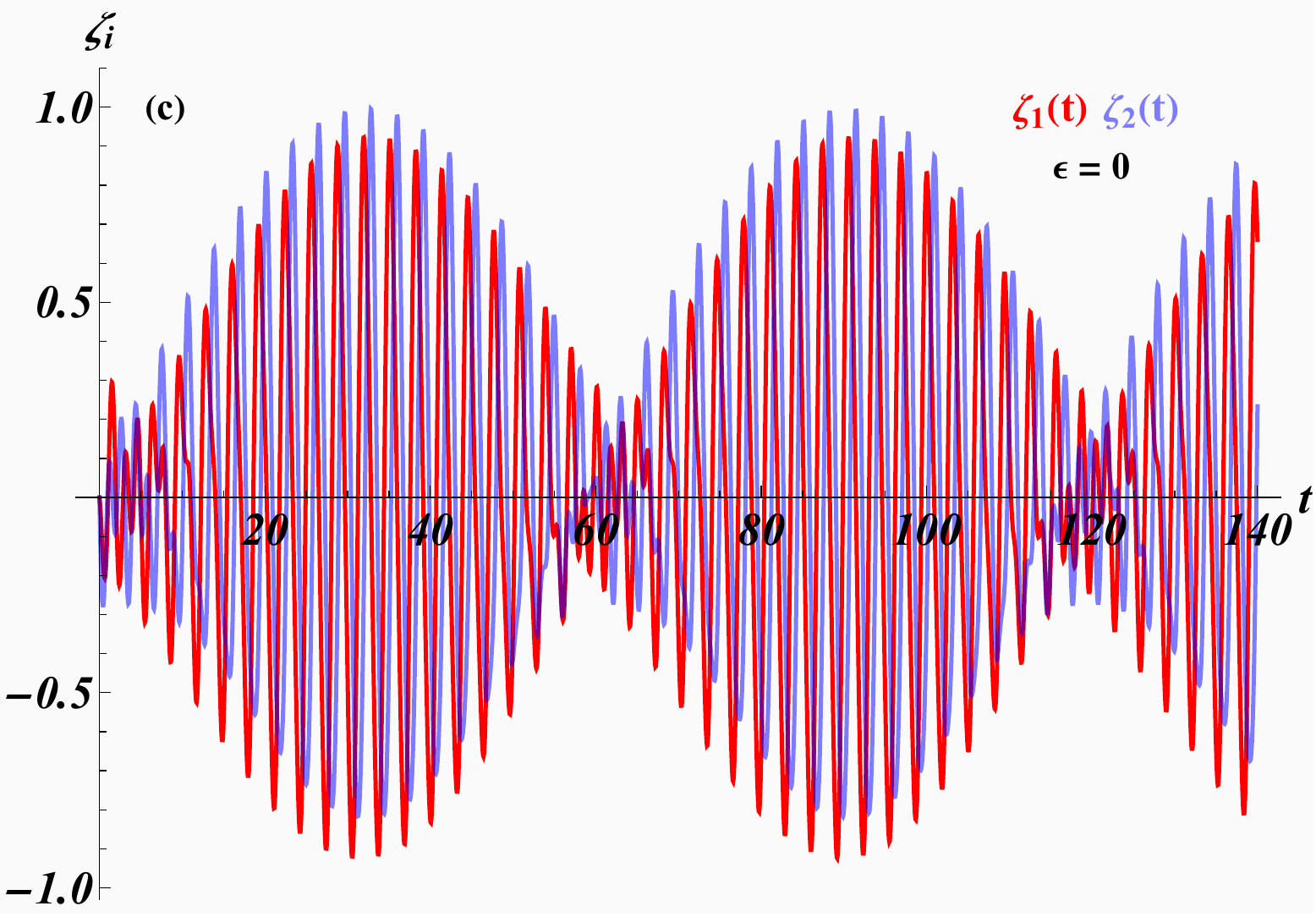}
\end{minipage}  
\begin{minipage}[b]{0.19\textwidth}              
	\includegraphics[width=\textwidth]{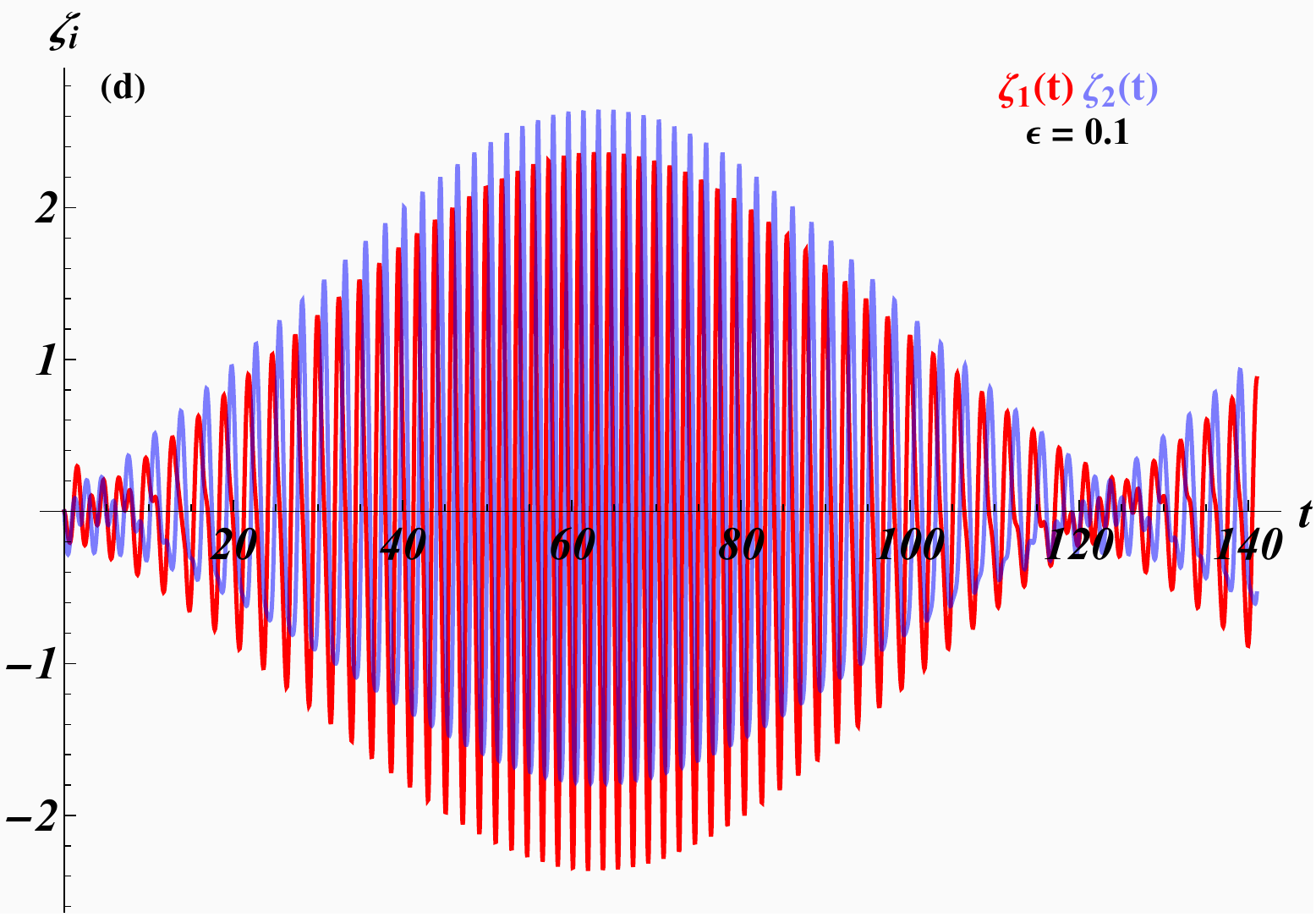}
\end{minipage}  
\begin{minipage}[b]{0.19\textwidth}              
	\includegraphics[width=\textwidth]{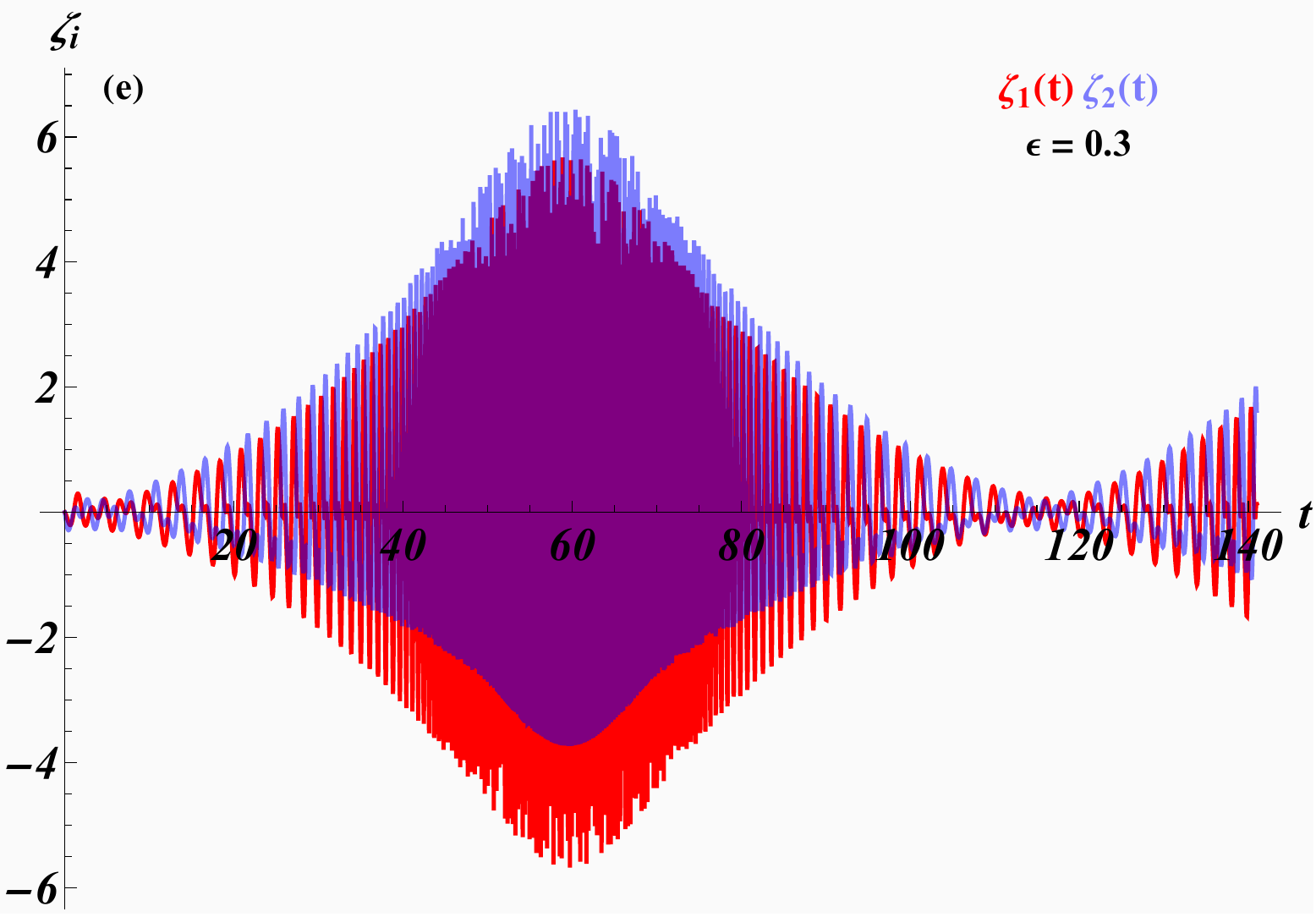}
\end{minipage}    
\centering    
\begin{minipage}[b]{0.19\textwidth}              
	\includegraphics[width=\textwidth]{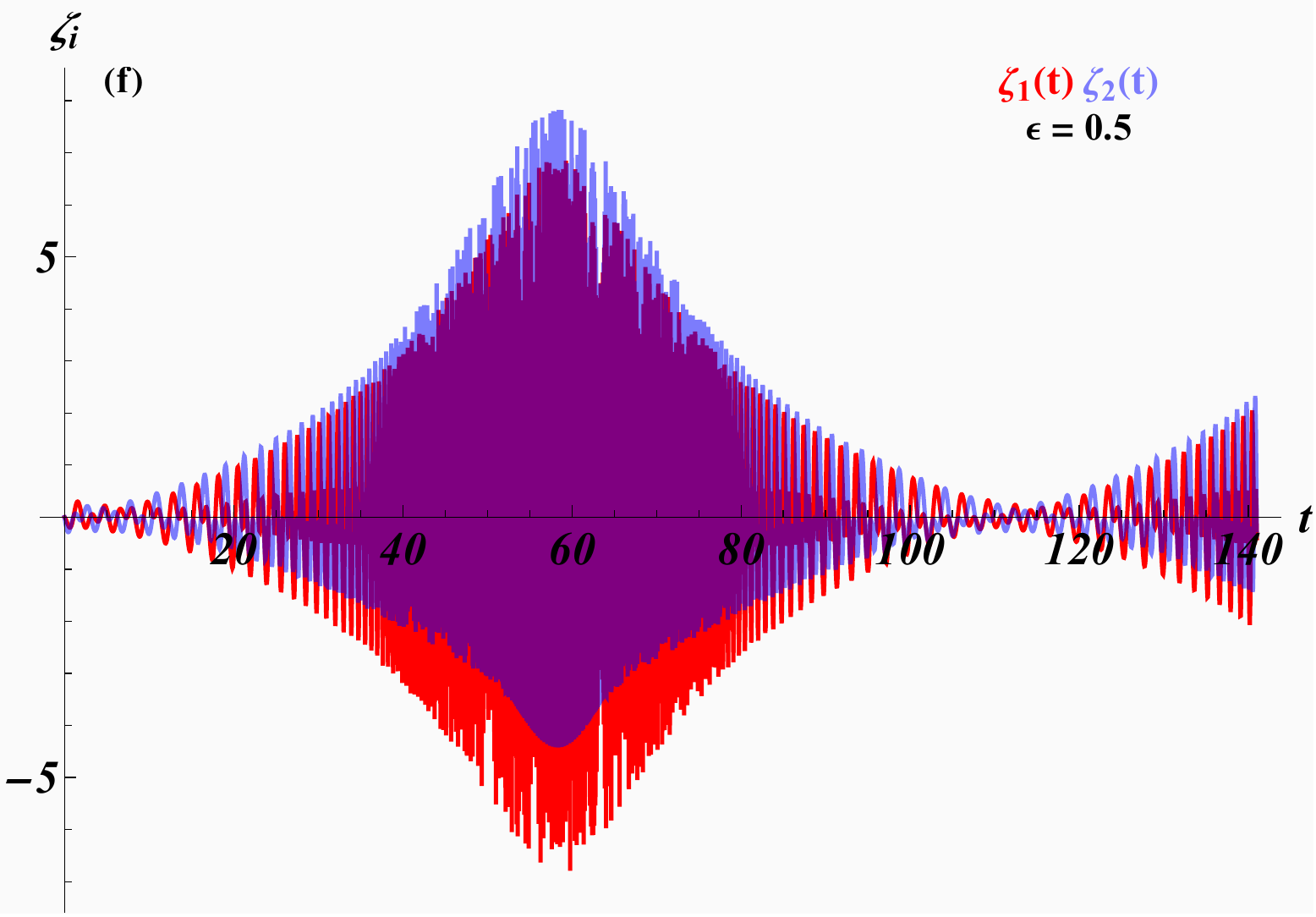}
\end{minipage}  
\begin{minipage}[b]{0.19\textwidth}              
	\includegraphics[width=\textwidth]{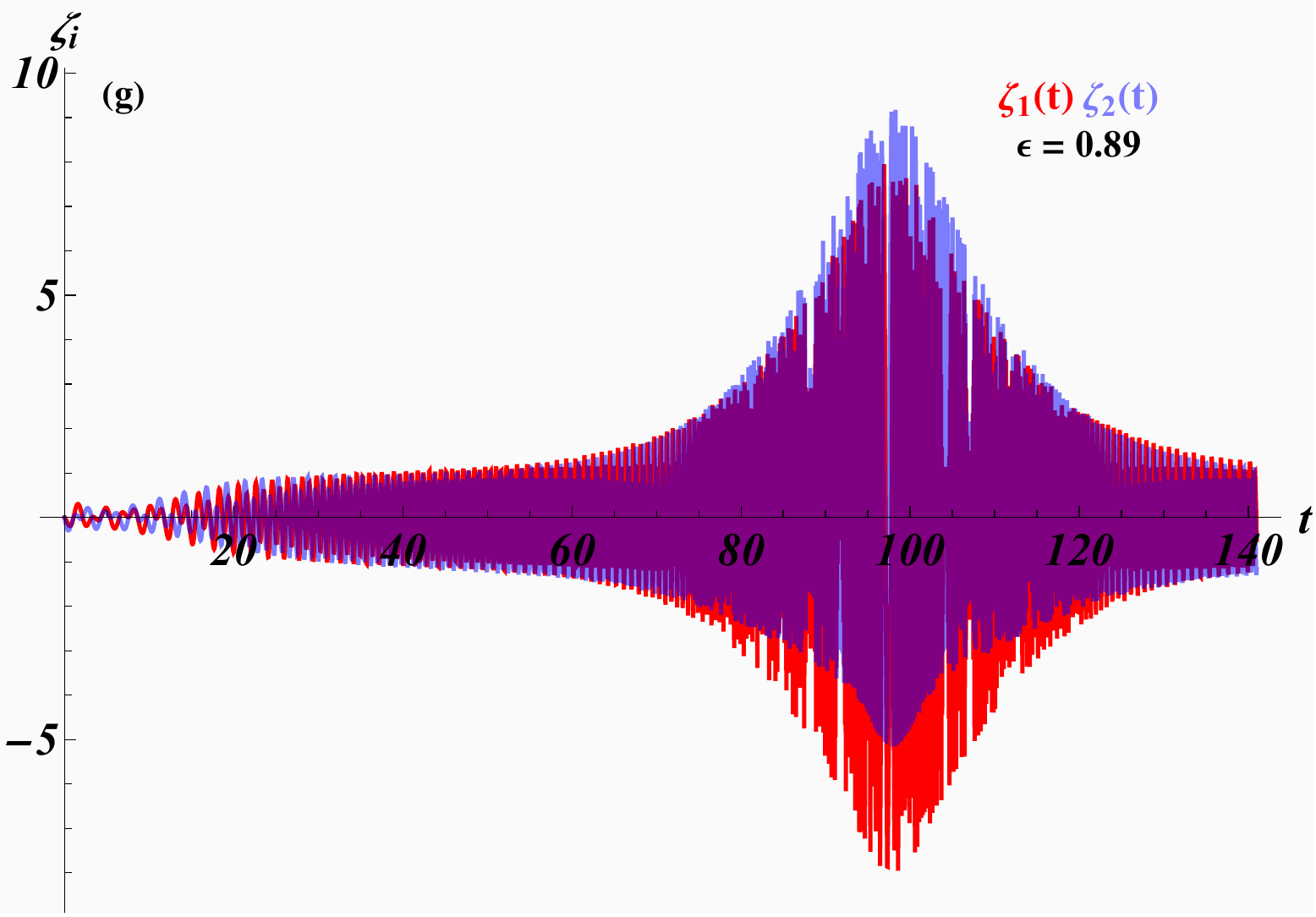}
\end{minipage}  
\begin{minipage}[b]{0.19\textwidth}              
	\includegraphics[width=\textwidth]{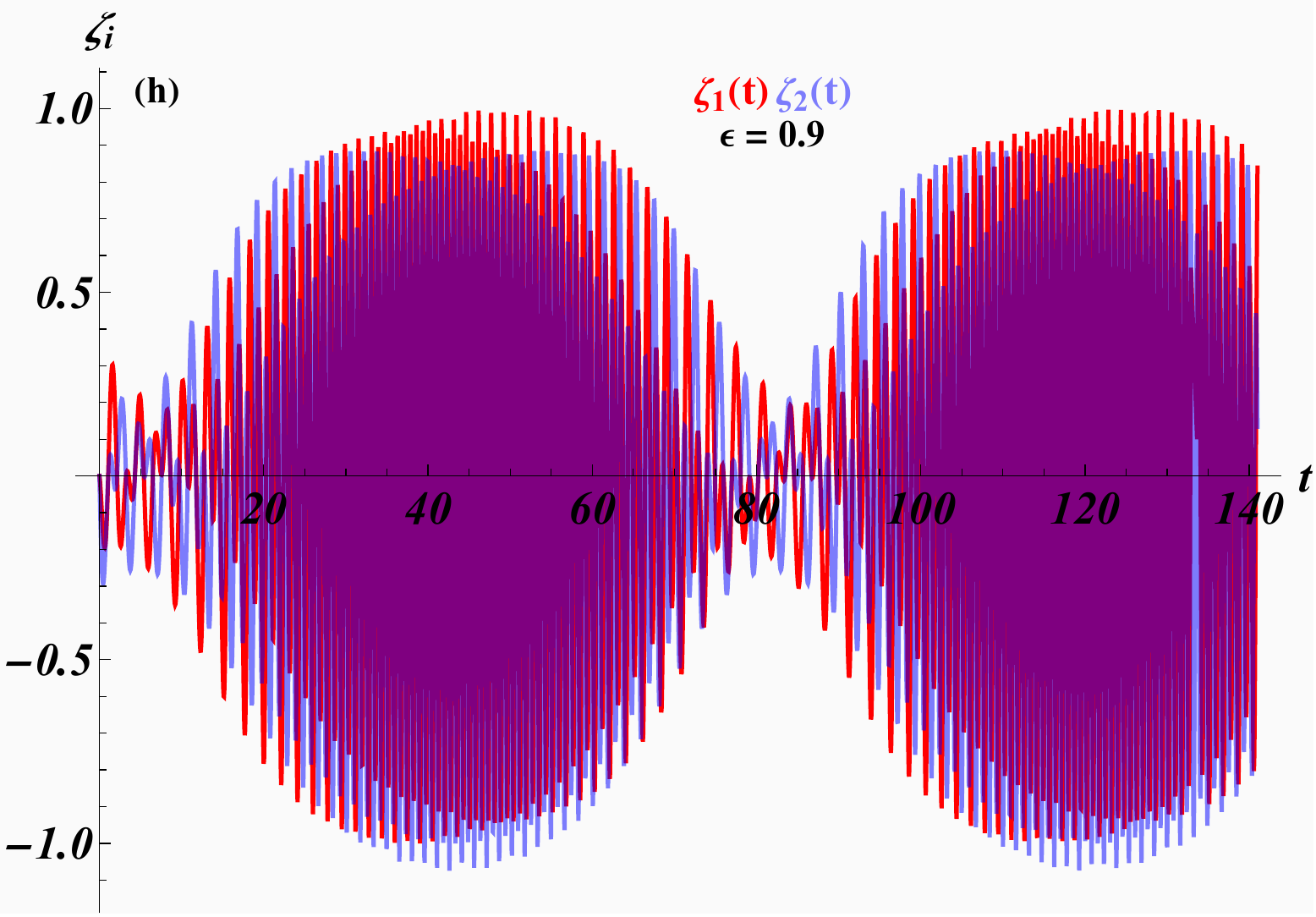}
\end{minipage}  
\begin{minipage}[b]{0.19\textwidth}              
	\includegraphics[width=\textwidth]{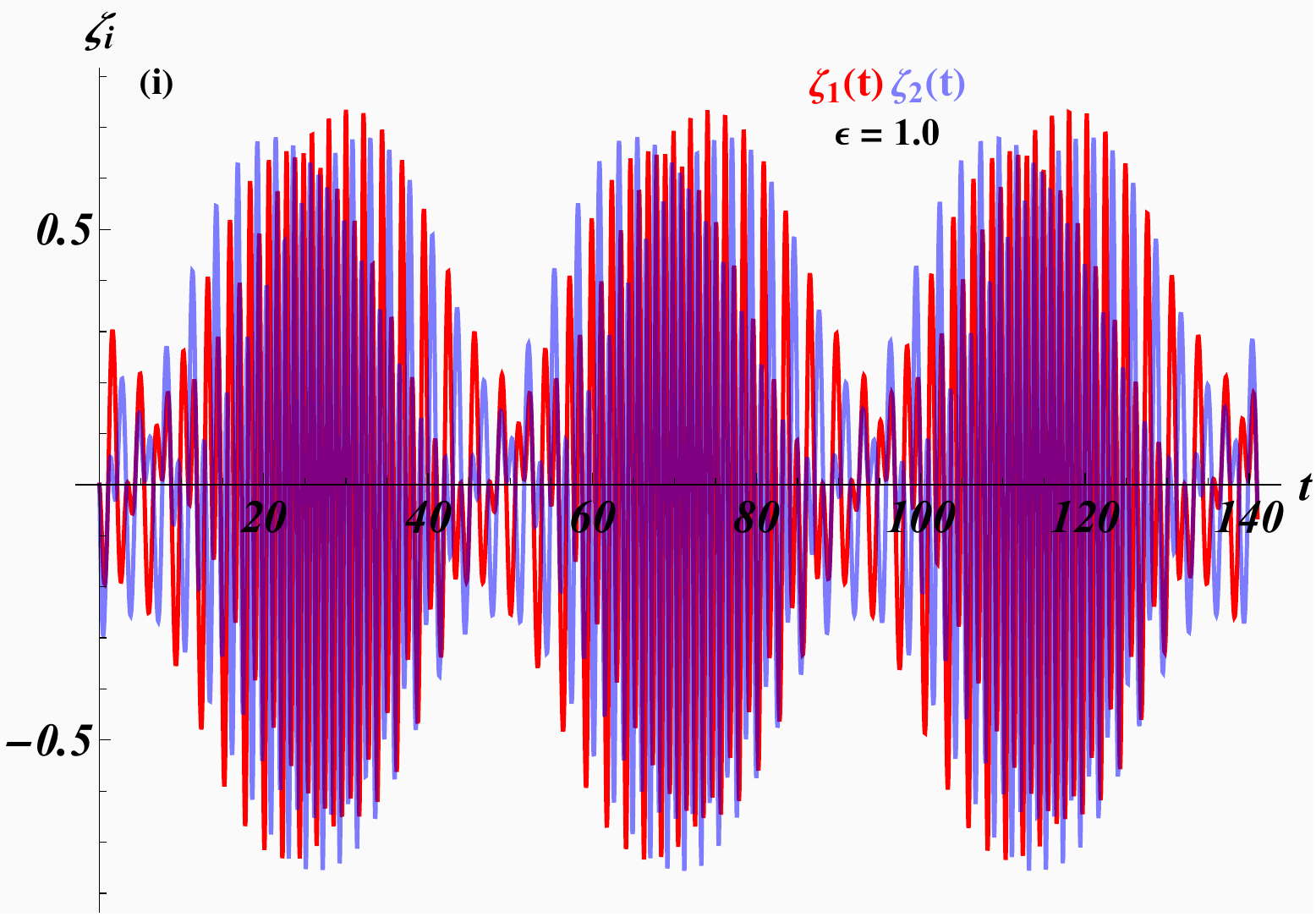}
\end{minipage}  
\begin{minipage}[b]{0.19\textwidth}              
	\includegraphics[width=\textwidth]{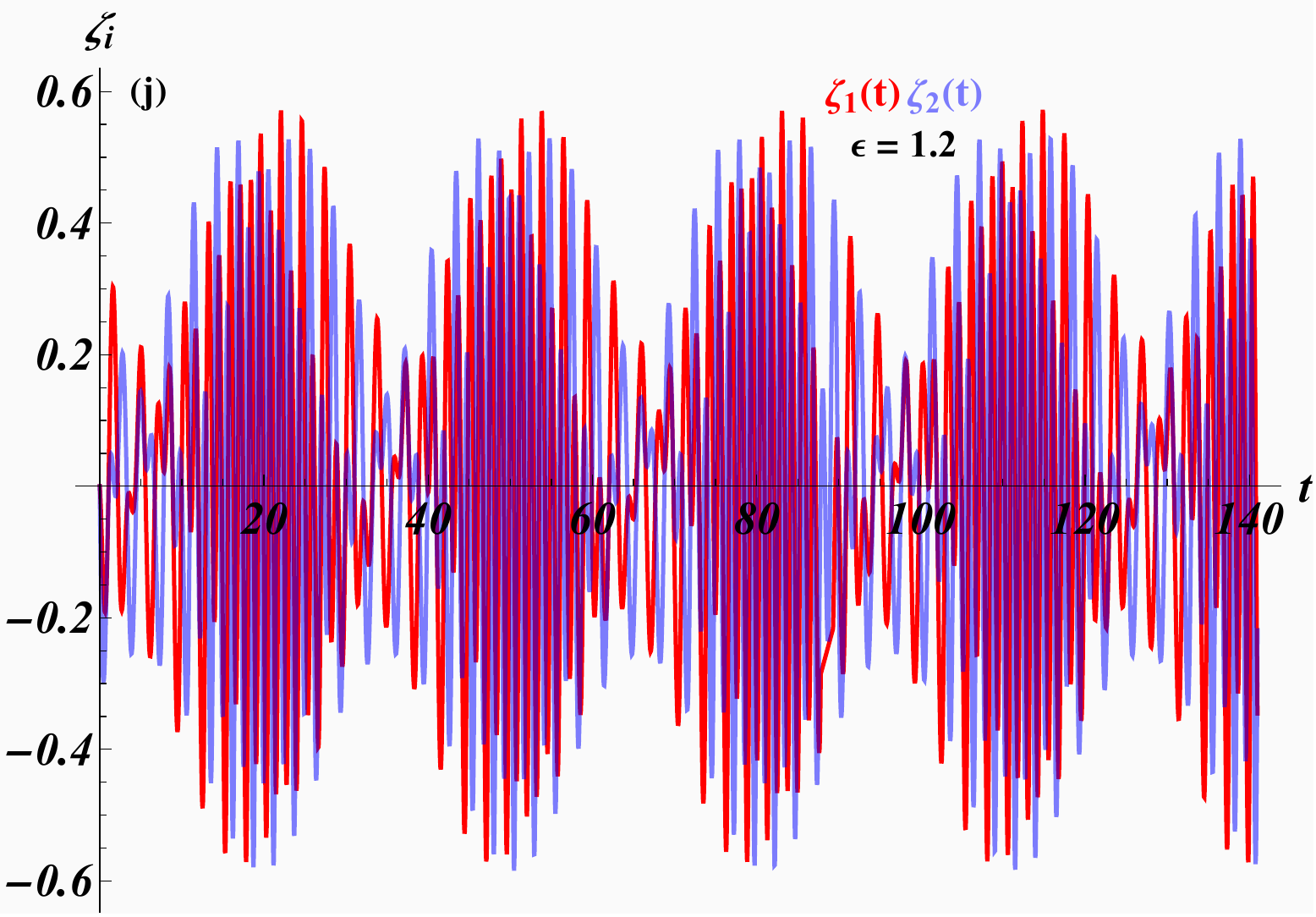}
\end{minipage}     
	\begin{minipage}[b]{0.19\textwidth}              
	\includegraphics[width=\textwidth]{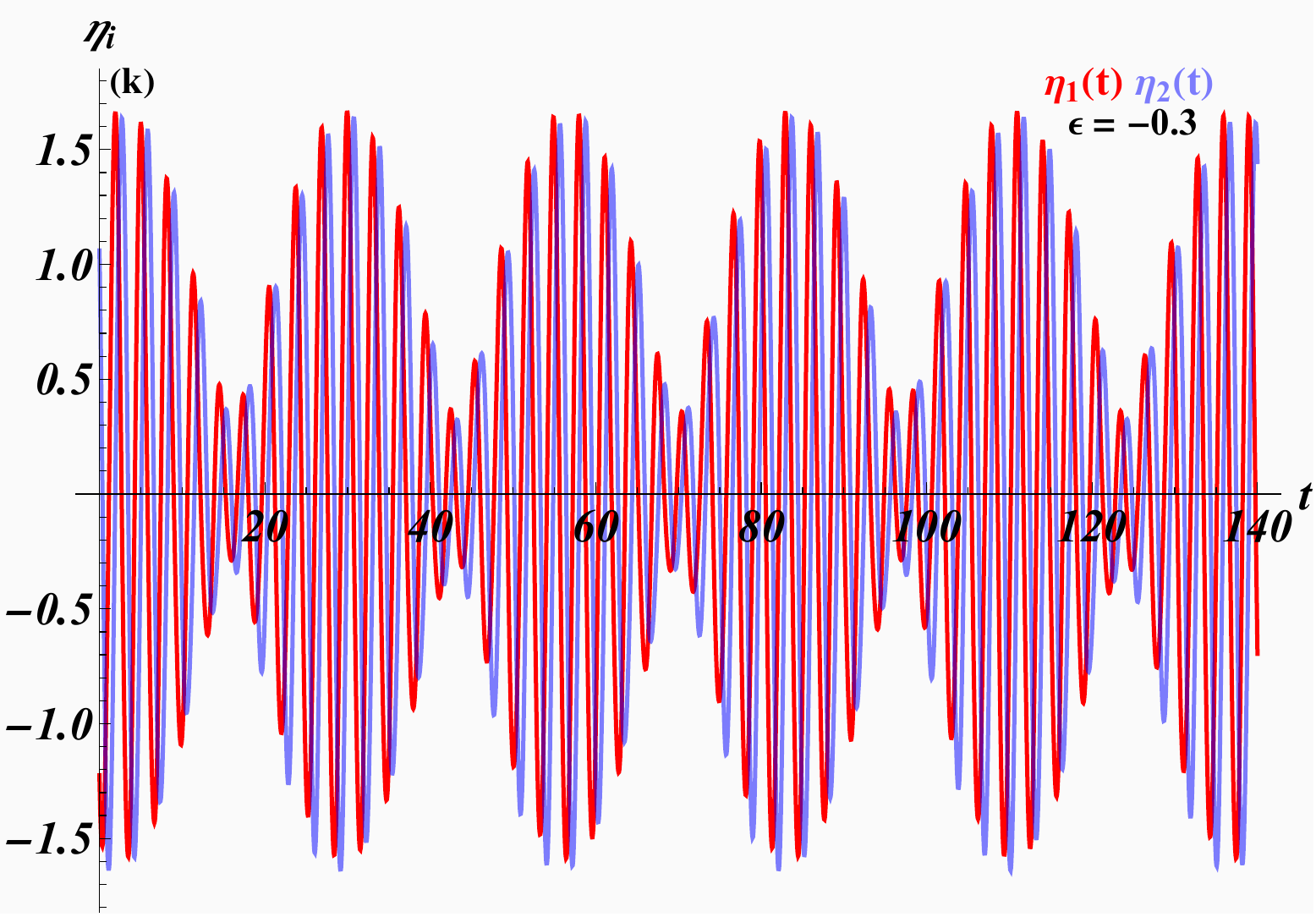}
\end{minipage}  
\begin{minipage}[b]{0.19\textwidth}              
	\includegraphics[width=\textwidth]{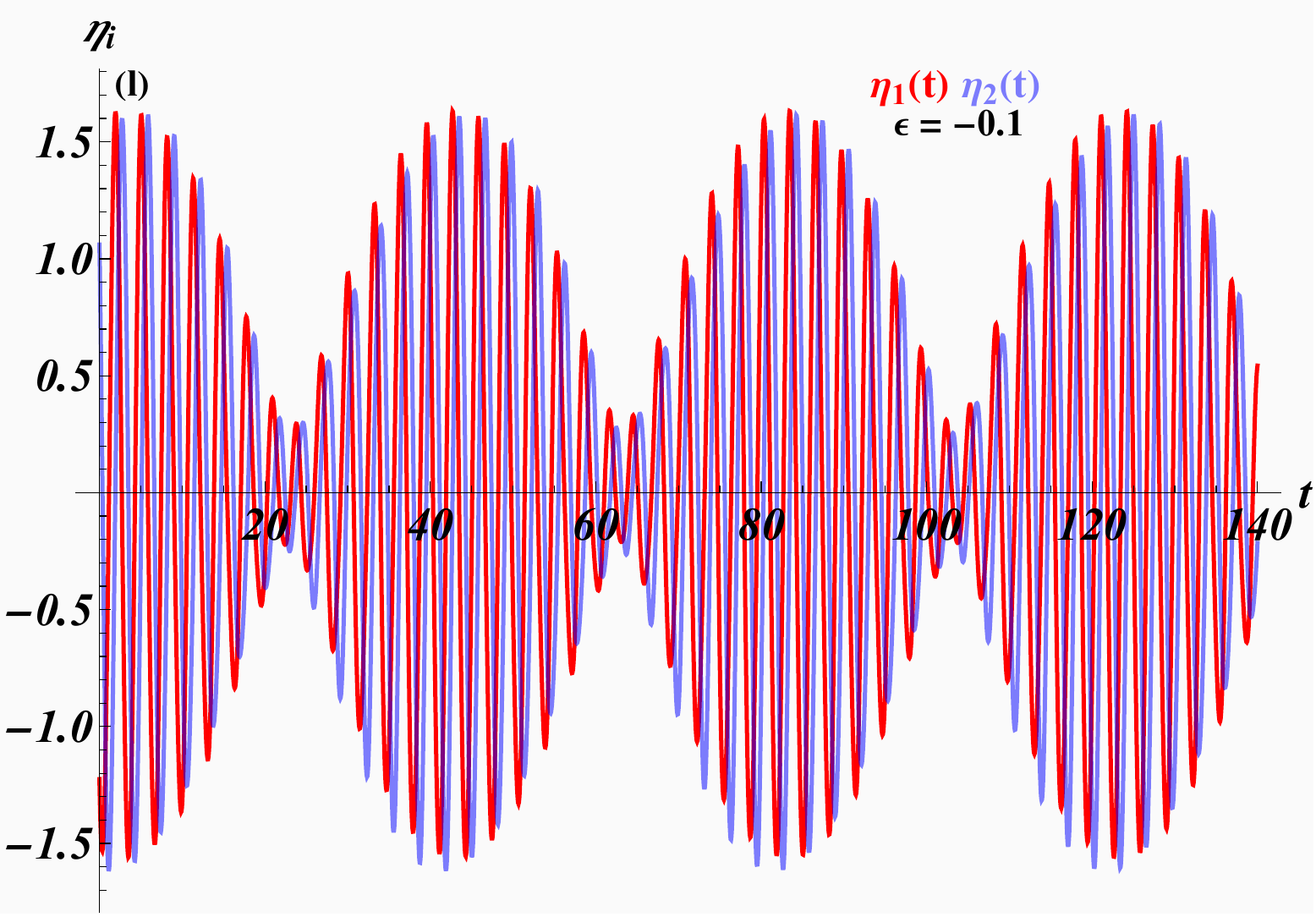}
\end{minipage}  
\begin{minipage}[b]{0.19\textwidth}              
	\includegraphics[width=\textwidth]{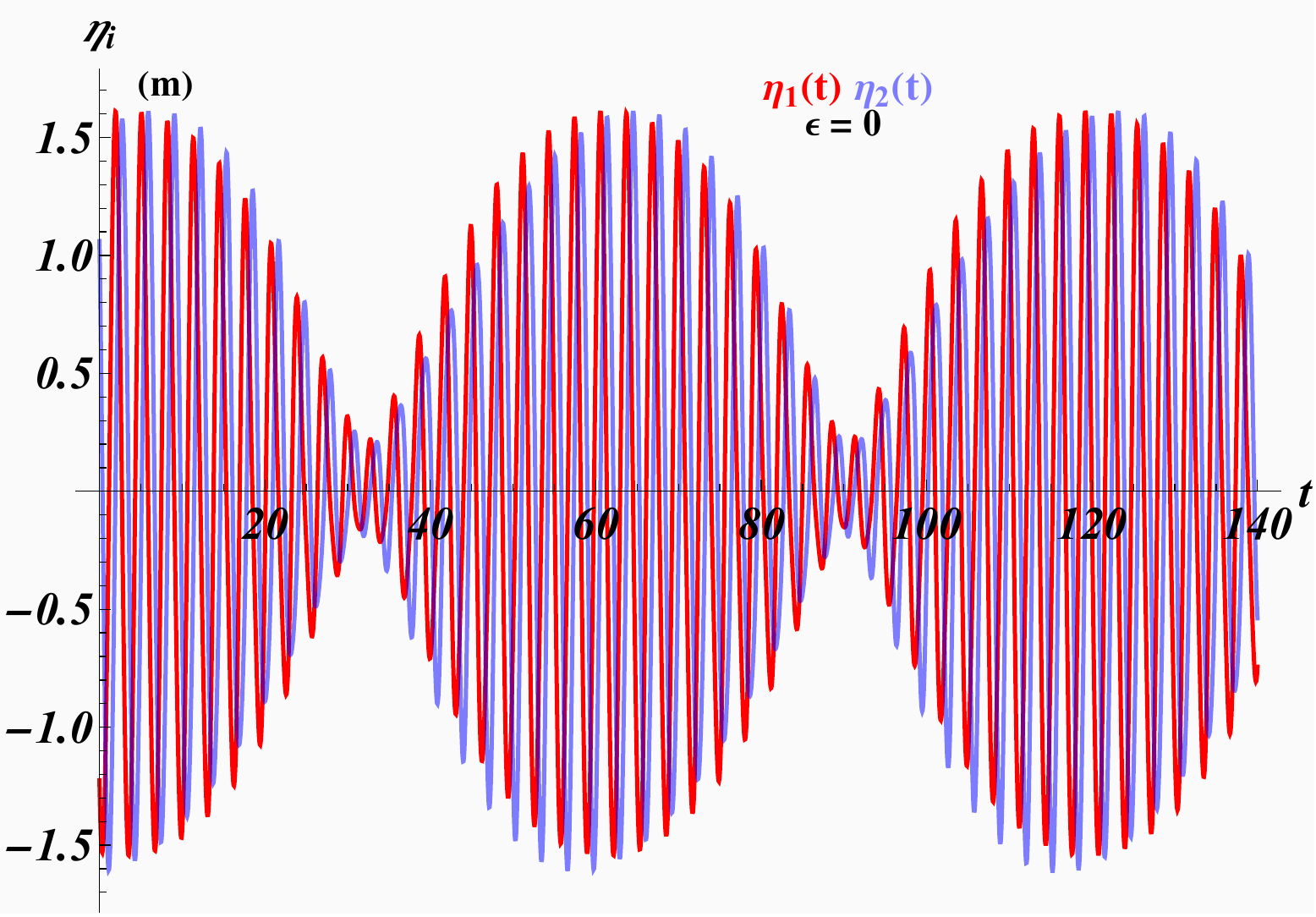}
\end{minipage}  
\begin{minipage}[b]{0.19\textwidth}              
	\includegraphics[width=\textwidth]{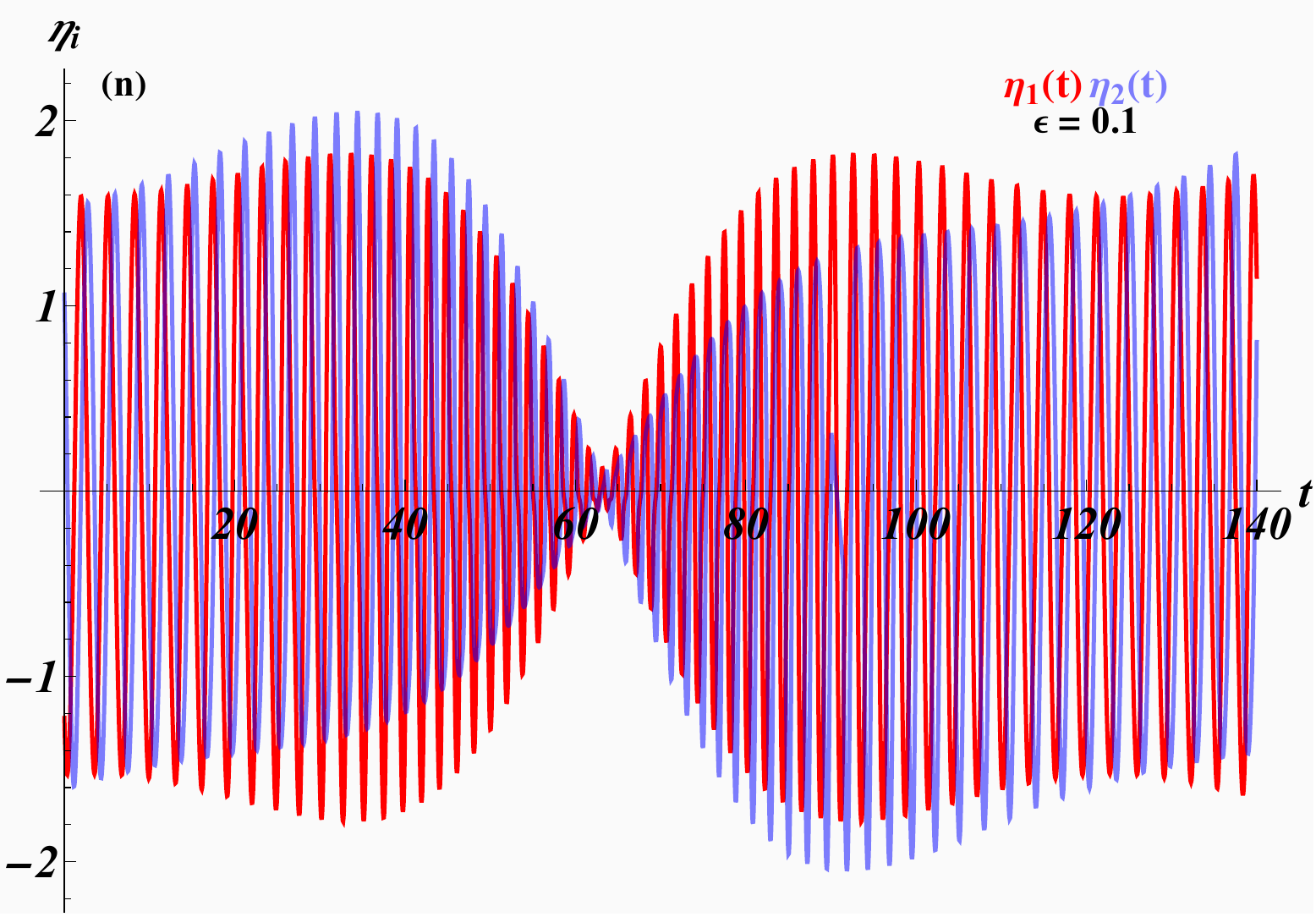}
\end{minipage}  
\begin{minipage}[b]{0.19\textwidth}              
	\includegraphics[width=\textwidth]{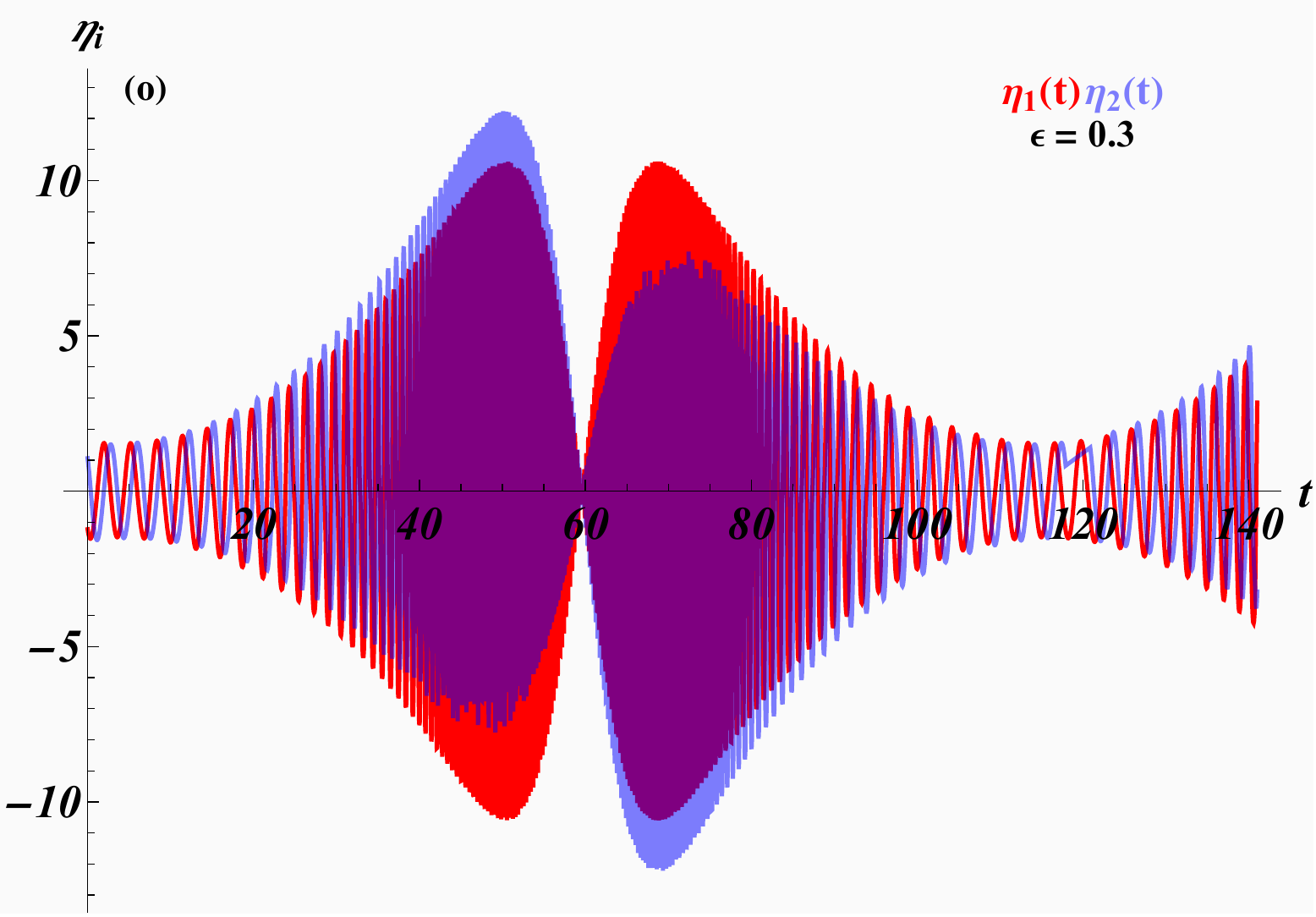}
\end{minipage}      
	\begin{minipage}[b]{0.19\textwidth}              
	\includegraphics[width=\textwidth]{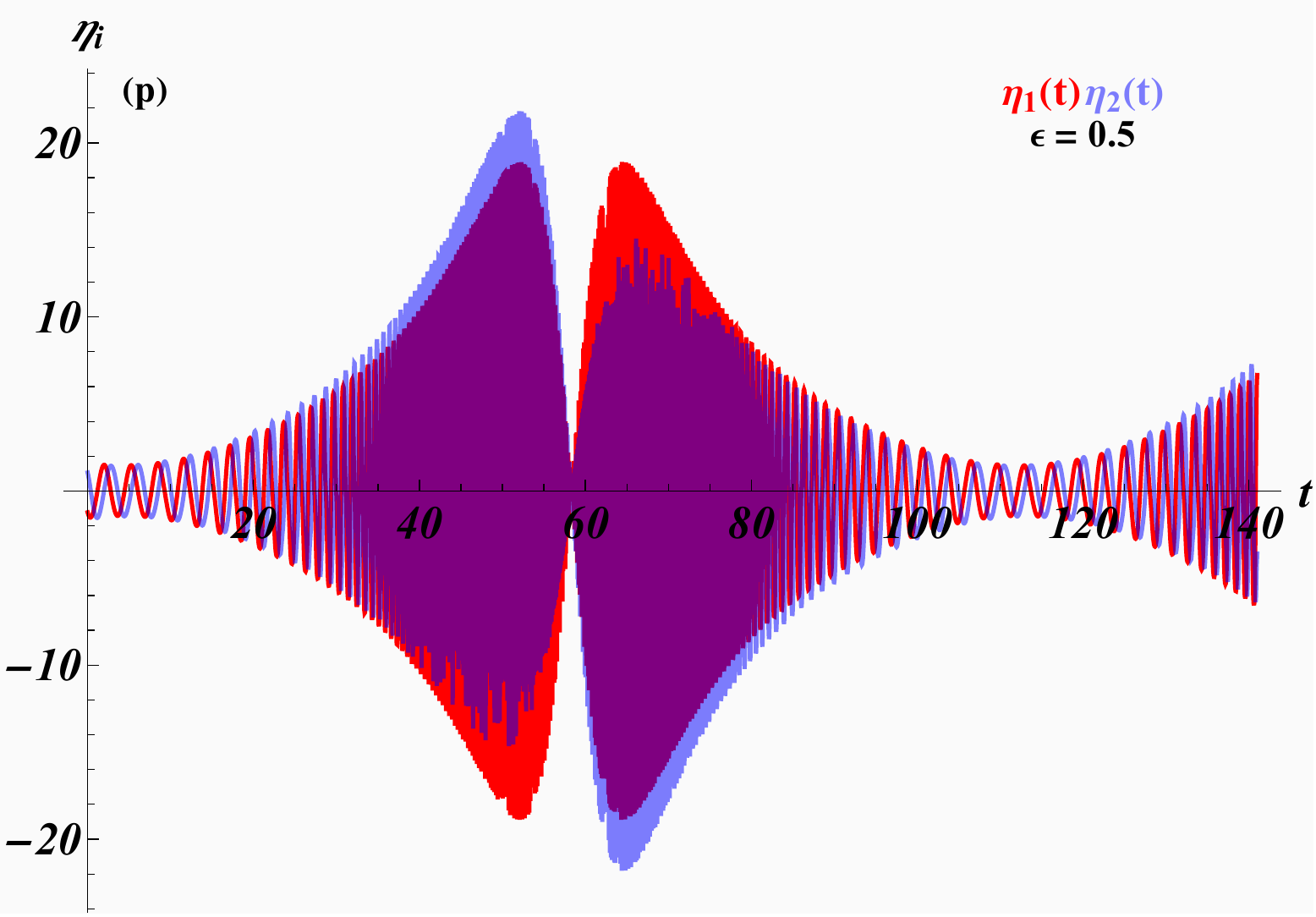}
\end{minipage}  
\begin{minipage}[b]{0.19\textwidth}              
	\includegraphics[width=\textwidth]{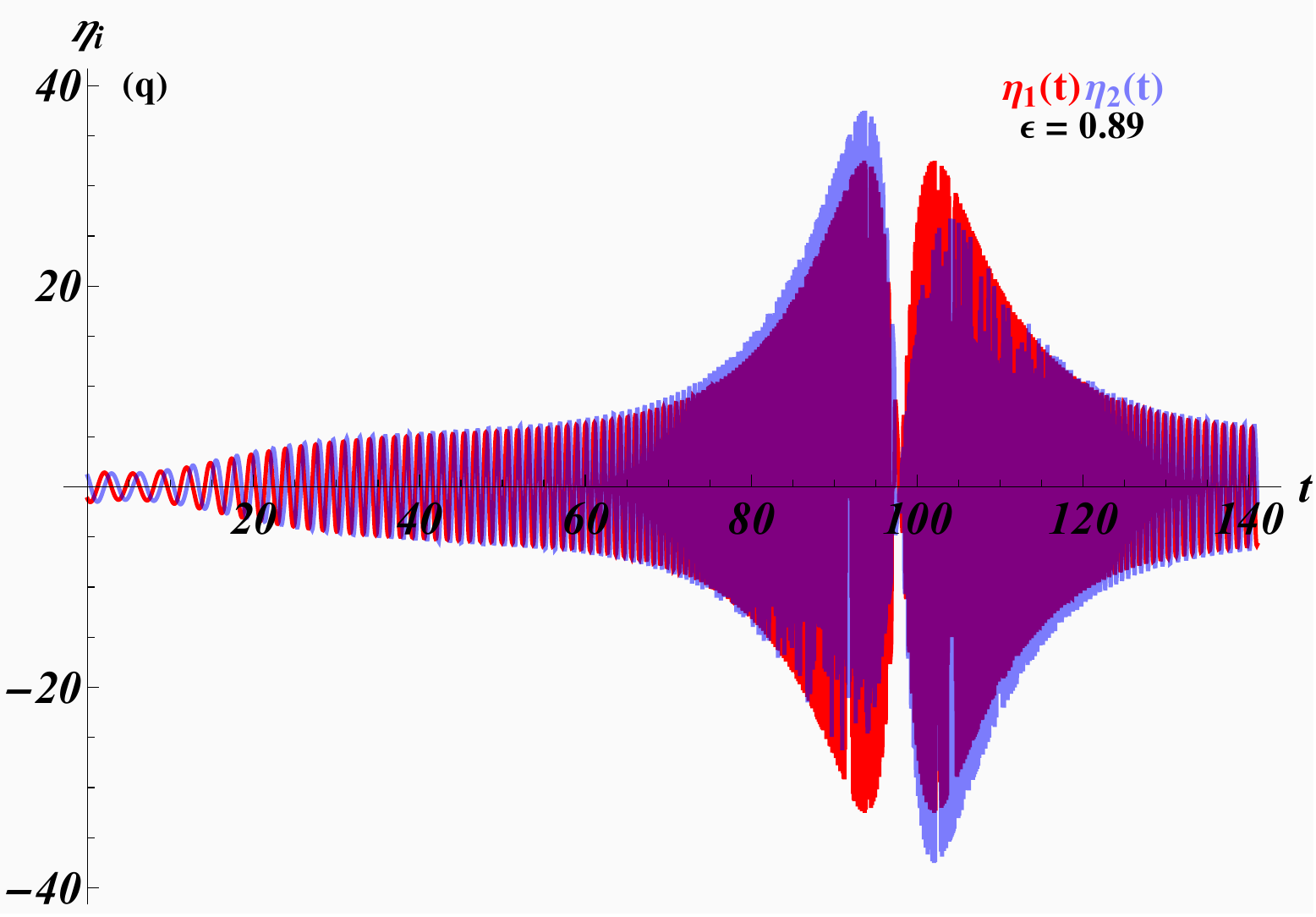}
\end{minipage}  
\begin{minipage}[b]{0.19\textwidth}              
	\includegraphics[width=\textwidth]{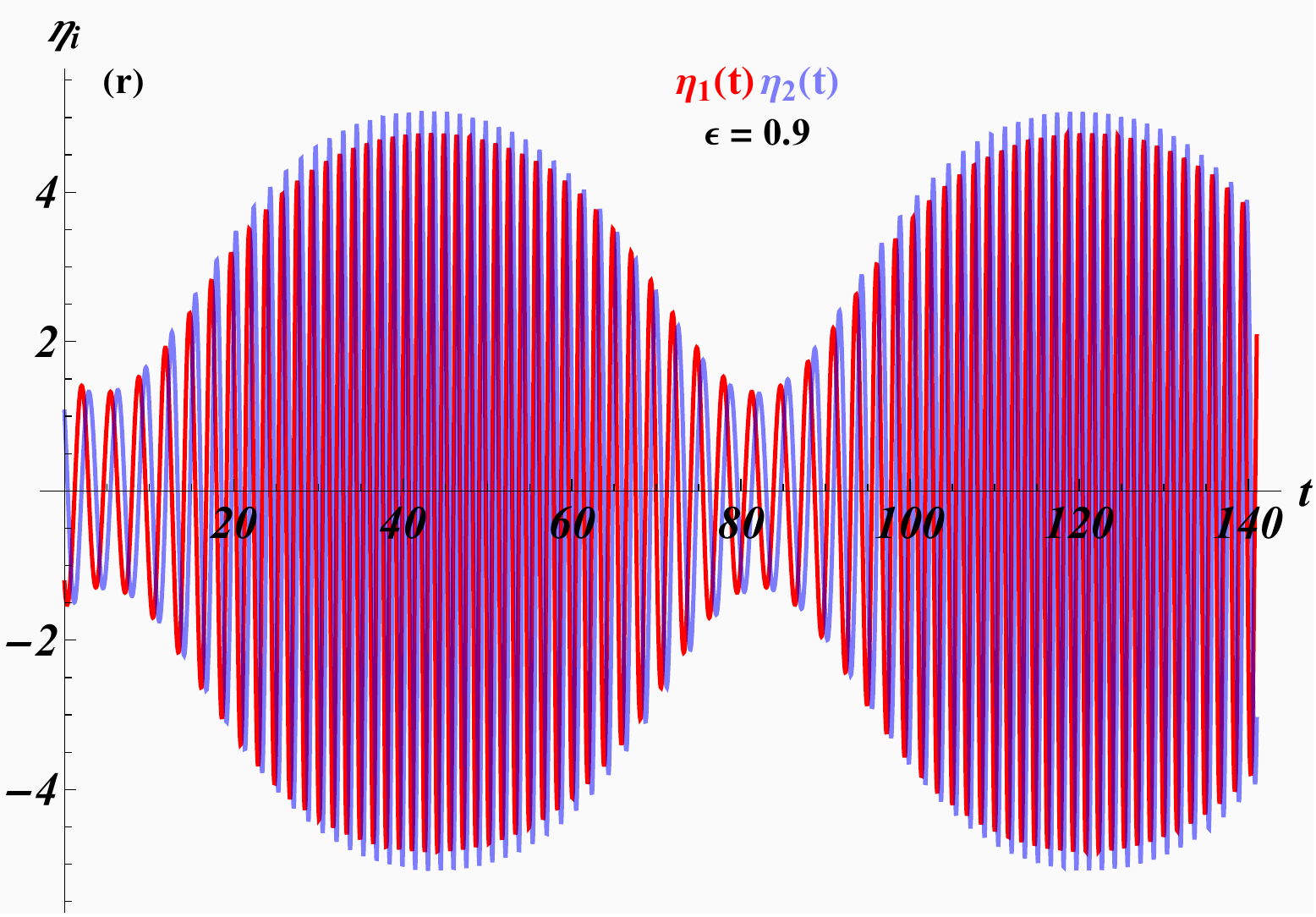}
\end{minipage}  
\begin{minipage}[b]{0.19\textwidth}              
	\includegraphics[width=\textwidth]{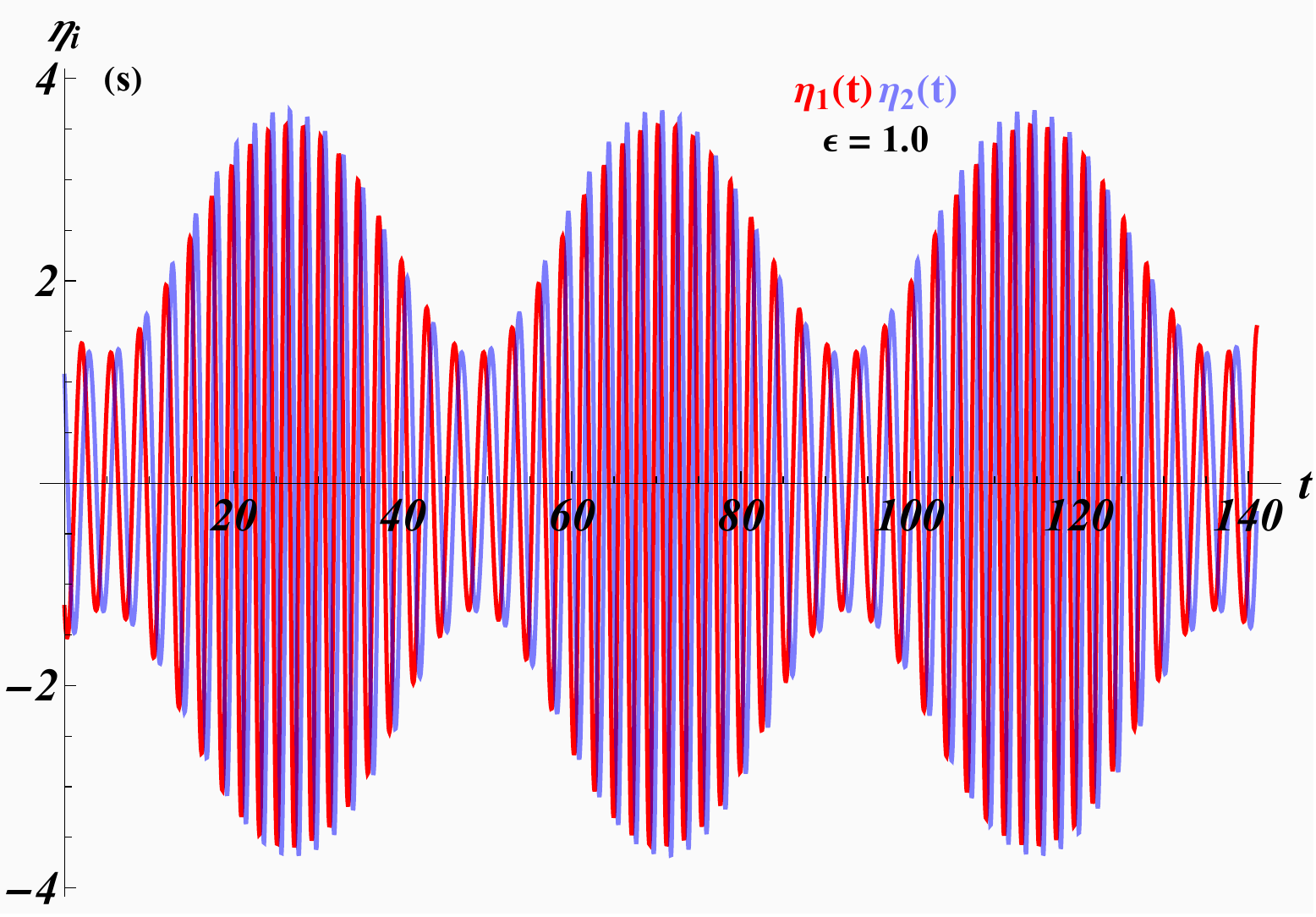}
\end{minipage}  
\begin{minipage}[b]{0.19\textwidth}              
	\includegraphics[width=\textwidth]{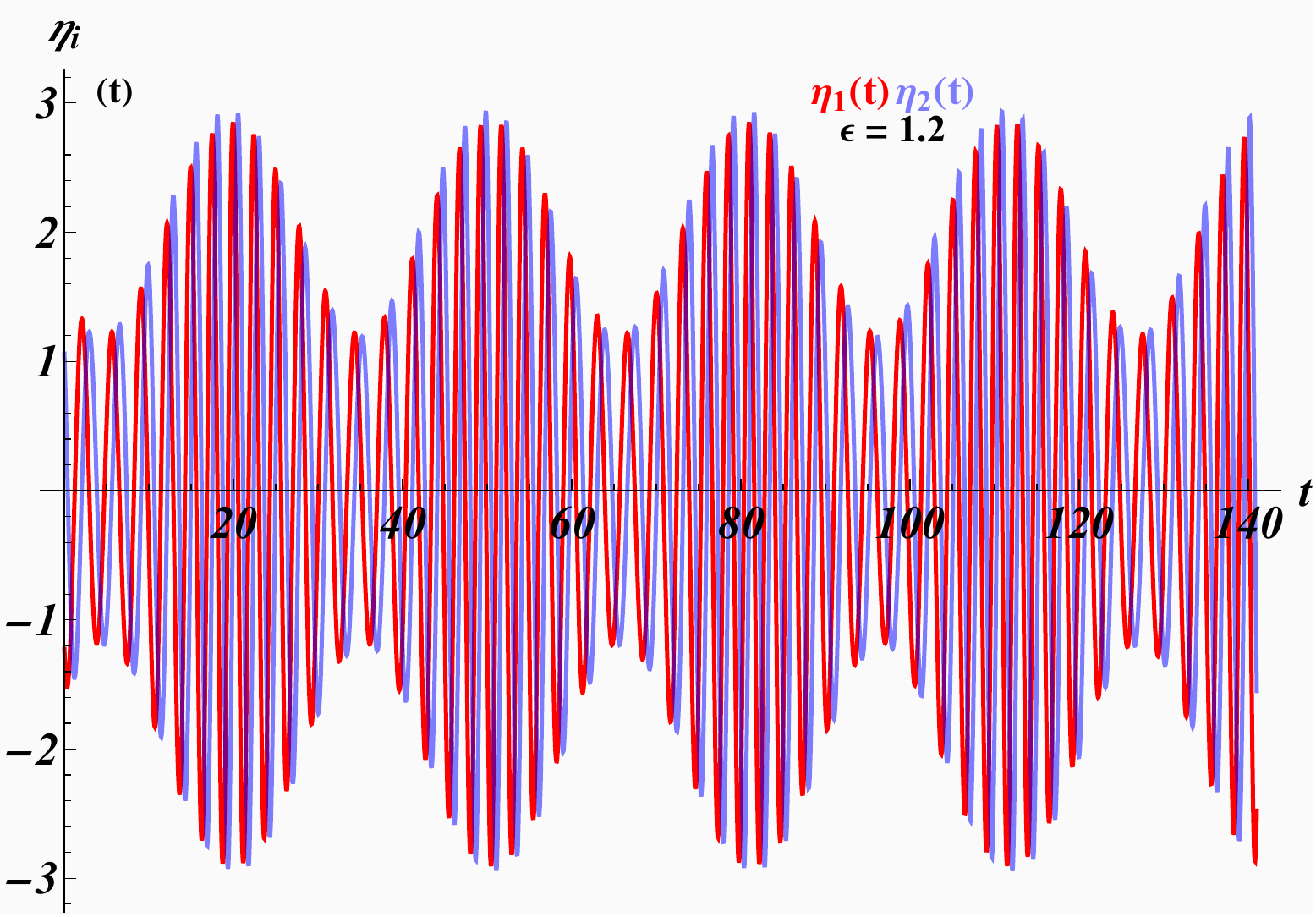}
\end{minipage}    
	\caption{Two-dimensional phase space components $\zeta_i$ and $\eta_i$ with $i=1,2$ for the deformed affine $A_2$-Toda lattice charge Hamiltonian $\tilde{Q}_3$ in (\ref{pertq33}) as functions of time $t$ for different values of $\epsilon$. The initial conditions are taken in all cases as $\zeta_1=\zeta_2=0$, $\eta_1=-3/\sqrt{6}$ and $\eta_2=3/2\sqrt{2}$.  } 
	\label{PerturbationA2}
\end{figure}

There are of course many more options to perturb the higher derivative charge Hamiltonians by adding small terms to them or even by deforming with additional terms of the same or larger magnitude. Here we only present one example to illustrate the general feature and to establish the robustness of some of the higher charge Hamiltonian trajectories. We leave a more systematic presentation to future investigations \cite{AFBTinprep}. We consider a Hamiltonian in form of a soft deformation of the $Q_3$-charge in the $A_2$-theory in the two-dimensional representation (\ref{A22cons}) by adding a two-dimensional harmonic oscillator potential
\begin{equation}
	 Q_3^p\left(\zeta_1, \zeta_2, \eta_1, \eta_2 \right) = Q_3\left( \zeta_1, \zeta_2, \eta_1, \eta_2 \right) + \epsilon \frac{1}{2} \left( \zeta_1^2 +\zeta_2^2 \right) . \label{pertq33}
\end{equation}
The solutions of the corresponding equations of motion  $\zeta_i(t)$, $\eta_i(t)$ with $i=1,2$ as functions of time are depicted in figure \ref{PerturbationA2} for several typical values of $\epsilon$. We still find the previously observed superposition of frequencies which vary with $\epsilon$. A transition between rather different types of qualitative behaviours is seen at $\epsilon_c \approx 0.8944$. When approaching this value from below the maximal values for $\zeta_i(t)$ and $\eta_i(t)$ increase smoothly and the functions become more localised, but once  $\epsilon_c$ is passed these values drop significantly and oscillations re-occur. At this point we do not have proper explanation of this behaviour.   

Naturally there are many more options to break the integrability of these systems \cite{AFBTinprep}.

\section{Conclusions}	

We investigated many more examples that support the conjecture, originally put forward by Smilga \cite{smilga2021benign}, that charges of integrable systems provide very promising candidates for higher derivative theories that possess benign ghost sectors in part or all of their parameter space. For our examples of affine Toda lattice theories associated to different algebras we found a multitude of possible scenarios. We demonstrated that proper choice of the dimension of the representation space and the choice of the initial conditions are crucial. Especially for the theories that used roots in their formulation represented in the same dimension as the rank of the underlying algebras we found solutions of oscillatory and benign oscillatory behaviour for higher charge Hamiltonian theories. We found no distinction between theories that are unbounded from below in the quantum theory, and therefore possess ghosts in their spectrum, and those that lack an odd parity symmetry with no ghost sectors. 

Moreover, these solutions were quite robust with regard to perturbations of the choice of the initial conditions. When deviating from this setup and using higher dimensional representations for the roots we found the coordinate solutions diverge as functions of time $t$ for all the $A_n$ examples investigated when the initial condition were taken differently from $Q_n=0$, due to the Poisson bracket relation (\ref{Poisson}). However, for the non-simply laced algebras $G_2$ and $B_3$ the trajectories for the higher charge Hamiltonians remained oscillatory even in the higher dimensional cases, with the difference that the former where very sensitive to changes of the initial conditions whereas the latter turned out to be stable, once again due to the Poisson bracket relation between the centre-of-mass coordinate and the charges. Based on the data generated here so far it is too early to extract more generic features that might be shared by some class of systems, e.g. simply laced versus non-simply laced etc. We leave these aspects for future investigations. 

We have also investigated some more extreme deformations of the integrable systems by a soft harmonic oscillator potential in section 5.2. As noted previously for different types of perturbations \cite{smilga2021benign} many of the benign trajectories found maintain this feature, but we also observed a critical point in the strength $\epsilon$ of these additional terms with an extreme sensitivity regarding the characteristic behaviour of the phase space trajectories. It seems worth to carry out some systematic investigations that clarify which type of perturbations, and even deformations, might be permitted in order to maintain the benign nature of the solutions \cite{AFBTinprep}.

Our results are summarised in table 1. 

\begin{center}
\begin{tabular}{l||l|l|l|l|l|l|l|l|l|}
            $Q$ (D of rep) $\backslash$ $\bf{g}$       & $A_2$ & $G_2$ & $B_3$ & $A_6$ & $A_2,p_1$ & $G_2,p_1$ & $B_3,p_1$ & $A_6,p_1$ & $A_2,p2$ \\ \hline \hline
       $H\, (r+1)$  & o     &   o   &  o   &   o    & d     &  d   &  o    &   d  &  \\
       $H\, (r)$  & o     &   o   &   o   &   o    & o     &  o   &   o   &   o  &  \\
       $Q_3 \, (r+1)$  & bd     &  $\times$   &   $\times$   &   bd    & bd     &  $\times$   &   $\times$   &   bd  &  \\
       $Q_3 \, (r)$  & bo     &   $\times$   &   $\times$   &   bo    & bo     &   $\times$   &   $\times$   &   bo  &  bo \\
       $Q_4 \, (r+1)$  &      & $\times$   &  o    &   d    &      & $\times$   &  o    &   d  &  \\
       $Q_4 \, (r)$  &      &  $\times$   &   d   &   d    &      &  $\times$   &   d   &   d  &  \\
        $Q_6 \, (r+1)$  &      & d   &  d    &   d    &      & d   &   o   &   d  &  \\
       $Q_6 \, (r)$  &      &  o   &  o   &   o    &      &  o   &   o   &   o  &   \\ 
\end{tabular}
\end{center}
\noindent {\bf {Table 1:} } Summary of results. 	
(o $\equiv$ oscillatory, d $\equiv$ divergent, bo $\equiv$ benign oscillatory, bd $\equiv$ benign divergent, $\times$ charge does not exist, r $\equiv$ rank of $\bf{g}$,  $p_1$ $\equiv$ perturbation with $Q_1(0) \neq 0$,  $p_2$ $\equiv$ perturbation with harmonic oscillator potential)

\medskip

\noindent \textbf{Acknowledgments:} BT is supported by a City, University of London Research Fellowship. AF thanks the Instituto de Ciencias F{\'{\i}}sicas y Matem{\'{a}}ticas of the 
Universidad Austral de Chile, where part of this work was completed for kind hospitality and Francisco Correa for financial support. AF would like to thank Andrei Smilga for useful discussions.

\newif\ifabfull\abfulltrue


\end{document}